\documentclass[aps,prd, notitlepage, onecolumn,superscriptaddress,floatfix,letterpaper,nofootinbib, longbibliography]{revtex4-1}

\usepackage{amsmath, amsthm, amssymb,slashed,mathtools,tabu}
\usepackage{pifont}

\usepackage[usenames, dvipsnames]{color}
\usepackage[svgnames]{xcolor}
\usepackage[colorlinks,citecolor=RoyalBlue, urlcolor=RoyalBlue, linkcolor=RoyalBlue]{hyperref} 

\usepackage[normalem]{ulem}

\newcommand{\cred}[1]{\textcolor{black}{#1}}

\usepackage{booktabs}

\definecolor{mygray}{gray}{0.6}

\usepackage{upgreek}
\usepackage{bbm}

\newenvironment{myfont}[2][]{\csname#2\endcsname[#1]}{}

\usepackage{slashed}
\usepackage[makeroom]{cancel}
\usepackage[normalem]{ulem}
\usepackage{soul}
\newcommand{\stkout}[1]{\ifmmode\text{\sout{\ensuremath{#1}}}\else\sout{#1}\fi}

\usepackage{sseq}
\usepackage[all,cmtip]{xy}
\usepackage{tikz-cd}
\usepackage{tikz}
\usetikzlibrary{matrix}
\usetikzlibrary{decorations.markings}
\usetikzlibrary{tikzmark,decorations.pathreplacing,positioning}
\usepackage{amsfonts}

\newcommand{\bea}{\begin{eqnarray}}
\newcommand{\eea}{\end{eqnarray}}
\def\be{\begin{equation}}
\def\ee{\end{equation}}

\newcommand{\e}{\hspace{1pt}\mathrm{e}}

\newcommand{\ii}{\hspace{1pt}\mathrm{i}\hspace{1pt}}

\definecolor{red}{rgb}{1,0,0}
\definecolor{blue}{rgb}{0,0,1}
\definecolor{dblue}{rgb}{0,0,0.4}
\definecolor{green}{rgb}{0,1,0}
\definecolor{black}{rgb}{0,0,0}
\definecolor{white}{rgb}{1,1,1}

\definecolor{brn}{rgb}{.8,.4,.0}
\definecolor{redo}{rgb}{1,.5,.0}
\definecolor{ddgrn}{rgb}{0,0.4,0}
\definecolor{dgrn}{rgb}{0,0.55,0}
\definecolor{dbl}{rgb}{0,0,0.5}

\usepackage[bbgreekl]{mathbbol}
\usepackage{amscd}

\newcommand{\Z}{\mathbb{Z}}

\newcommand{\R}{\mathbb{R}}

\newcommand{\dd}{\hspace{1pt}\mathrm{d}}
\newcommand{\<}{\langle} 
\renewcommand{\>}{\rangle} 

\newcommand{\Refe}[1]{Ref.~[\onlinecite{#1}]}
\newcommand{\Eq}[1]{Eq.~(\ref{#1})} 
\newcommand{\eq}[1]{(\ref{#1})}

\newcommand{\Tr}{{\rm Tr}}

\newcommand{\prt}{\partial}

\newcommand{\bpm}{\begin{pmatrix}}
\newcommand{\epm}{\end{pmatrix}}
\newcommand{\bmm}{\begin{matrix}}
\newcommand{\emm}{\end{matrix}}

\newcommand{\cD}{ {\cal D} }

\newcommand{\cL}{ {\cal L} } 
\newcommand{\cP}{ {\cal P} }

\newcommand{\cT}{ {\cal T} }

\newcommand{\al}{\alpha} 
\newcommand{\bt}{\beta}

\def\CM{{\cal M}}
\def\cM{{\cal M}}

\def\CO{{\cal O}}

\def\Z{{\mathbb{Z}}}

\def\R{{\mathbb{R}}}

\def\Tr{{\mathrm{Tr}}}

\def \Z{\mathbb{Z}}

\newcommand {\emptycomment}[1]{}

\def\TP{\mathrm{TP}}

\newcommand{\Spin}{{\rm Spin}}
\newcommand{\U}{{\rm U}}
\newcommand{\SU}{{\rm SU}}

\usepackage{centernot}
\newcommand{\nn}{{\nonumber}}
\newcommand{\Sec}[1]{Sec.~\ref{#1}}

\newcommand{\diag}{{\rm diag}}
\usepackage{enumitem} 
\usepackage{mathtools,amssymb,varwidth}

\newcommand{\Fig}[1]{Fig.~\ref{#1}} 
\newcommand{\Table}[1]{Table \ref{#1}}

\usepackage{datetime}

\usepackage{comment}

\newcommand{\rP}{{\rm P}}
\newcommand{\rT}{{\rm T}}
\newcommand{\rR}{{\rm R}}
\newcommand{\rF}{{\rm F}}

\newcommand{\rE}{{\rm E}}

\def\rI{\mathrm{I}}

\newcommand{\SMG}{\text{SMG}}

\begin{document}


\title{CT or P Problem and Symmetric Gapped 
Fermion Solution 
}

\author{Juven Wang}
\email[]{jw@cmsa.fas.harvard.edu}

\affiliation{Center of Mathematical Sciences and Applications, Harvard University, MA 02138, USA}

\begin{abstract} 

An analogous ``Strong CP problem''
 is identified in a toy model in two dimensional spacetime: 
 a general 1+1d abelian U(1) anomaly-free chiral fermion and chiral gauge theory
with a generic theta instanton term $\frac{\theta}{2 \pi} \int F$.
The theta term alone violates the charge-conjugation-time-reversal CT and the parity P discrete symmetries.
The analogous puzzle here is the CT or P problem in 1+1d: 
Why can the $\bar{\theta}$ angle (including the effect of $\theta$ and the complex phase of a mass matrix)
be zero or small for a natural reason?
We show that this CT or P problem can be solved by a Symmetric Mass Generation mechanism 
(SMG, namely generating a mass or energy gap while preserving an anomaly-free symmetry).
This 1+1d toy model mimics several features of the 3+1d Standard Model: chiral matter content, confinement, and 
Anderson-Higgs-induced mass by Yukawa-Higgs term. 
One solution replaces some chiral fermion's Higgs-induced mean-field mass with SMG-induced non-mean-field mass.
Another solution enriches this toy model by introducing several new physics beyond the Standard Model: 
a parity-reflection PR discrete symmetry maps between the chiral and mirror fermions 
as fermion doubling localized on two domain walls at high energy, and 
SMG dynamically generates mass to the mirror fermion while still preserving the anomaly-free chiral symmetry 
at an intermediate energy scale,
much before the Higgs mechanism generates mass to the chiral fermion at lower energy.
Without loss of generality, an arguably simplest 1+1d U(1) symmetric 
anomaly-free chiral fermion/gauge theory (e.g., Weyl fermions with $3_L$-$4_L$-$5_R$-$0_R$ U(1) charges)
is demonstrated in detail. As an analogy to the superfluid-insulator {or order-disorder} quantum phase transition,
in contrast to the conventional Peccei-Quinn solution sitting {in} the (quasi-long-range-order) superfluid-like {ordered} phase,
our solution is {in} the SMG insulator {disordered} phase.




\end{abstract}



\maketitle
\tableofcontents

\newpage
\section{Introduction, the Problem, and the Summary}

One of the open puzzles in the 3+1d Standard Model (SM) 
is the strong CP problem \cite{Smith1957ht, Bakerhepex0602020, Abel2020PRL2001.11966, Dine2000TASI0011376, Hook2018TASI1812.02669}.
The corresponding strong gauge force SU(3) should naturally allow
a theta $\theta_3$ term inserted into the SM path integral with a weight factor
$\exp(\ii \theta_3 n^{(3)})$,
with the appropriately quantized instanton number
$n^{(3)} \equiv \frac{1}{8 \pi^2} \int \Tr[F^{(3)} \wedge F^{(3)}] \in \Z$ 
\cite{BelavinBPST1975, tHooft1976instanton} 
of a nonabelian SU(3) field strength $F^{(3)}$.
The Cabibbo-Kobayashi-Maskawa (CKM) matrix CP violating angle $\delta_{\rm CP}$ is experimentally verified to be order 1.
N\"aively, due to Naturalness without fine-tuning \cite{tHooft1979ratanomaly}, the other CP violating theta angle 
\bea 
\label{eq:theta3}
{\bar{\theta}_{3}} 
\equiv
{\theta_{3}}
+\arg (\det 
(M_{u}
M_{d}))
\eea
shall also be generically order 1.
(Here $M_{u}$ and $M_{d}$ are
two rank-3 matrices specifying the Yukawa-Higgs coupling, 
for $u,c,t$-type quarks and $d,s,b$-type quarks respectively.)
A generic theoretical ${\bar{\theta}_{3}} \in [0, 2 \pi)$ can violate CP thus T (except at ${\bar{\theta}_{3}}=0$ or $\pi$),
but 
the surprising experimental fact 
measured by the real-world neutron electric dipole moment
shows that 
$| {\bar{\theta}_{3}} | < 10^{-10} \simeq 0$  
is nearly zero 
\cite{Smith1957ht, Bakerhepex0602020, Abel2020PRL2001.11966},
making the quantum chromodynamics (QCD) 
almost a CP and T invariant --- this puzzle is known as the strong CP problem.
Since there is no particular reason (not even an anthropic reason) 
for the SM to have ${\bar{\theta}_{3}} \simeq 0$,
the typical strong CP solutions
proposed in the past literature tend to
modify or enlarge the SM 
to include some assumptions: 
(1) some of quarks (e.g., up quark) is massless \cite{tHooft1976ripPRL},
(2) extra continuous U(1) symmetry, then based on dynamical arguments on the spontaneous symmetry breaking which relaxes to ${\bar{\theta}_{3}} \simeq 0$, 
e.g., Peccei-Quinn symmetry with axions \cite{PecceiQuinn1977hhPRL,PecceiQuinn1977urPRD, Weinberg1977ma1978PRL, Wilczek1977pjPRL}
(3) extra discrete P or CP symmetry imposed at a high energy 
\cite{Nelson1983zb1984PLB,Barr1984qx1984PRL, BabuMohapatra1989, BabuMohapatra1989rbPRD, BarrChangSenjanovic1991qxPRL}.
The purpose of this present work is to propose a new type of CP problem's solution 
by involving the Symmetric Mass Generation (SMG) mechanism \cite{WangYou2204.14271}
--- fermions can become massive (also known as gapped) by a symmetric deformation from a massless (aka gapless) theory, 
without involving any symmetry breaking within an anomaly-free symmetry group.
In the context of the global symmetry $G$ being anomaly-free, 
this condition is well-known as the 't Hooft anomaly free in $G$ \cite{tHooft1979ratanomaly}.

In this work, we investigate the analogous problem 
in a 1+1d toy model to mimic the strong CP problem in the 3+1d SM.
Our toy model is a chiral fermion theory coupled to U(1) gauge field, as 
a chiral gauge theory version of the modified Schwinger's 1+1d quantum electrodynamics (QED) \cite{Schwinger1962PR2}. 
Then we provide a solution to this problem in 1+1d.
(In a companion work  \cite{StrongCPtoappear-2212.14036}, 
we will propose the similar SMG solution to the Strong CP problem for the 3+1d models, including the SM.)
The plan of this article goes as follows.

First,  in \Sec{sec:problem}, we define the analogous problem called
CT or P problem in this 1+1d toy model with chiral fermions and with a seemly generic nonzero $\theta$ term, 
written as  $\frac{\theta}{2 \pi} \int F$ for some abelian field strength $F$.
Under the standard unitary C, P, and anti-unitary T active transformations on the fields (e.g., \cite{WangCPT2109.15320}), 
the two components of 1+1d abelian gauge field $A=(A_0,A_1)$ and its field strength
$F_{01}$ at the spacetime coordinates $x^\mu=(x^0,x^1)=(t,x)$
are transformed as
\bea
\Z_2^{\rm C} &:& A_0 (t,x) \mapsto -A_0 (t,x), \quad A_1 (t,x)   \mapsto -A_1 (t,x), \quad F_{01}(t,x) \mapsto - F_{01}(t,x).\nn \\ 
\label{eq:CPT}
\Z_2^{\rm P} &:& A_0 (t,x) \mapsto + A_0 (t,-x), \quad A_1 (t,x)   \mapsto -A_1 (t,-x), \quad F_{01}(t,x) \mapsto - F_{01}(t,-x).\\ 
\Z_2^{\rm T} &:& A_0 (t,x) \mapsto + A_0 (-t,x), \quad A_1 (t,x)   \mapsto -A_1 (-t,x), \quad F_{01}(t,x) \mapsto + F_{01}(-t,x),
\quad \ii \mapsto - \ii.\nn
\eea
The only nonzero field strength $F_{01}=-F_{01} = \prt_0 A_1 - \prt_1 A_0 \equiv E_1$ is the electric field along the spatial $x$ direction.
The electric field $E_1$ is C odd, P odd, and T even.
So the $\frac{\theta}{2 \pi} \int F$ violates both CT and P for a generic $\theta \in [0, 2 \pi)$ within the $2 \pi$ periodicity, 
unless at the special $\theta=0$ or $\pi$ that preserves CT and P.\footnote{In 1+1d, the theta term of abelian field strength $\frac{\theta}{2 \pi} \int F$ is
C odd, P odd, and T even. In contrast to 3+1d, the theta term of non-abelian Yang-Mills field strength $\frac{\theta}{8 \pi^2} \int \Tr[F \wedge F]$
 is C even, P odd, and T odd. 
We could say that the small 1+1d theta term implies the C, P, CT, or PT problem.
We could also say that the small 3+1d theta term implies the P, T, CP, or CT problem; typically it is called the Strong CP problem for the SU(3) strong force.
 }
Hence we shall call this a CT problem or P problem in 1+1d:
How can the analogous 1+1d $\bar\theta$ angle be zero or small 
for a natural reason?

\newpage 

{The conceptual idea of our solution focus on the re-examination and re-interpretation of the role of the mass matrix $M$ in \eq{eq:theta3}.
We point out that the previous studies and solutions to the P problem in the past literature 
only or mainly rely on the mean-field mass such that mass matrix $M$ is obtained via the 
mean-field expectation value $\< M\>$. 
So what \eq{eq:theta3} really means schematically is 
\bea 
\label{eq:theta}
{\bar{\theta}_{}} 
\equiv
{\theta_{}}
+\arg (\det 
\< M\>).
\eea
In other words, the experimental measurement of the ${\bar{\theta}_{}}$ (such as the neutron electric dipole moment in the SM)
really involves the mean-field mass matrix $\< M\>$.
There the mass matrix $M$ is schematically obtained from the fermion bilinear or quadratic term in the Lagrangian 
\bea
\label{eq:mass-term}
\xi_{\rm I} M_{\rm IJ} \psi_{\rm J} + {\rm h.c.},
\eea 
where $\xi$ and $\psi$ are fermion fields in appropriate representations. The $M_{\rm IJ}$ may receive a contribution from the dynamical fields like Higgs $\phi_H$, etc.,
such that some $\< M_{\rm IJ}\>\neq 0$.
The whole $\xi_{\rm I} M_{\rm IJ} \psi_{\rm J}$ is a singlet scalar in a trivial representation of the Lorentz group.}

{Now, our key new idea is that SMG mechanism \cite{WangYou2204.14271} can go beyond the mean-field mass,
such that the SMG deformation
\bea \label{eq:SMG-mass}
\xi_{\rm I} \CO_{\rm SMG, IJ} \psi_{\rm J} + {\rm h.c.},
\eea 
receives no mean-field value in $\< \CO_{\rm SMG, IJ}\>=0$ but this \eq{eq:SMG-mass} can still symmetrically give 
\emph{non-mean-field mass} energy gaps to the full set of fermions by preserving an anomaly-free symmetry $G$. 
The \eq{eq:SMG-mass} may involve the \emph{multi-fermion interaction} and \emph{disordered mass field interaction}, beyond the quadratic fermion interaction.
The reason that $\< \CO_{\rm SMG, IJ}\>=0$ is due to that the condensation of these $\CO_{\rm SMG, IJ}$ operators 
would often break the $G$ symmetry (so nonzero $\< \CO_{\rm SMG, IJ}\>$ often implies no SMG). In our work, we shall generalize the mass matrix 
in \eq{eq:theta} and \eq{eq:mass-term} to $\mathbf{M}$ so
to include both the \emph{mean-field} mass $M=\< M \>$ (e.g., from Higgs) and the SMG's \emph{non-mean-field} mass $\< \CO_{\rm SMG, IJ}\>=0$: 
\bea
\mathbf{M}_{\rm IJ} &=&M_{\rm IJ}+\CO_{\rm SMG, IJ},\cr
\<\mathbf{M}_{\rm IJ} \>&=& \< M_{\rm IJ} \> , \quad \text{ while } \<\CO_{\rm SMG, IJ}\>=0,\cr
\xi_{\rm I} \mathbf{M}_{\rm IJ} \psi_{\rm J} + {\rm h.c.}  &=& \xi_{\rm I} (M_{\rm IJ} + \CO_{\rm SMG, IJ}) \psi_{\rm J} + {\rm h.c.}, \cr 
{\bar{\theta}_{}} 
&\equiv&
{\theta_{}}
+\arg (\det 
\< \mathbf{M}\>)
=
{\theta_{}}
+\arg (\det 
\< M\>).
\eea 
Then our solution to the CT or P problem requires at least \emph{any one} of the fermions (call this fermion $\zeta$) in the full theory  
to receive no mean-field mass at all (so there is at least one zero eigenvalue of the mean-field mass matrix $\< M_{\rm IJ} \>$) 
but can still be massive or gapped due to the SMG contribution
($\CO_{\rm SMG, IJ}\neq 0$ but $\<\CO_{\rm SMG, IJ}\>=0$).
Thus at least one of the eigenvalues of the mean-field mass matrix $\< M\>$ being zero
implies that the $\det \<\mathbf{M} \>= \det \< M\> =0$.
In that case,
${\bar{\theta}_{}} 
={\theta_{}}$ and next we can do the chiral transformation only on this specific fermion $\zeta \to \e^{\ii \upalpha } \zeta$ that has no mean-field mass.
This chiral transformation will send ${\theta_{}} \mapsto {\theta_{}} - \upalpha$,
while this 
also sends
$M_{\rm IJ} + \CO_{\rm SMG, IJ} \mapsto M_{\rm IJ}(\upalpha) + \CO_{\rm SMG, IJ}(\upalpha)$ presumably with $\upalpha$ dependence.
However, the $\< M_{\rm IJ}\>=\<M_{\rm IJ}(\upalpha)\>$ is not changed because the $\zeta$ has no mean-field mass.
The mean-field $\<\CO_{\rm SMG, IJ}(\upalpha)\>=0$ anyway so it does not contribute to the ${\bar{\theta}}$.
So we end up redefining ${\bar{\theta}_{}}$ by a chiral transformation, 
with $\det \<\mathbf{M} (\upalpha) \>=0$ still, 
so ${\bar{\theta}_{}}= {\theta_{}} - \upalpha =0$ can be appropriately chosen to be zero.
This provides the solution of the CT or P problem: The ${\bar{\theta}_{}}$ is zero for the entire theory.
The ${\bar{\theta}_{}}=0$ is in principle solved at some energy scale, then we will provide arguments how ${\bar{\theta}_{}}$ remains zero or small at the IR low energy theory.
}

{\bf Order to Disorder the ${\theta_{}}$ and mass field}:
{There is another interpretation to look at our solution setting ${\bar{\theta}_{}}=0$. 
We are looking at the \emph{disordered} phase of the dynamical ${\bar{\theta}_{}}$ 
(which includes the \emph{disordered} dynamical ${\theta_{}}$ and dynamical complex phase of mass field 
$\arg (\det \< \mathbf{M}\>)$ \cite{DisorderTheta}).
Instead, Peccei-Quinn solution with axions \cite{PecceiQuinn1977hhPRL,PecceiQuinn1977urPRD, Weinberg1977ma1978PRL, Wilczek1977pjPRL}
looked at the \emph{ordered} phase of the dynamical ${\bar{\theta}_{}}$ (i.e., the small fluctuation around the vacuum expectation value $\< {\bar{\theta}_{}}\>$ gives rise to axion mode). The \emph{ordered} phase to the \emph{disordered} phase of the ${\bar{\theta}_{}}$ is analogous to the (algebraic-)superfluid-to-insulator type of phase transition  \cite{Fisher1989zza}.\footnote{Even more precisely, the continuous Peccei-Quinn symmetry
would \emph{not} be a global symmetry once the internal symmetry of the gauge group $G$ is dynamically gauged. 
Due to the mixed anomaly between the $G$ and Peccei-Quinn symmetry,
the classical Peccei-Quinn symmetry is broken down by the Adler-Bell-Jackiw (ABJ) anomaly 
\cite{Adler1969gkABJ, Bell1969tsABJ} to its discrete subgroup.
So rigorously speaking, neither \emph{superfluid} nor \emph{algebraic superfluid} exists 
as Peccei-Quinn symmetry-breaking phase in the $G$ gauge theory.
Nonetheless, at least in the weakly gauge or the global symmetry limit of $G$,
we will explain the physical intuitive picture of the (algebraic-)superfluid to insulator transition analogy in \Sec{sec:Conclusion}.
}
When the ${\bar{\theta}_{}}$ is in the ordered phase, it makes sense to ask the value of $ \< \exp(\ii{\bar{\theta}_{}}) \>$, which determines the
orientation of the ${\bar{\theta}_{}}$-clock and how it affects the CT or P breaking.
However, when the ${\bar{\theta}_{}}$ is in the disordered phase, 
it makes no sense to extract the mean-field value of $\< \exp(\ii{\bar{\theta}_{}}) \>=0$, which says the
${\bar{\theta}_{}}$-clock is fully disoriented, with no CT or P breaking.
The \emph{disordered} phase of the dynamical ${\bar{\theta}_{}}$ also gives the non-mean-field mass gap to a set of fermions via the SMG. 
}

{We will implement this particular no-mean-field massive fermion $\zeta$ in two approaches.
In  \Sec{sec:Another-Solution}, we only need the original \emph{chiral fermion} theory, without adding any \emph{mirror fermion} sector.
In that case, some of the original \emph{chiral} fermion say $\zeta$ receives its full mass from the SMG, not from the Higgs condensation.
In \Sec{sec:Solution}, instead, we will choose the $\zeta$ as a set of \emph{mirror fermions} being fully gapped by SMG.
The \emph{mirror fermion} sector is the \emph{fermion doubling} of the original \emph{chiral fermion} theory.
These two approaches provide two different solutions to the CT or P problem.
Let us summarize in more detail below. 
}

In \Sec{sec:Another-Solution}, 
we provide {one solution to this 1+1d CT or P problem of the chiral fermion theory via replacing its gapped Higgs phase to the SMG phase}.
Namely, once we assume the gapped chiral fermion theory is not due to a mean-field mass or Higgs condensation transition, 
but instead due to a certain interacting SMG deformation,
then the $\theta$ angle can be absorbed to the SMG interaction terms, thus the $\bar\theta$ angle is zero.
{In general, we can also have a mixed scenario such that both the Higgs condensation and SMG contribute to the mass gap
of the set of fermions. In the mixed scenario, as long as there is any one of the fermions gaining its full mass gap only from \emph{non-mean-field} SMG 
but without any \emph{mean-field} Higgs contribution 
(while the remaining fermions can gain mass from both Higgs condensation and SMG), then the full theory can still have
$\bar\theta=0$.}

In \Sec{sec:Solution}, we provide another one of CT or P solutions to the 1+1d toy model to argue why $\bar\theta \simeq 0$
by introducing the mirror fermion doubling to the chiral fermion theory, and gapping the mirror fermion by SMG \cite{WangYou2204.14271}.
%
%
In particular, here our 1+1d P solution consists of the following inputs:\\
1. {\bf Fermion doubling} at high energy above an energy scale $\rE > \Lambda_{\rm SMG}$, 
due to {Nielsen-Ninomiya (NN)} argument  \cite{NielsenNinomiya1981hkPLB}.
Our model contains a chiral fermion sector and a mirror fermion sector on two 1+1d branes (or two domain walls) separated by a finite-width 2+1d bulk,
imposed at the high energy.\\
2. {\bf Parity-reflection symmetry} $\Z_2^{\rm PR}$ is imposed at a high energy $\rE > \Lambda_{\rm SMG}$, 
reminiscent of the {parity solution} of the strong CP problem in 3+1d 
\cite{BabuMohapatra1989, BabuMohapatra1989rbPRD, BarrChangSenjanovic1991qxPRL, CraigGarciaKoszegiMcCune2012.13416}.
Both chiral and mirror fermion sectors are mapped to each other under the $\Z_2^{\rm PR}$ symmetry. However, our approach is 
still \emph{different} from the parity solution 
\cite{BabuMohapatra1989, BabuMohapatra1989rbPRD, BarrChangSenjanovic1991qxPRL, CraigGarciaKoszegiMcCune2012.13416} 
for at least two reasons: (1) Chiral and mirror fermions are fermion doublings with opposite chiralities \cite{NielsenNinomiya1981hkPLB}
(applicable in even dimensional spacetime in general). 
(2) We do \emph{not} double copy the gauge group of the chiral sector to the mirror sector as 
\Refe{BabuMohapatra1989, BabuMohapatra1989rbPRD, BarrChangSenjanovic1991qxPRL, CraigGarciaKoszegiMcCune2012.13416}  did.
Our chiral and mirror fermions couple to the exactly same set of gauge fields.\\
3. {\bf Symmetric Mass Generation (SMG)} \cite{WangYou2204.14271} fully gaps the mirror fermion sector but still fully preserves the chiral symmetry on both the chiral and mirror fermion sides.
The idea is an incarnation of the Eichten-Preskill approach \cite{Eichten1985ftPreskill1986}, 
but in the modern setup enriched by the SMG deformation \cite{WangYou2204.14271}.
We will follow the explicit 1+1d multi-fermion interaction construction in \cite{Wang2013ytaJW1307.7480, Wang2018ugfJW1807.05998, ZengZhuWangYou3450SMG2202.12355}.
The SMG is sometimes called the Kitaev-Wen mechanism \cite{FidkowskifSPT1,FidkowskifSPT2,Wen2013ppa1305.1045,WangWen2018cai1809.11171} 
or a mass gap without mass term \cite{You2014oaaYouBenTovXu1402.4151, YX14124784, BenTov2014eea1412.0154, BenTov2015graZee1505.04312}.
In general, we can induce the SMG energy gap (or mass gap) without any explicit or spontaneously $G$-symmetry-breaking mass term 
if and only if $G$ is 't Hooft anomaly-free,
although such SMG deformation often requires to introduce 
either dangerously 
irrelevant operators with nonperturbative coupling  \cite{Wang2013ytaJW1307.7480, Wang2018ugfJW1807.05998, ZengZhuWangYou3450SMG2202.12355}, 
or new matter or gauge sectors brought down from the high-energy or short distance 
\cite{NSeiberg-Strings-2019-talk, JW2008.06499, RazamatTong2009.05037, Tong2104.03997, Wang2106.16248}. 
Thus SMG deformations are largely overlooked in the previous literature.
In general, in order to check the quantum field theory (QFT) is anomaly-free, 
we require classification of perturbative local and nonperturbative global anomalies
by one-higher dimensional invertible topological field theories (iTFTs), 
which becomes systematic especially due to the work of Freed-Hopkins \cite{2016arXiv160406527F}.  

In \Sec{sec:Conclusion}, we conclude by explaining some physics intuitions behind our solutions.
%
{In \Sec{sec:transition},
we explain in more detail the \emph{ordered} phase to the \emph{disordered} phase 
of the ${\bar{\theta}_{}}$ as an approximate analogy to the (algebraic-)superfluid to insulator type of quantum phase transition \cite{Fisher1989zza}.} 
In \Sec{sec:thought-exp},
we demonstrate the interplay between the SMG  \cite{WangYou2204.14271} and 
 Laughlin's style of the flux insertion thought experiment \cite{Laughlin1981PRB}, namely threading a magnetic flux through 
the hole of the annulus or cylinder strip to induce the electric field and the 1+1d boundary anomalous current transport through the 2+1d bulk 
\cite{Wang2013ytaJW1307.7480, SantosWang1310.8291}.

\newpage
\section{A 1+1d model with the CT or P problem}
\label{sec:problem}

In this section, we will show that the CT or P problem is a general issue for a 1+1d theory 
with U(1) gauge field.\footnote{Here U(1) field can be either treated as a \emph{background gauge field} (i.e., the path integral $Z[A]$ depends on the specific choices of gauge bundle and connection $A$) or as a \emph{dynamically gauge field} (i.e., summed over the gauge bundle and connection in the path integral
$Z= \int [DA] Z[A]$), depending on the context that we will illuminate.}
Our CT or P solution will be applicable to the most general 1+1d U(1) anomaly-free theory involving chiral fermions or chiral bosons 
following the setup in \cite{Wang2013ytaJW1307.7480, Wang2018ugfJW1807.05998}.

In particular, for simplicity but without loss of generality,
we shall focus on arguably the simplest example of 1+1d U(1) symmetric chiral fermion theory
known as the $3_L$-$4_L$-$5_R$-$0_R$ model or the 3-4-5-0 model. 
This 1+1d Lorentz invariant and U(1) symmetric chiral fermion 3-4-5-0 theory
has two left-moving and two right-moving 
complex Weyl fermions $\psi_{L,3},\psi_{L,4},\psi_{R,5},\psi_{R,0}$
with U(1) charges 3,4 and 5,0. Relevant data of this 1+1d theory is organized in \Table{table:3450charge}.
(The model can be straightforward generalized to other 1+1d models, which will be reserved for future work.)

\begin{table}[!h]
\begin{tabular}{l |cccc | cc | l}
\hline
& \multicolumn{4}{c|}{ Weyl fermion} & 
\multicolumn{2}{c|}{
$\begin{array}{c} \text{disorder}\\ 
\text{scalar}\\
\end{array}$  
}  & \; Higgs \\
\cline{2-8}
& $\psi_{L,3}$ & $\psi_{L,4}$ & $\psi_{R,5}$ &  $\psi_{R,0}$ & $\phi_1$ & $\phi_2$ & \;  $\phi_H$ \\
\hline
$\U(1)_{\rm 1st}$ \;\;\,\, $q_{{1}}$ & 3 & 4 & 5 & 0  & $-4$  & $-2$  & \;  1\\
$\U(1)_{\rm 2nd}$ \quad $q_{{2}}$ & 0 & 5 & 4 & 3  &  $-2$ &  $-4$ & \;  $-1$\\
\hline
$\U(1)_{\rm 3rd}$ \quad $q_{{3}} =K^{} q_1$ & 3 & 4 & $-5$ & 0  &  &  \\
$\U(1)_{\rm 4th}$ \quad $q_{{4}} =K^{} q_2$ & 0 & 5 & $-4$ & $-3$  &  & \\
\hline
$\U(1)_{\rm vector}$\;$q_{\rm v}=  \frac{1}{3} (q_1 + q_2)$ & 1 &  3 & 3 & 1  & &  & \;  0 \\
\hline
\multicolumn{8}{c}{Yukawa-Higgs $\phi_H$ Vector mass term's power exponent}   \\
\hline
$m_{3,0}$ : $\phi_H^3 \psi_{L,3}^\dagger \psi_{R,0}$ & $-1$ & 0 & 0 & $1$  & 0  & 0 & \;  3 \\
$m_{4,5}$ : $\phi_H^\dagger \psi_{L,4}^\dagger \psi_{R,5}$ & 0 & $-1$ & 1 & 0  & 0  & 0  & \;  $-1$\\
\hline
\multicolumn{8}{c}{Multi-fermion (power exponent) or sine-Gordon $\sum_{\alpha=1}^2  g_{\alpha}\cos(\ell_{\alpha,{\rm I} }\varphi_{\rm I})$ (coefficient)  interactions}  \\
\hline
$\ell_{1} =  \frac{1}{3} (q_3 -2 q_4)$ & 1 & $-2$  & $1$  & $2$ &   &  \\
$\ell_{2}=  \frac{1}{3} (2 q_3 - q_4)$ & 2 & 1 & $-2$ & $1$ &  & \\
\hline
\multicolumn{8}{c}{Disorder Yukawa-scalar $\phi_1$ and $\phi_2$ term's power exponent}  \\
\hline
$\phi_1^2 \psi_{L,3} \psi_{R,5}$ & 1 & $0$ & 1 & 0  & 2  & 0 \\
$\phi_1^\dagger \psi_{L,4}^\dagger \psi_{R,0}$ & 0 & $-1$ & 0 & 1 & $-1$  & 0 \\
$\phi_2^\dagger \psi_{L,3}\psi_{R,5}^\dagger$ & 1 & 0 & $-1$ & 0  & 0  & $-1$ \\
$\phi_2^2 \psi_{L,4} \psi_{R,0}$ & 0 & 1 & 0 & 1  & 0  & 2 \\
\hline
\end{tabular}  
\caption{
{Without loss of generality, 
we present an arguably simplest 1+1d U(1) symmetric 
anomaly-free chiral fermion theory, the $3_L$-$4_L$-$5_R$-$0_R$ or 3-4-5-0 model
with two left-moving and two right-moving Weyl fermions \eq{eq:free} and
a choice of $\U(1)_{\rm 1st} \times \U(1)_{\rm 2nd} \times\U(1)_{\rm 3rd}  \times \U(1)_{\rm 4th}$ symmetry charges.
The field contents, their quantum numbers, and the coefficient or power exponent of interactions are organized above.\\
$\bullet$ See Appendix \ref{appendix:A} for the mathematical structure behind the general class of these 1+1d $\U(1)$ symmetric models.\\
$\bullet$ The Yukawa-Higgs vector mass term
$\phi_H^3 \psi_{L,3}^\dagger \psi_{R,0} + 
\phi_H^\dagger \psi_{L,4}^\dagger \psi_{R,5}
+\text{h.c.}$ of \eq{eq:YH} breaks the $\U(1)^4$ symmetry down to  $\U(1)_{\rm vector}$ in the $\phi_H$ Higgs condensed phase ($\<\phi_H \> \neq 0$).
We \emph{cannot} rotate the CT and P violating $\theta$ angle away in the chiral sector alone
without introducing a compensating complex phase to the vector mass term.\\
$\bullet$ The Symmetric Mass Generation (SMG) to fully gap this chiral fermion theory can be achieved by
the point-splitting multi-fermion interactions 
$\big( {g}_{1}   ({\psi}_{L,3}) 
(  {\psi}_{L,4}^\dagger)^2
( {\psi}_{R,5}  )(  {\psi}_{R,0})^2 
+  {g}_{2}     ( {\psi}_{L,3} )^2
({\psi}_{L,4})
( {\psi}_{R,5}^\dagger )^2
(  {\psi}_{R,0})+\text{h.c.} \big)
$ in \eq{eq:multi-fermion}, whose bosonized counterpart
${\psi \equiv : \exp(\pm \ii \varphi ):}$ 
contains two cosine sine-Gordon interactions 
$g_{1}\cos(\ell_{1}\varphi_{})+ g_{2}\cos(\ell_{2}\varphi_{})$
with the $\ell_{1}$ and $\ell_{2}$ vectors as in \eq{eq:interaction-cosine-term}.
Alternatively, we can introduce the disorder scalars $\phi_1$ and $\phi_2$
as 
$(\phi_1^2 \psi_{L,3} \psi_{R,5}
+\phi_1^\dagger \psi_{L,4}^\dagger \psi_{R,0}
+\phi_2^\dagger \psi_{L,3}\psi_{R,5}^\dagger 
+\phi_2^2 \psi_{L,4} \psi_{R,0}
+\text{h.c.}
+\frac{1}{{\tilde{g}_1}} \phi_1^\dagger\phi_1
+\frac{1}{{\tilde{g}_2}} \phi_2^\dagger\phi_2)$ in \eq{eq:disorder}, 
where disorder scalars can be integrated out to derive the multi-fermion interactions \eq{eq:multi-fermion}.
The coupling is related by $g_\alpha \sim {\tilde{g}_\alpha}^2$.
}
}
 \label{table:3450charge}
\end{table}

The free massless fermion part of action is
 \bea \label{eq:free}
 S_{\Psi,\rm free}=\int  \dd t \dd x  \big(
 \ii \psi^\dagger_{L,3} (\partial_t-\partial_x) \psi_{L,3}
 + \ii\psi^\dagger_{L,4} (\partial_t-\partial_x) \psi_{L,4}
+ \ii \psi^\dagger_{R,5} (\partial_t+\partial_x) \psi_{R,5}
+\ii\psi^\dagger_{R,0} (\partial_t+\partial_x) \psi_{R,0}\big).
\eea
Let us define the multiplet 
$\Psi_{\rm I} \equiv \Big(\begin{smallmatrix}
\psi_{L,3}\\
\psi_{L,4}\\
\psi_{R,5}\\
\psi_{R,0}
\end{smallmatrix}\Big)_{\rm I}$
and a diagonal unimodular rank-4 matrix $K=\diag(1,1,-1,-1)$ and ${\rm I,J} \in \{1,2,3,4\}$,
so \Eq{eq:free} is equivalent to
$ S_{\Psi,\rm free}=\int  \dd t \dd x \ii \Psi_{\rm I}^\dagger (\delta_{\rm IJ} \partial_t-K_{\rm IJ}^{-1} \partial_x) \Psi_{\rm J} 
=\int  \dd t \dd x \ii  \Psi_{\rm I}^\dagger (\partial_t-K_{\rm II}^{-1} \partial_x) \Psi_{\rm I}$.
Here the repeated indices are summed, namely $\sum_{\rm I,J}$ and $\sum_{\rm I}$ respectively. 
The corresponding charges of fermions are listed in \Table{table:3450charge}.
Let us systematically enumerate the properties and comment on this 1+1d model. 
\begin{enumerate}[leftmargin=-0mm, label=\textcolor{blue}{\arabic*}., ref={\arabic*}]
\item {\bf Symmetry}:
The free massless fermion part of the theory \eq{eq:free}
has a spacetime Lorentz symmetry $\Spin$ group and
an enlarged internal symmetry $\U(2)_L \times \U(2)_R$.
The Spin group is a double cover of the special orthogonal group SO, graded by the fermion parity $\Z_2^{\rF}: \Psi \mapsto - \Psi$.
We can choose
$\U(2)_L \times \U(2)_R \supset \U(1)_{\rm 1st} \times \U(1)_{\rm 2nd} \times\U(1)_{\rm 3rd}  \times \U(1)_{\rm 4th}$
to contain the maximal torus $\U(1)^4$ abelian subgroup. 
The spacetime Spin and internal $\U(1)^4$ symmetry combined together is ${\Spin \times \U(1)^4}$ because the fermions as the spacetime spinors of Spin group are allowed to 
have both odd or even charges of U(1).
(However, the spacetime-internal symmetry would be $\frac{\Spin \times \U(1)^4}{\Z_2^{\rF}} 
\equiv {\Spin \times_{\Z_2^{\rF}} \U(1)^4}$ if the fermions must have odd charges of U(1) while the bosons must have even charges of U(1).)
The active Lorentz symmetry Spin(1,1) and U(1) symmetry transformations map on fermions respectively as:
\bea
\Spin &:& \Psi_{\rm I} (x^{\mu} ) \mapsto   \Lambda_{\frac{1}{2}} \Psi_{\rm I} (\Lambda^{\mu}{}_\nu x^{\nu} ).  \\
\U(1)&:& \Psi_{\rm I} (x^{\mu} ) \mapsto  \e^{\ii q_{\rm I} \upalpha } \Psi_{\rm I} (x^{\mu} ), \quad \upalpha \in [0, 2 \pi).
\eea
The $ \Lambda_{\frac{1}{2}}$ and $\Lambda^{\mu}{}_\nu$ are respectively
the standard Lorentz transformations on the spin-$\frac{1}{2}$ left or right-handed Weyl spinor 
and the spin-1 Lorentz vector. The U(1) can be any choice from \Table{table:3450charge},
the $q_{\rm I}$ has ${\rm I} \in \{1,2,3,4\}$ assigned to four Weyl fermions $\psi_{L,3}, \psi_{L,4}, \psi_{R,5}, \psi_{R,0}$ respectively. 

Moreover, 
following \eq{eq:CPT}, 
the active discrete C-P-T symmetry transformations act on Weyl fermion fields as:
\bea
\Z_2^{\rm C} &:& \psi_L (t,x) \mapsto -\psi_L^* (t,x), \quad \psi_R (t,x) \mapsto +\psi_R^* (t,x).\nn \\ 
\label{eq:CPT-psi}
\Z_2^{\rm P} &:& \psi_L (t,x) \mapsto + \psi_R (t,-x), \quad \psi_R (t,x)   \mapsto +\psi_L (t,-x).\\ 
\Z_2^{\rm T} &:& \psi_L (t,x) \mapsto + \psi_R (-t,x), \quad \psi_R (t,x)   \mapsto +\psi_L (-t,x).
\nn
\eea
The free fermion 3-4-5-0 theory preserves the charge conjugation C, but violates the parity P (thus being chiral)
and time-reversal T symmetry,
hence it is also named the $3_L$-$4_L$-$5_R$-$0_R$ chiral fermion theory.
\item {\bf Anomaly}:
Recent developments of the classification of invertible topological field theories via cobordism groups based on 
Kapustin et al.\cite{Kapustin2014tfa1403.1467, Kapustin1406.7329} 
and Freed-Hopkins et al.' work 
\cite{2016arXiv160406527F, WanWang2018bns1812.11967} 
help to systematically classify the anomalies of quantum field theories.
The mathematically rigorous anomaly classification of the spacetime-internal ${\Spin \times \U(1)^4}$ symmetry is obtained 
by the cobordism group $\TP_{3}^{\Spin \times \U(1)^4}=\Z^{11}$ \cite{WanWang2018bns1812.11967}.
The nonperturbative global anomalies are classified by the finite subgroup part (namely the torsion subgroup) of the bordism group $\Omega_{3}^{\Spin \times \U(1)^4}=0$.
The perturbative local anomalies are classified by the integer $\Z$ classes (namely the free subgroup) of  the bordism group $\Omega_{4}^{\Spin \times \U(1)^4}=\Z^{11}$.
Follow the computation in \Refe{WanWang2018bns1812.11967}, it can be shown that 
the $\TP_{3}^{\Spin \times \U(1)^4}$ only contains the $\Z$ perturbative local anomalies
 --- either the gravitational local anomaly captured by the Feynman diagram 
 $\includegraphics[height=0.32in]{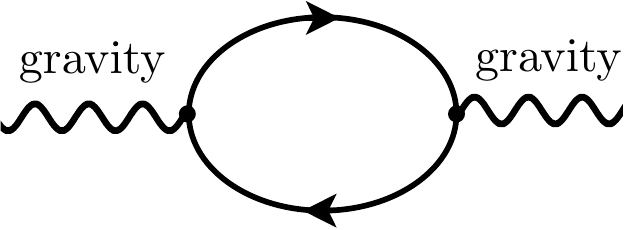}$
or the U(1) local anomalies captured by the Feynman diagram  $\includegraphics[height=0.32in]{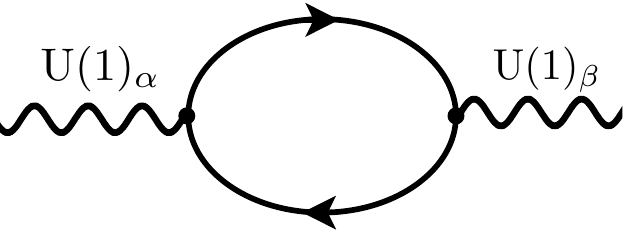}$.
The gravitational local anomaly coefficient is the chiral central charge $c_- \equiv c_L - c_R$ which is anomaly-free ($c_-=2-2=0$) for the 3-4-5-0 model.
The U(1) local anomalies can be expressed as a symmetric bilinear form using the $K$ matrix shown below.

\item 
The one-loop Feynman graph of two-point correlators of the background $A$ gauge fields of $\U(1)$ charges $q_\al$ and $q_\bt$
have the anomaly coefficients 
\bea \label{eq:U1anomaly-free}
q^{\intercal}_\al K^{} q_\bt \equiv \sum_{\rm I,J} q_{\al,\rm I} K^{}_{\rm IJ} q_{\bt,\rm J}
\eea 
with the transpose ${\intercal}$.
The $\U(1)_{\rm 1st} \times \U(1)_{\rm 2nd}$ are self and mutual 't Hooft anomaly-free,
$q_1^{\intercal} K^{} q_1=q_2^{\intercal} K^{} q_2=q_1^{\intercal} K^{} q_2=0$.
Similarly, 
the $\U(1)_{\rm 3rd} \times \U(1)_{\rm 4th}$ are also self and mutual 't Hooft anomaly-free
$q_3^{\intercal} K^{} q_3=q_4^{\intercal} K^{} q_4=q_3^{\intercal} K^{} q_4=0$.
But mutually between 
$\U(1)_{\rm 1st} \times \U(1)_{\rm 2nd}$ and $\U(1)_{\rm 3rd} \times \U(1)_{\rm 4th}$,
there are mixed anomalies, because
$q_1^{\intercal} K^{} q_3=q_2^{\intercal} K^{} q_4= 50$
and $q_1^{\intercal} K^{} q_4=q_2^{\intercal} K^{} q_3=40$. 
\item {\bf Analogy to the 3+1d SM}: 
We now make an analogy between the 3+1d's electroweak Higgs symmetry breaking to the 1+1d model with Higgs 
symmetry breaking.\footnote{Below we write the SM's Lie algebra $su(3) \times su(2) \times u(1)_{Y}$ because the SM's Lie group has the four compatible versions, 
$(\SU(3) \times \SU(2) \times \U(1)_{Y})/\Z_p$ with $p=1,2,3,6$.}
The 3+1d SM chiral fermion with the chiral $su(2) \times u(1)_{Y}$ 
symmetry for the $su(2)$ chiral weak force and the chiral hypercharge $Y$
is analogous to 
the 1+1d chiral fermion model with the chiral $\U(1)_{\rm 1st} \times \U(1)_{\rm 2nd}$ symmetry.

In 3+1d, the SM Higgs condensation breaks the chiral electroweak $su(2) \times u(1)_{Y}$ down to the vector $u(1)_{\rm EM}$ subgroup, 
since $su(2) \times u(1)_{Y} \supset u(1)_{\rm EM}$.
The electroweak Higgs carries nonzero $su(2)$ and $u(1)_{Y}$ charges, but carries a zero $u(1)_{\rm EM}$.
Similarly, in 1+1d, we can introduce a scalar Higgs $\phi_H$
and the Yukawa-Higgs (YH) term with a Higgs potential
\bea \label{eq:YH}
S_{\rm YH}= \int  \dd t \dd x 
(\phi_H^3 \psi_{L,3}^\dagger \psi_{R,0} + 
\phi_H^\dagger \psi_{L,4}^\dagger \psi_{R,5}
+\text{h.c.} -\mu_H |\phi_H|^2 - \lambda_H |\phi_H|^4 ) 
\eea to the action.
We denote the first 
YH term as the $m_{3,0}$ term and the second 
YH term as the $m_{4,5}$ term.
Kinematically, the 
YH term preserves the 
$\U(1)_{\rm 1st}  \times \U(1)_{\rm 2nd} \supset \U(1)_{\rm vector}$. 
The $\phi_H$ carries
nonzero $\U(1)_{\rm 1st}$ and $\U(1)_{\rm 2nd}$ charges, but carries a zero $\U(1)_{\rm vector}$ charge.
Dynamically, when $\mu_H <0$, 
the Higgs condensation
$\< \phi_H \> \neq 0$
breaks
$\U(1)_{\rm 1st}  \times \U(1)_{\rm 2nd}$ down to the $\U(1)_{\rm vector}$ subgroup
and
induces
a vector mass term to all the fermions.

Notice that $\psi_{R,0}$ has a $\U(1)_{\rm 1st}$ charge 0 in the 1+1d toy model, 
thus it is somehow analogous to the neutrino of the 3+1d SM.
But in this analogy,
$\psi_{R,0}$ carries a $\U(1)_{\rm 2nd}$  charge 3 and a $\U(1)_{\rm vector}$ charge 1, 
so it is different from the SM's \emph{right-handed neutrino} that is neutral to all SM gauge force $su(3) \times su(2) \times u(1)_Y$ and $u(1)_{\rm EM}$.
The $\psi_{R,0}$ carries a $\U(1)_{\rm vector}$ charge 1,
thus it is also different from the SM's \emph{left-handed neutrino} that is neutral to the $u(1)_{\rm EM}$.
The mass term of $\psi_{R,0}$ (pairing with the charged $\psi_{L,3}$) is also different from the unsettled neutrino mass mechanism in the SM 
(e.g., Dirac, Majorana, or other causes from the interacting energy gap above the topological field theory 
\cite{JW2006.16996,JW2012.15860}).
\item We can dynamically gauge $\U(1)_{\rm 1st} \times \U(1)_{\rm 2nd}$ by replacing the background field $A$ to dynamical gauge fields
$a_1$ and $a_2$. Namely, we promote 
the spacetime derivative $\partial_t + v_x \partial_x$ 
(with $v_x=-1$ for the left-moving $\psi_L$ while $v_x=+1$ for the right-moving $\psi_R$)
to the covariant derivative $D_t + v_x D_x \equiv (\partial_t - \ii q a_t) + v_x (\partial_x - \ii q a_x)$
where the charge $q$ of $\U(1)_{\rm 1st} \times \U(1)_{\rm 2nd}$ can be read from \Table{table:3450charge}.\footnote{In
our case, the velocity $v_{x, \rm I}$ for the I-th Weyl fermion is determined by its unimodular $K$ matrix component: 
$v_{x, \rm I}= - K_{\rm I I}^{-1}$.}
We also map $ S_{\Psi,\rm free} \mapsto S_{\Psi,\rm gauge}$
by replacing $\partial_t + v_x \partial_x \mapsto D_t + v_x D_x$ in the action.
Under the $\U(1)_{\rm 1st} \times \U(1)_{\rm 2nd}$ symmetry transformations,
$\Psi_{\rm I} \mapsto \e^{\ii q_{{\rm 1},\rm I}  \upalpha_{\rm 1st}} \Psi_{\rm I}$
and $\Psi_{\rm I} \mapsto \e^{\ii  q_{{\rm 2},\rm I} \upalpha_{\rm 2nd}} \Psi_{\rm I}$ with the quantized $q_{{\rm 1},\rm I}, q_{{\rm 2},\rm I}$ charges in \Table{table:3450charge},
the corresponding currents are conserved even at the quantum level when $\U(1)_{\rm 1st} \times \U(1)_{\rm 2nd}$ are dynamically gauged:
\be \label{eq:1st2ndcurrent}
J^{\mu}_{\rm 1st}
\equiv  
\Psi_{\rm I}^\dagger
q_{{\rm 1},\rm I}
\tau^\mu_{ L/R} 
\Psi_{\rm I}
\equiv
\sum_{\rm I,J}
\Psi_{\rm I}^\dagger
\Big(\begin{smallmatrix}
3 \tau^\mu_{ L} & & &\\
 & 4 \tau^\mu_{ L}& &\\
  & & 5 \tau^\mu_{ R} &\\
   & & & 0 \tau^\mu_{ R}\\
\end{smallmatrix}\Big)_{\rm IJ}
\Psi_{\rm J}, \quad
J^{\mu}_{\rm 2nd} 
\equiv  
\Psi_{\rm I}^\dagger
q_{{\rm 2},{\rm I}}
\tau^\mu_{ L/R} 
\Psi_{\rm I}
\equiv
\sum_{\rm I,J}
\Psi_{\rm I}^\dagger
\Big(\begin{smallmatrix}
0 \tau^\mu_{ L} & & &\\
 & 5 \tau^\mu_{ L}& &\\
  & & 4 \tau^\mu_{ R} &\\
   & & & 3 \tau^\mu_{ R}\\
\end{smallmatrix}\Big)_{\rm IJ}
\Psi_{\rm J}.
\ee
Here in 1+1d, 
we denote
\bea \label{eq:tau-mu}
\tau^\mu_{ L/R} \equiv 
\left\{
\begin{array}{l} 
\tau^\mu_{ L} \equiv \bar\tau^\mu =(1,-1),\quad  \mu=0,1. \\
\tau^\mu_{ R} \equiv \tau^\mu=(1,1), \quad \mu=0,1.
\end{array}
\right.
\eea

On the other hand, under the classical $\U(1)_{\rm 3rd} \times \U(1)_{\rm 4th}$ symmetry transformations,
$\Psi_{\rm I} \mapsto \e^{\ii q_{{\rm 3},\rm I}  \upalpha_{\rm 3rd}} \Psi_{\rm I}$
and $\Psi_{\rm I} \mapsto \e^{\ii  q_{{\rm 4},\rm I} \upalpha_{\rm 4th}} \Psi_{\rm I}$
with the quantized $q_{{\rm 3},\rm I}, q_{{\rm 4},\rm I}$ charges in \Table{table:3450charge}.
Spoiled by the mixed anomaly between 
$\U(1)_{\rm 1st} \times \U(1)_{\rm 2nd}$ and $\U(1)_{\rm 3rd} \times \U(1)_{\rm 4th}$,
we can verify their corresponding currents
\be \label{eq:3rd4thcurrent}
J^{\mu}_{\rm 3rd}
\equiv  
\Psi_{\rm I}^\dagger
q_{{\rm 3},\rm I}
\tau^\mu_{ L/R} 
\Psi_{\rm I}
\equiv
\sum_{\rm I,J}
\Psi_{\rm I}^\dagger
\Big(\begin{smallmatrix}
3 \tau^\mu_{ L} & & &\\
 & 4 \tau^\mu_{ L}& &\\
  & & -5 \tau^\mu_{ R} &\\
   & & & 0 \tau^\mu_{ R}\\
\end{smallmatrix}\Big)_{\rm IJ}
\Psi_{\rm J}, \quad
J^{\mu}_{\rm 4th} 
\equiv  
\Psi_{\rm I}^\dagger
q_{{\rm 4},{\rm I}}
\tau^\mu_{ L/R} 
\Psi_{\rm I}
\equiv
\sum_{\rm I,J}
\Psi_{\rm I}^\dagger
\Big(\begin{smallmatrix}
0 \tau^\mu_{ L} & & &\\
 & 5 \tau^\mu_{ L}& &\\
  & & -4 \tau^\mu_{ R} &\\
   & & & -3 \tau^\mu_{ R}\\
\end{smallmatrix}\Big)_{\rm IJ}
\Psi_{\rm J},
\ee
are not conserved, 
when $\U(1)_{\rm 1st} \times \U(1)_{\rm 2nd}$ are dynamically gauged
by summing over their gauge connections and bundles in the path integral measure 
$\int [\cD a_1][\cD a_2]$.

The $S_{\rm YH}$ also transforms to 
$S_{\rm YH}^{(\upalpha)}$ 
with the 
modified YH term 
$\e^{-\ii 3 (\upalpha_{\rm 3rd}+\upalpha_{\rm 4th} )} \phi_H^3 \psi_{L,3}^\dagger \psi_{R,0} + 
\e^{-\ii 9 (\upalpha_{\rm 3rd}+\upalpha_{\rm 4th} )} \phi_H^\dagger \psi_{L,4}^\dagger \psi_{R,5}
+\text{h.c.}$
Under the variation $\upalpha_{\rm 3rd}$ and $\upalpha_{\rm 4th}$, 
overall the path integral $Z$ is transformed to 
\bea \label{eq:Z3450-vary}
Z &\mapsto& \int [\cD a_1][\cD a_2] [\cD \psi][\cD \bar\psi] \e^{\ii (
S_{\Psi,\rm gauge} + S_{\rm YH}^{(\upalpha)}
+ \int  (\frac{1}{2 e^2} F^{\rm 1st} \wedge \star F^{\rm 1st})
+ \int  (\frac{1}{2 e^2} F^{\rm 2nd} \wedge \star F^{\rm 2nd})
+ \int (\upalpha_{\rm 3rd}  \dd \star  J_{\rm 3rd} 
+ \upalpha_{\rm 4th} \dd \star  J_{\rm 4th})
)} 
\cr
&& \e^{\ii \int \frac{1}{2 \pi} 
\big(
(\theta_{\rm 1st} -\upalpha_{\rm 3rd} ( q_{{\rm 1}}^{\intercal} \cdot K^{}  \cdot q_{{\rm 3}})
-\upalpha_{\rm 4th} ( q_{{\rm 1}}^{\intercal} \cdot K^{}  \cdot q_{{\rm 4}})) 
F^{\rm 1st}+ 
(\theta_{\rm 2nd} 
-\upalpha_{\rm 3rd} ( q_{{\rm 2}}^{\intercal} \cdot K^{}  \cdot q_{{\rm 3}})
-\upalpha_{\rm 4th} ( q_{{\rm 2}}^{\intercal} \cdot K^{}  \cdot q_{{\rm 4}})) 
F^{\rm 2nd}
\big)}.
\eea
Note that 
$\dd \star J = (\prt_\mu J^{\mu}) \dd t \dd x$,
and
$\frac{\theta}{2 \pi} F \equiv \frac{\theta}{2 \pi} \dd a 
=\frac{\theta}{4 \pi} \epsilon^{\mu \nu} F_{\mu \nu} \dd^2 x
=\frac{\theta}{4 \pi} \epsilon^{\mu \nu} (\prt_{\mu} a_\nu -\prt_{\nu} a_\mu) \dd^2 x
=\frac{\theta}{2 \pi} \epsilon^{\mu \nu} \prt_{\mu} a_\nu \dd^2 x$.
We take $F=F^{\rm 1st}$ or $F^{\rm 2nd}$ for the field strength of  $\U(1)_{\rm 1st}$ or $\U(1)_{\rm 2nd}$ gauge field.
The periodicity constrains $\theta_{\rm 1st}, \theta_{\rm 2nd}, \upalpha_{\rm 3rd}, \upalpha_{\rm 4th}  \in [0, 2 \pi)$.

Here the Maxwell term
$\int \frac{1}{2 e^2} F \wedge \star F =
\int \dd t \dd x (-\frac{1}{4 e^2} F_{\mu \nu}F^{ \mu \nu})
=\int \dd t \dd x (-\frac{1}{2 e^2} F^{}_{0 1}F^{ 0 1})$
with a minus sign $(-1)$ on the right-hand side due to the Lorentzian volume form.
It is equal to $\int \dd t \dd x \frac{1}{2 e^2} (F^{ }_{0 1})^2=\int \dd t \dd x \frac{1}{2 e^2}  (E_1)^2$,
the spacetime integral of the electric field square.
The $e$ is the unit charge coupling for the $\U(1)$ gauge theory.

For the moment, let us turn off the 
YH term $S_{\rm YH} =0$ (thus $S_{\rm YH}^{(\upalpha)}=0$), 
so we can analyze the largest chiral symmetry.\\
$\bullet$ When $\U(1)_{\rm 1st} \times \U(1)_{\rm 2nd}$ and $\U(1)_{\rm 3rd} \times \U(1)_{\rm 4th}$,
are all treated as global symmetries,
they have mixed 't Hooft anomaly because the aforementioned
$q_1^{\intercal} K^{} q_3=q_2^{\intercal} K^{} q_4= 50$
and $q_1^{\intercal} K^{} q_4=q_2^{\intercal} K^{} q_3=40$.
This can be signaled by coupling to non-dynamical background fields.
't Hooft anomaly implies that there is an obstruction to gauge $\U(1)^4$ altogether. 
But we can gauge either of the two anomaly-free subgroups, which we choose $\U(1)_{\rm 1st} \times \U(1)_{\rm 2nd}$.\\
$\bullet$ 
When $\U(1)_{\rm 1st} \times \U(1)_{\rm 2nd}$ are dynamically gauged by coupling to the dynamical $a_1$ and $a_2$,  
the $J^{\mu}_{\rm 3rd}$ and $J^{\mu}_{\rm 4th}$ are indeed
non-conserved
due to the Adler-Bell-Jackiw (ABJ) anomaly 
\cite{Adler1969gkABJ, Bell1969tsABJ}.
\Eq{eq:Z3450-vary} shows that:\\ 
If we gauge $\U(1)_{\rm 1st}$ only, ABJ anomaly breaks $\U(1)_{\rm 3rd} \times \U(1)_{\rm 4th}$  to the discrete $\Z_{50,{\rm 3rd}} \times \Z_{40,{\rm 4th}}$.\\
If we gauge $\U(1)_{\rm 2nd}$ only, ABJ anomaly breaks $\U(1)_{\rm 3rd} \times \U(1)_{\rm 4th}$ to  $\Z_{40,{\rm 3rd}} \times \Z_{50,{\rm 4th}}$.\\
If we gauge both $\U(1)_{\rm 1st} \times \U(1)_{\rm 2nd}$, ABJ anomaly breaks $\U(1)_{\rm 3rd} \times \U(1)_{\rm 4th}$ to $\Z_{10,{\rm 3rd}} \times \Z_{10,{\rm 4th}}$.
\item There is no CT or P problem if there is no 
YH or quadratic mass term.
Namely, we can always use the anomalous symmetry transformation 
with appropriate two variables $\upalpha_{\rm 3rd}$ and $\upalpha_{\rm 4th}$
to solve two constraints 
$\theta_{\rm 1st} -\upalpha_{\rm 3rd} ( q_{{\rm 1}}^{\intercal} \cdot K^{}  \cdot q_{{\rm 3}})
-\upalpha_{\rm 4th} ( q_{{\rm 1}}^{\intercal} \cdot K^{}  \cdot q_{{\rm 4}})=0$
and 
$\theta_{\rm 2nd} 
-\upalpha_{\rm 3rd} ( q_{{\rm 2}}^{\intercal} \cdot K^{}  \cdot q_{{\rm 3}})
-\upalpha_{\rm 4th} ( q_{{\rm 2}}^{\intercal} \cdot K^{}  \cdot q_{{\rm 4}})=0$. 
The $\theta_{\rm 1st}$ and $\theta_{\rm 2nd}$ can be redefined by the anomalous chiral transformations to 0, 
so the massless theory has no CT nor P violation. 

\item {\bf CT or P problem}:
Generic models with 
YH or mass terms suffer from the P problem.
For example, even in the Higgs condensed phase, 
there is a corresponding $\theta_{\rm v}$ term of a vector gauge field $a_{\rm v}$ for 
the unbroken $\U(1)_{\rm vector}$ gauge group. 
We attempt to do the generic chiral rotations on every chiral fermion to rotate the $\theta_{\rm v}$ away:
\bea \label{eq:upchi}
 (\psi_{L,3}, \psi_{L,4},\psi_{R,5},\psi_{R,0}) \mapsto 
(\e^{\ii \upalpha_{L,3}} \psi_{L,3}, \e^{\ii \upalpha_{L,4}}\psi_{L,4}, \e^{\ii \upalpha_{R,5}}\psi_{R,5}, \e^{\ii \upalpha_{R,0}}\psi_{R,0}),
\eea
which we define this transformation as $\Psi_{\rm I} \mapsto \e^{\ii \upchi_{\rm I} \delta_{\rm IJ}} \Psi_{\rm J}$ with 
$\upchi^{\intercal} \equiv ({\upalpha_{L,3}} , {\upalpha_{L,4}} , {\upalpha_{R,5}} , { \upalpha_{R,0}} )$.
Under \Eq{eq:upchi},
the path integral $Z$ is transformed accordingly,
\bea \label{eq:Z3450-vary-upchi}
Z &\mapsto& \int [\cD a_{\rm v}] [\cD \psi][\cD \bar\psi] \e^{\ii (
S_{\Psi,\rm gauge} + S_{\rm YH}^{(\upchi)}
+
\int 
(\frac{1}{2 e^2} F^{\rm v} \wedge \star F^{\rm v})
+ 
\int (\upchi^{\intercal} \cdot \dd \star  J_{\upchi}) 
)
 } 
 \e^{\ii \int \frac{1}{2 \pi} 
(\theta_{\rm v} -
 ( q_{{\rm v}}^{\intercal} \cdot K^{}  \cdot \upchi)
) 
F^{\rm v} 
}.\\
\label{eq:YH-upchi}
S_{\rm YH}^{(\upchi)}
& \equiv&
\int  \dd t \dd x 
(\phi_H^3  \e^{\ii (-\upalpha_{L,3} + \upalpha_{R,0} )} \psi_{L,3}^\dagger \psi_{R,0} + 
\phi_H^\dagger \e^{\ii (-\upalpha_{L,4} + \upalpha_{R,5} )} \psi_{L,4}^\dagger \psi_{R,5}
+\text{h.c.} -\mu_H |\phi_H|^2 - \lambda_H |\phi_H|^4 ).
\eea
Here $\upchi^{\intercal} \cdot \dd \star  J_{\upchi} 
\equiv \big(
{\upalpha_{L,3}} 
\prt_\mu(
\psi_{L,3}^\dagger
\tau^\mu_{ L} 
\psi_{L,3})
+  {\upalpha_{L,4}}
\prt_\mu (
\psi_{L,4}^\dagger
\tau^\mu_{ L} 
\psi_{L,4}) 
+ 
{\upalpha_{R,5}}
\prt_\mu (
\psi_{R,5}^\dagger
\tau^\mu_{ R} 
\psi_{R,5}) 
+ 
{ \upalpha_{R,0}}
\prt_\mu (
\psi_{R,0}^\dagger
\tau^\mu_{ R} 
\psi_{R,0})  
\big) \dd t \dd x$.

To identify the P problem of this model, we need to identify the associated invariant $\bar \theta$ angle.
The rank-2 mass matrix $M$ is defined by rewriting the 
YH term in \Eq{eq:YH-upchi}
as
$(\psi_{L,3}^\dagger, \psi_{L,4}^\dagger) 
M 
(\begin{smallmatrix}
\psi_{R,5}\\
\psi_{R,0}
\end{smallmatrix}) +\text{h.c.}$
So, we have 
$\arg (\det 
M) \mapsto
\arg (\<\phi_H \>^2)
-\upalpha_{L,3}-\upalpha_{L,4} + \upalpha_{R,5}  + \upalpha_{R,0} 
=\arg (\<\phi_H \>^2) - 
(1, 1, 1, 1) \cdot K^{}  \cdot \upchi$.
{For a single Higgs field, 
we can absorb the complex phase of $\<\phi_H \>$ to the chiral rotation,
thus without losing generality, 
below we choose  the real condensate $\<\phi_H \> \in \R$, so
$\arg (\<\phi_H \>^2)=0$.}

We can define
$\arg (\diag M) \mapsto \arg (\diag M) + (
-\upalpha_{L,3}+ \upalpha_{R,0}, -\upalpha_{L,4} + \upalpha_{R,5}  
)$ as a 2-component vector where angular value arg is evaluated along the diagonal of $M$.
Then analogous to the italic-form charge $q_{\rm v} \equiv  (1,  3, 3, 1)$,
we define the 2-component bold text-form charge ${\bf q}_{\rm v} \equiv (1,3)$ associated with the diagonal two component of $M$.
\bea
\theta_{\rm v}
&\mapsto&
\theta_{\rm v} -
 ( q_{{\rm v}}^{\intercal} \cdot K^{}  \cdot \upchi)
 =\theta_{\rm v} - \upalpha_{L,3} -3 \upalpha_{L,4} +3 \upalpha_{R,5} + \upalpha_{R,0}. 
\nn \\
\arg (\det 
M)  
&\mapsto&
\arg (\det M) 
- (1, 1, 1, 1) \cdot K^{}  \cdot \upchi.
\nn \\
{\bf q}_{{\rm v}}^{\intercal} \cdot \arg (\diag M) &\mapsto&
{\bf q}_{{\rm v}}^{\intercal} \cdot \arg (\diag M)
 -
( q_{{\rm v}}^{\intercal} \cdot K^{}  \cdot \upchi).\nn\\
\label{eq:bar-theta}
{\bar{\theta}_{\rm v}} &\mapsto& {\bar{\theta}_{\rm v}} 
\equiv
{\theta_{\rm v}}-
{\bf q}_{{\rm v}}^{\intercal} \cdot
\arg (\diag M).
\eea
Then
${\bar{\theta}_{\rm v}}\equiv
{\theta_{\rm v}}-
{\bf q}_{{\rm v}}^{\intercal} \cdot
\arg (\diag M)$ is invariant under any chiral transformation for any value of $\upchi$.
For this 1+1d model, 
the Naturalness says that a generic ${\bar{\theta}_{\rm v}}$ is nonzero.
However, if we live in this 1+1d model and experimentally test that miraculously ${\bar{\theta}_{\rm v}}\simeq 0$,
then we encounter {\bf the CT or P problem in this 1+1d model}.

In \Sec{sec:Solution}, we will provide {a solution to this 1+1d CT or P problem via the symmetric mass generation}. 
Below we clarify more on the properties of this 1+1d P problem.

\item Let us rephrase the above remark differently. Among the four parameters
$(\upalpha_{L,3},\upalpha_{L,4}, \upalpha_{R,5}, \upalpha_{R,0})$, 
we only have two available degrees of freedom from the anomalous $\U(1)_{\rm 3rd} \times \U(1)_{\rm 4th}$ symmetry
to rotate the $\theta_{\rm v}$. 
We should perform the chiral transformation here as the field redefinition even before we dynamically gauge $\U(1)_{\rm 1st} \times \U(1)_{\rm 2nd}$.
Under the field redefinition $(\upalpha_{\rm 3rd}, \upalpha_{\rm 4th}) \in \U(1)_{\rm 3rd} \times \U(1)_{\rm 4th}$,
\bea
\theta_{\rm v}
&\mapsto&
\theta_{\rm v}
-\upalpha_{\rm 3rd} ( q_{{\rm v}}^{\intercal} \cdot K^{}  \cdot q_{{\rm 3}})
-\upalpha_{\rm 4th} ( q_{{\rm v}}^{\intercal} \cdot K^{}  \cdot q_{{\rm 4}})
=\theta_{\rm v}
{- 30 (\upalpha_{\rm 3rd}+\upalpha_{\rm 4th} )}.\nn \\
\arg (m_{3,0}) \equiv \vartheta_{3,0}&\mapsto& \vartheta_{3,0} {- 3 (\upalpha_{\rm 3rd}+\upalpha_{\rm 4th} )}.\nn \\
\arg (m_{4,5}) \equiv \vartheta_{4,5} &\mapsto& \vartheta_{4,5} {- 9 (\upalpha_{\rm 3rd}+\upalpha_{\rm 4th} )}.\nn \\
\label{eq:bar-theta-}
{\bar{\theta}_{\rm v}} &\mapsto& {\bar{\theta}_{\rm v}} 
\equiv
{\theta_{\rm v}}-
{\bf q}_{{\rm v}}^{\intercal} \cdot
\arg (\diag M)
= {\theta_{\rm v}}-\vartheta_{3,0} - 3 \vartheta_{4,5}.
\eea
The three complex phases from 
the $\theta_{\rm v}$, the $\vartheta_{3,0}$ and $\vartheta_{4,5}$ 
from the first and second 
YH term ($m_{3,0}$ and $m_{4,5}$) terms are generally nonzero.
With only a single parameter $(\upalpha_{\rm 3rd}+\upalpha_{\rm 4th} )$ available, 
we cannot rotate all of three generic complex phases to zero,
nor can we rotate the invariant linear combination 
${\bar{\theta}_{\rm v}}\equiv
{\theta_{\rm v}}-
{\bf q}_{{\rm v}}^{\intercal} \cdot
\arg (\diag M)$ to zero. This is precisely {\bf the CT or P problem in the 1+1d model}.
In short, the generic nonzero $\theta_{\rm v}$, $\vartheta_{3,0}$, and $\vartheta_{4,5}$
in a generic path integral cannot be all set to zeros by any field redefinition or any chiral transformation:
\be 
\hspace{-4mm}
\label{eq:CT-problem}
 \int [\cD a_{\rm v}] [\cD \psi][\cD \bar\psi] \e^{\ii (
S_{\Psi,\rm gauge} + 
 \int  \dd t \dd x 
(\phi_H^3 \e^{\ii \vartheta_{3,0} }\psi_{L,3}^\dagger \psi_{R,0} + 
\phi_H^\dagger \e^{\ii \vartheta_{4,5} } \psi_{L,4}^\dagger \psi_{R,5}
+\text{h.c.} - \mu_H |\phi_H|^2 - \lambda_H |\phi_H|^4 ) 
+
\int 
(\frac{1}{2 e^2} F^{\rm v} \wedge \star F^{\rm v})
)} 
 \e^{\ii \int \frac{1}{2 \pi} 
\theta_{\rm v}  
F^{\rm v} 
}.
\ee

\item {{\bf No Peccei-Quinn solution and no truly long-range order axion}: One may wonder whether we can imitate the Peccei-Quinn solution 
\cite{PecceiQuinn1977hhPRL,PecceiQuinn1977urPRD, Weinberg1977ma1978PRL, Wilczek1977pjPRL} 
to solve this P problem in 1+1d. 
In that case, Peccei-Quinn symmetry would be chosen from the $\U(1)_{\rm 3rd} \times \U(1)_{\rm 4th}$, which has mixed anomaly with $\U(1)_{\rm 1st} \times \U(1)_{\rm 2nd}$.
However, in 1+1d, 
there is no spontaneous continuous symmetry-breaking (here $\U(1)_{\rm 3rd} \times \U(1)_{\rm 4th}$), no Goldstone bosons and no truly long-range order, 
due to the Coleman-Mermin-Wagner-Hohenberg theorem \cite{Coleman1973Goldstone, MerminWagner1966PhysRevLett, Hohenberg1967PhysRev}.
So any new scalar field that we introduce to the model with a modified anomalous U(1) symmetry 
(known as the Peccei-Quinn symmetry \cite{PecceiQuinn1977hhPRL,PecceiQuinn1977urPRD}) 
cannot be spontaneously broken. 
So it is not feasible to solve this 1+1d P problem via the conventional Peccei-Quinn solution.} 

{However, even if 
there is no truly \emph{long-range order} in 1+1d (i.e., any correlator such
as $\< \CO(x_i) \CO(x_j) \>$ cannot approach to a constant when $|x_i-x_j| \gg $ the correlation length or $|x_i-x_j|  \to \infty$, 
for any local bosonic operators $\CO(x)$ built from elementary fields on $x$),
there still allows a so-called \emph{algebraic long-range order} or \emph{quasi-long-range order}
(namely $\< \CO(x_i) \CO(x_j) \> \sim  |x_i-x_j|^{-\alpha}$ for some real number $\alpha$).
So, although there is \emph{no} continuous symmetry-breaking \emph{superfluid} nor truly \emph{long-range order} in 1+1d,
there is still \emph{algebraic superfluid} with \emph{algebraic long-range order} 
modified version of Peccei-Quinn solution in 1+1d.
The correlator $\< \CO(x_i) \CO(x_j) \> \sim  |x_i-x_j|^{-\alpha} = \exp(-\alpha \log( |x_i-x_j|))$ still implies a large correlation length with gapless modes.}

{Even more precisely, the continuous Peccei-Quinn symmetry (here $\U(1)_{\rm 3rd} \times \U(1)_{\rm 4th}$)
would \emph{not} be a global symmetry once the 
$\U(1)_{\rm 1st} \times \U(1)_{\rm 2nd}$ are dynamical gauged. 
The classical $\U(1)_{\rm 3rd} \times \U(1)_{\rm 4th}$ symmetry are broken down by the Adler-Bell-Jackiw (ABJ) anomaly 
\cite{Adler1969gkABJ, Bell1969tsABJ} to its discrete subgroup.
So rigorously speaking, neither \emph{superfluid} nor \emph{algebraic superfluid} exists in the $\U(1)_{\rm 1st} \times \U(1)_{\rm 2nd}$ gauge theory,
because there is no continuous $\U(1)_{\rm 3rd} \times \U(1)_{\rm 4th}$ symmetry to begin with 
in the $\U(1)_{\rm 1st} \times \U(1)_{\rm 2nd}$ gauge theory.
Nonetheless, at least in the weakly gauge or the global symmetry limit of $\U(1)_{\rm 1st} \times \U(1)_{\rm 2nd}$ symmetry,
we find the physical intuitive picture on the (algebraic-)superfluid to insulator transition \cite{Fisher1989zza} very helpful.
If the Peccei-Quinn solution corresponds to the (algebraic-)superfluid solution,
then our solution, demonstrated in \Sec{sec:Another-Solution} and \Sec{sec:Solution}, corresponds to the insulator solution to the CT or P problem.
We will explain the (algebraic-)superfluid to insulator transition analogy in \Sec{sec:Conclusion}.}

\end{enumerate}

\newpage
\section{A CT or P solution via including the SMG to the Higgs phase for the chiral fermions}
\label{sec:Another-Solution}


Earlier we shape the analogous ``Strong CP problem'' as the CT or P problem in 1+1d in \eq{eq:CT-problem}
with the Yukawa-Higgs interaction and Anderson-Higgs mechanism to gap the chiral fermion sector. 
This 1+1d Higgs field is meant to mimic the phenomena of the 3+1d Standard Model's Higgs field with vacuum expectation values giving masses to
the Standard Model's confirmed fermions including quarks and leptons.

However, suppose we live in this 1+1d world and perform the experiment to only confirm that the fermions are massive and gapped,
but we do not yet confirm all of the mass generating mechanisms.
Instead, we may alternatively assume the chiral fermions in the chiral sector are gapped with some energy gap 
by  Symmetric Mass Generation (SMG, not by the Higgs mechanism).
Namely, the path integral of our current interest is not \eq{eq:Z3450-vary}
but instead this new one:
\bea \label{eq:Z-full-chiral-SMG}
 Z= \int  [\cD \psi][\cD \bar\psi]
 \e^{\ii (
S_{\Psi,\rm free} + 
g S_{\text{multi-}\Psi}
+ \int  (\frac{1}{2 e^2} F^{\rm 1st} \wedge \star F^{\rm 1st})
+ \int  (\frac{1}{2 e^2} F^{\rm 2nd} \wedge \star F^{\rm 2nd})
)} 
 \e^{\ii \int_{} \frac{1}{2 \pi} 
(\theta_{\rm 1st}   
F^{\rm 1st} +
\theta_{\rm 2nd}   
F^{\rm 2nd}  ) 
}.
\eea
Below we only need the chiral fermion sector and do not require the mirror fermion sector.\\

\noindent
{\bf SMG by multi-fermion interactions}:
We follow the 1+1d SMG by the multi-fermion interactions to design 
$g S_{\text{multi-}\Psi}$ \cite{Wang2013ytaJW1307.7480, Wang2018ugfJW1807.05998, ZengZhuWangYou3450SMG2202.12355, Tong2104.03997}
that satisfies a mathematically rigorous gapping condition \cite{Haldane1995Stability, KapustinSaulina1008.0654KS, Wang2015Boundary, Levin1301.7355}
that gap all fermions:
\bea
\hspace{-6mm}
\label{eq:multi-fermion-SMG}
g S_{\text{multi-}\Psi} &\equiv& 
 \int  \dd t \dd x \; \big(
 {g}_{1}   ({\psi}_{L,3}) 
(  {\psi}_{L,4}^\dagger)^2
( {\psi}_{R,5}  )(  {\psi}_{R,0})^2 
+  {g}_{2}     ( {\psi}_{L,3} )^2
({\psi}_{L,4})
( {\psi}_{R,5}^\dagger )^2
(  {\psi}_{R,0})+\text{h.c.} \big).
\eea
The multi-fermion interactions are derived based on the 
bosonization method \cite{Wang2013ytaJW1307.7480, Wang2018ugfJW1807.05998}.\footnote{Since the fermions are Grassmann numbers, 
all higher power (larger than 1) of 
any fermion needs to be point-splitted, such as 
${\psi}_{R,0}^2 \equiv {\psi}_{R,0} (\prt_t + \prt_x)  {\psi}_{R,0}$
or ${\psi}_{R,0} (D_t + D_x)  {\psi}_{R,0}$ in the covariant derivative form, etc. 
On the lattice, we have to split ${\psi}_{R,0}^2(x) \equiv {\psi}_{R,0}(x) {\psi}_{R,0}(x+ \epsilon)$ where $\epsilon$ is the lattice constant.
These terms preserve the Lorentz spacetime Spin group, and the internal $\U(1)_{\rm 1st}$ and  $\U(1)_{\rm 2nd}$ symmetries.}
The schematic coupling $g$ controls the coupling ${g}_{1}$ and  ${g}_{2}$.\\

\noindent
{\bf SMG by disorder-scalar interactions}: 
We can also rewrite the multi-fermion interactions $g S_{\text{multi-}\Psi}$
\eq{eq:multi-fermion-SMG}
by integrating out the disorder scalar $\int [\cD \phi_1][\cD \phi_2]$ of the following action \cite{ZengZhuWangYou3450SMG2202.12355}: 
\bea
\label{eq:disorder-SMG}
\hspace{-6mm}
&&g S_{\text{disorder}} \equiv \int  \dd t \dd x \;  \Big(
\phi_1^2 \psi_{L,3} \psi_{R,5}
+\phi_1^\dagger \psi_{L,4}^\dagger \psi_{R,0}
+\phi_2^\dagger \psi_{L,3}\psi_{R,5}^\dagger 
+\phi_2^2 \psi_{L,4} \psi_{R,0}
+\text{h.c.}
+\frac{1}{{\tilde{g}_1}} \phi_1^\dagger\phi_1
+\frac{1}{{\tilde{g}_2}} \phi_2^\dagger\phi_2\Big)
.
\eea
We do not have to introduce explicit kinetic terms for these scalar fields ($\phi_1, \phi_2$).
We also shall include any random configuration of these scalar fields in the spacetime when doing the path integral $\int [\cD \phi_1][\cD \phi_2]$.
The coupling is related by $g_\alpha \sim {\tilde{g}_\alpha}^2$. These interaction terms are also known as the gapping term to gap the
``1+1d Luttinger liquid theory'' in condensed matter.
There are additional density-density interactions, such as $\psi_L^\dagger \psi_L \psi_R^\dagger \psi_R$,
whose effect is only to renormalize the effective ``speed of light'' in the Luttinger liquid theory.
We shall set the renormalized speed of light as the $c=1$ of the 1+1d relativistic field theory.\footnote{Notice that 
our disorder scalar term is different from the formulation of \Refe{Chen2013Model3451211.6947}
that includes different and \emph{too many} types of disorder-scalar interactions (of Yukawa-Dirac and Yukawa-Majorana terms) 
that are \emph{not compatible} with the topological gapping conditions \cite{Haldane1995Stability, KapustinSaulina1008.0654KS, Wang2015Boundary, Levin1301.7355}.
\Refe{Chen2013Model3451211.6947}'s formulation \emph{cannot},
but \Refe{Wang2013ytaJW1307.7480, Wang2018ugfJW1807.05998, ZengZhuWangYou3450SMG2202.12355}'s formulation \emph{can}
fully symmetrically gap the chiral fermion theory.}

How could we distinguish the Anderson-Higgs mass mechanism \eq{eq:YH} vs SMG mechanism \eq{eq:Z-full-chiral-SMG} in this world of the 1+1d chiral fermion sector?
We can distinguish them by examining their {\bf masses}, their {\bf symmetry} properties, and the {\bf energy-momentum dispersion relations} ---
these are falsifiable signatures by numerics or experiments:

\noindent
$\bullet$ The {\bf Anderson-Higgs mass} from
$\< \phi_H^3\> \psi_{L,3}^\dagger \psi_{R,0} +  \< \phi_H^\dagger \> \psi_{L,4}^\dagger \psi_{R,5} +\text{h.c.}$
demands that the $\psi_{L,3}$ and $\psi_{R,0}$ paired up to be Dirac fermion and get the Dirac mass.
It also demands that the $\psi_{L,4}$ and $\psi_{R,5}$ paired up to be another Dirac fermion and get another Dirac mass.
So in this scenario,  $\psi_{L,3}$ mass $m_3$ and $\psi_{R,0}$ mass $m_0$ are the same: $m_3 = m_0 \equiv m_{3,0}$,
while $\psi_{L,4}$ mass $m_4$ and $\psi_{R,5}$ mass $m_5$ are the same: $m_4 = m_5 \equiv m_{4,5}$.
The Higgs condensation also breaks $\U(1)_{\rm 1st} \times \U(1)_{\rm 2nd}$ down to $\U(1)_{\rm vec}$.
The dispersion relations above the mass gap are $\rE(p_x) = \sqrt{m_{3,0}^2 +p_x^2}$ for the $\psi_{L,3}$ and $\psi_{R,0}$'s momentum $p_x$,
and $\rE(p_x) =\sqrt{m_{4,5}^2 +p_x^2}$
for the $\psi_{L,4}$ and $\psi_{R,5}$'s momentum $p_x$.\\

\noindent
$\bullet$ The {\bf SMG} (by the multi-fermion \eq{eq:Z-full-chiral-SMG} and the disorder scalar \eq{eq:disorder-SMG})
actually demands that the disorder scalar $\phi_1$ pairing up
$\psi_{L,3}$ and $\psi_{R,5}$ while the disorder scalar $\phi_2$ pairing up
$\psi_{L,4}$ and $\psi_{R,0}$. It can be numerically verified that
$\psi_{L,3}$ and $\psi_{R,5}$ have the same mass $m_3 = m_5 \equiv m_{3,5}$,
while $\psi_{L,4}$ and $\psi_{R,0}$ have another mass $m_4 = m_0 \equiv m_{4,0}$ \cite{ZengZhuWangYou3450SMG2202.12355}.
The SMG of course preserves the full $\U(1)_{\rm 1st} \times \U(1)_{\rm 2nd}$.
The SMG dispersion relations above the energy gap suggest 
that $\rE(p_x) = \sqrt{m_{3,5}^2 +p_x^2}$ for the $\psi_{L,3}$ and $\psi_{R,5}$'s momentum $p_x$,
and $\rE(p_x) =\sqrt{m_{4,0}^2 +p_x^2}$
for the $\psi_{L,4}$ and $\psi_{R,0}$'s momentum $p_x$. 
Surprisingly, 
the SMG dispersion relations above the energy gap still suggest the same form as the mean-field mass gap \cite{ZengZhuWangYou3450SMG2202.12355}.
But SMG and the mean-field Anderson-Higgs mass show different mass gap size structures.\\

By assuming the gapped chiral fermions \eq{eq:Z-full-chiral-SMG} is due to SMG,
the variation of \eq{eq:Z-full-chiral-SMG}, under
the $\U(1)_{\rm 3rd} \times \U(1)_{\rm 4th}$ symmetry transformations
$\Psi_{\rm I} \mapsto \e^{\ii q_{{\rm 3},\rm I}  \upalpha_{\rm 3rd}} \Psi_{\rm I}$
and $\Psi_{\rm I} \mapsto \e^{\ii  q_{{\rm 4},\rm I} \upalpha_{\rm 4th}} \Psi_{\rm I}$,
leads to
\bea \label{eq:Z-full-chiral-SMG-vary}
Z &\mapsto& \int [\cD \psi][\cD \bar\psi] \e^{\ii (
S_{\Psi,\rm free} + 
g S^{(\upalpha)}_{\text{multi-}\Psi}
+ \int  (\frac{1}{2 e^2} F^{\rm 1st} \wedge \star F^{\rm 1st})
+ \int  (\frac{1}{2 e^2} F^{\rm 2nd} \wedge \star F^{\rm 2nd})
+ \int (\upalpha_{\rm 3rd}  \dd \star  J_{\rm 3rd} 
+ \upalpha_{\rm 4th} \dd \star  J_{\rm 4th})
)} 
\cr
&& \e^{\ii \int \frac{1}{2 \pi} 
\big(
(\theta_{\rm 1st} -\upalpha_{\rm 3rd} ( q_{{\rm 1}}^{\intercal} \cdot K^{}  \cdot q_{{\rm 3}})
-\upalpha_{\rm 4th} ( q_{{\rm 1}}^{\intercal} \cdot K^{}  \cdot q_{{\rm 4}})) 
F^{\rm 1st}+ 
(\theta_{\rm 2nd} 
-\upalpha_{\rm 3rd} ( q_{{\rm 2}}^{\intercal} \cdot K^{}  \cdot q_{{\rm 3}})
-\upalpha_{\rm 4th} ( q_{{\rm 2}}^{\intercal} \cdot K^{}  \cdot q_{{\rm 4}})) 
F^{\rm 2nd}
\big)}.
\eea
Of course, this $\U(1)_{\rm 1st} \times \U(1)_{\rm 2nd}$ anomaly-free theory
can be dynamically gauged by summing over the U(1) gauge connections and bundles in $\int [\cD a_1][\cD a_2]$.
What makes the difference between the previous path integral \eq{eq:Z3450-vary} and our present path integral \eq{eq:Z-full-chiral-SMG-vary}?\\

\noindent
$\bullet$ 
Previously, the Yukawa-Higgs term has a mean-field mass $S_{\rm YH}^{(\upalpha)}$, thus varying $\upalpha$ to absorb $\theta$ to 0 would result in 
a compensating complex phase $\upalpha$ gained in the mean-field mass. Thus the 1+1d CT or P problem 
\emph{cannot} yet be solved by varying $\upalpha$ in \eq{eq:Z3450-vary}
in the chiral sector. \\

\noindent
$\bullet$  Now we only have the multi-fermion $g S^{(\upalpha)}_{\text{multi-}\Psi}$
or the equivalent disorder scalar $g S^{(\upalpha)}_{\text{disorder}}$ formulation with all configuration summed $\int [\cD \phi_1][\cD \phi_2]$.
The chiral fermions do not have any mean-field mass, but only have an interacting energy gap.
Varying $\upalpha$ to absorb $\theta$ to 0 would result in 
a compensating complex phase $\upalpha$ in the $g S^{(\upalpha)}_{\text{multi-}\Psi}$ and $g S^{(\upalpha)}_{\text{disorder}}$.
But this is just a shift of $\upalpha$ in the random disorder sample of disorder scalars. This shift of $\upalpha$ has no mean-field consequence.
It is easier to explain in the disorder scalar formulation, where $g S_{\text{disorder}} $ in \eq{eq:disorder} maps to
\begin{multline}
g S_{\text{disorder}}  \mapsto 
\int  \dd t \dd x \; g \;  \Big(
\e^{\ii (-2 \upalpha_{\rm 3rd} - 4 \upalpha_{\rm 4th} )} 
\phi_1^2 \psi_{L,3} \psi_{R,5}
+
\e^{\ii (-4 \upalpha_{\rm 3rd} - 8 \upalpha_{\rm 4th})} 
\phi_1^\dagger \psi_{L,4}^\dagger \psi_{R,0}\\
+\e^{\ii (8 \upalpha_{\rm 3rd} + 4 \upalpha_{\rm 4th} )} \phi_2^\dagger \psi_{L,3}\psi_{R,5}^\dagger 
+\e^{\ii (4 \upalpha_{\rm 3rd} + 2 \upalpha_{\rm 4th} )} \phi_2^2 \psi_{L,4} \psi_{R,0}
+\text{h.c.}
+\phi_1^\dagger\phi_1
+\phi_2^\dagger\phi_2\Big).
\end{multline}
The SMG no-mean field mass condition requires
$\< \phi_1 \> =\< \phi_2 \> =\< \phi^2_1 \> =\< \phi^2_2 \> = 0$,
so does
\bea
\< \e^{\ii (-2 \upalpha_{\rm 3rd} - 4 \upalpha_{\rm 4th} )} 
\phi_1^2  \> =
 \e^{\ii (-2 \upalpha_{\rm 3rd} - 4 \upalpha_{\rm 4th} )} 
\<\phi_1^2  \>=0.&\quad\quad
\< \e^{\ii (-4 \upalpha_{\rm 3rd} - 8 \upalpha_{\rm 4th})} 
\phi_1^\dagger \> =
\e^{\ii (-4 \upalpha_{\rm 3rd} - 8 \upalpha_{\rm 4th})} 
\<\phi_1^\dagger \> =0.
\cr
\< \e^{\ii (8 \upalpha_{\rm 3rd} + 4 \upalpha_{\rm 4th} )} \phi_2^\dagger 
 \> =
\e^{\ii (8 \upalpha_{\rm 3rd} + 4 \upalpha_{\rm 4th} )} 
 \<\phi_2^\dagger  \> =0.&
 \quad\quad
\< \e^{\ii (4 \upalpha_{\rm 3rd} + 2 \upalpha_{\rm 4th} )} \phi_2^2\> =
\e^{\ii (4 \upalpha_{\rm 3rd} + 2 \upalpha_{\rm 4th} )}\<\phi_2^2\> =0.
\eea
Thus, we can use $\upalpha_{\rm 3rd}$ and $\upalpha_{\rm 4th}$ to rotate
$\theta_{\rm 1st}$ and $\theta_{\rm 2nd}$ to zeros in \eq{eq:Z-full-chiral-SMG-vary},
while leave those $\upalpha$ undetectable in the disorder non-mean field mass term. 
Namely, $\bar\theta_{\rm 1st}=\theta_{\rm 1st}=0$ and $\bar\theta_{\rm 2nd}=\theta_{\rm 2nd}=0$.

Of course, in reality in the SM, we do have fermions already gaining mean-field mass from the Higgs condensation.
So what we could do better is to include both Higgs \eq{eq:Z3450-vary} and SMG contribution \eq{eq:Z-full-chiral-SMG-vary}.
The variation of the combined path integral, under
the $\U(1)_{\rm 3rd} \times \U(1)_{\rm 4th}$ symmetry transformations
$\Psi_{\rm I} \mapsto \e^{\ii q_{{\rm 3},\rm I}  \upalpha_{\rm 3rd}} \Psi_{\rm I}$
and $\Psi_{\rm I} \mapsto \e^{\ii  q_{{\rm 4},\rm I} \upalpha_{\rm 4th}} \Psi_{\rm I}$,
becomes
\bea \label{eq:Z3450-vary-Higgs-SMG}
Z &\mapsto& \int [\cD a_1][\cD a_2] [\cD \psi][\cD \bar\psi] \e^{\ii (
S_{\Psi,\rm gauge} + S_{\rm YH}^{(\upalpha)}
+ g S^{(\upalpha)}_{\text{multi-}\Psi}
+ \int  (\frac{1}{2 e^2} F^{\rm 1st} \wedge \star F^{\rm 1st})
+ \int  (\frac{1}{2 e^2} F^{\rm 2nd} \wedge \star F^{\rm 2nd})
+ \int (\upalpha_{\rm 3rd}  \dd \star  J_{\rm 3rd} 
+ \upalpha_{\rm 4th} \dd \star  J_{\rm 4th})
)} 
\cr
&& \e^{\ii \int \frac{1}{2 \pi} 
\big(
(\theta_{\rm 1st} -\upalpha_{\rm 3rd} ( q_{{\rm 1}}^{\intercal} \cdot K^{}  \cdot q_{{\rm 3}})
-\upalpha_{\rm 4th} ( q_{{\rm 1}}^{\intercal} \cdot K^{}  \cdot q_{{\rm 4}})) 
F^{\rm 1st}+ 
(\theta_{\rm 2nd} 
-\upalpha_{\rm 3rd} ( q_{{\rm 2}}^{\intercal} \cdot K^{}  \cdot q_{{\rm 3}})
-\upalpha_{\rm 4th} ( q_{{\rm 2}}^{\intercal} \cdot K^{}  \cdot q_{{\rm 4}})) 
F^{\rm 2nd}
\big)}.
\eea
In this case, as long as \emph{any} of the fermions do not get the mean-field mass at all 
(say, \emph{some} of the fermions get the entire mass from the SMG), then we could still solve the 1+1d CT or P problem.
For example, if we turn off the Yukawa-Higgs coupling between either $\psi_{L,3}^\dagger \psi_{R,0}$ or  
$\psi_{L,4}^\dagger \psi_{R,5}$, then we could still give all these 3-4-5-0 fermions' remaining mass by SMG.
Similarly, we could leave the original 3-4-5-0 chiral fermions with the Higgs-induced \emph{mean-field} mass,
but add a new family of 3-4-5-0 chiral fermions hidden above the high energy gapped by \emph{non-mean-field} SMG.
Then we could still do an appropriate chiral transformation on the purely SMG fermion (that has no mean-field mass),
which results in a redefinition of $\bar\theta$ to zero.
This solves the 1+1d CT or P problem without introducing the mirror fermion sector, 
which concludes \Sec{sec:Another-Solution} as a solution.

\newpage
\section{Another CT or P Solution via Symmetric Mass Generation on the mirror fermions}
\label{sec:Solution}

\begin{figure}[!h] 
\centering
\includegraphics[height=2.5in]{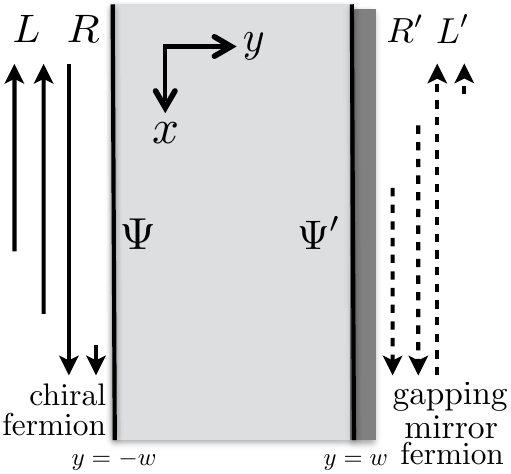} 
\caption{Our model has a chiral fermion $\Psi$ theory and a mirror fermion $\Psi'$ theory sharing the same internal symmetry,
placed on two domain walls, respectively shown on a spatial manifold
(e.g., two ends of a 2+1d regularizable manifold $\CM^2 \times I^1_{y}$). 
The mirror fermion is gapped by SMG at some energy scale around $\Lambda_{\SMG}$.
The solid-line arrows represent the chiral fermion modes $3_L$-$4_L$-$5_R$-$0_R$, 
and the dash-line arrows represent the mirror fermion modes $3_R$-$4_R$-$5_L$-$0_L$.
The shaded area means to schematically imply the dash-line mirror fermions are gapped by SMG and become absent from the low energy IR physics. 
}
\label{fig:1}
\end{figure}

\begin{figure}[!h] 
\centering
{{(a)}} \includegraphics[height=2.6in]{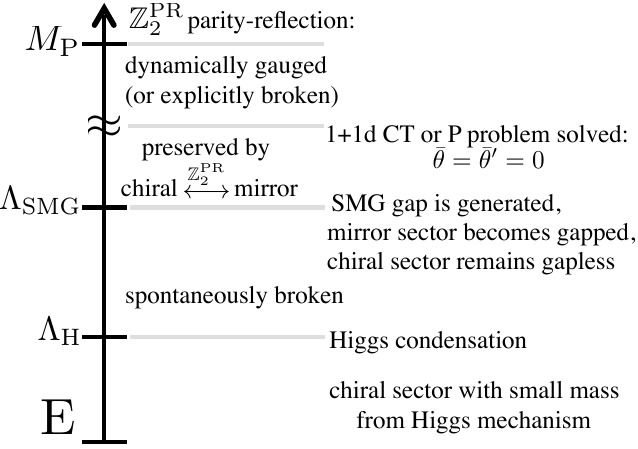}  
{{(b)}} \includegraphics[height=2.6in]{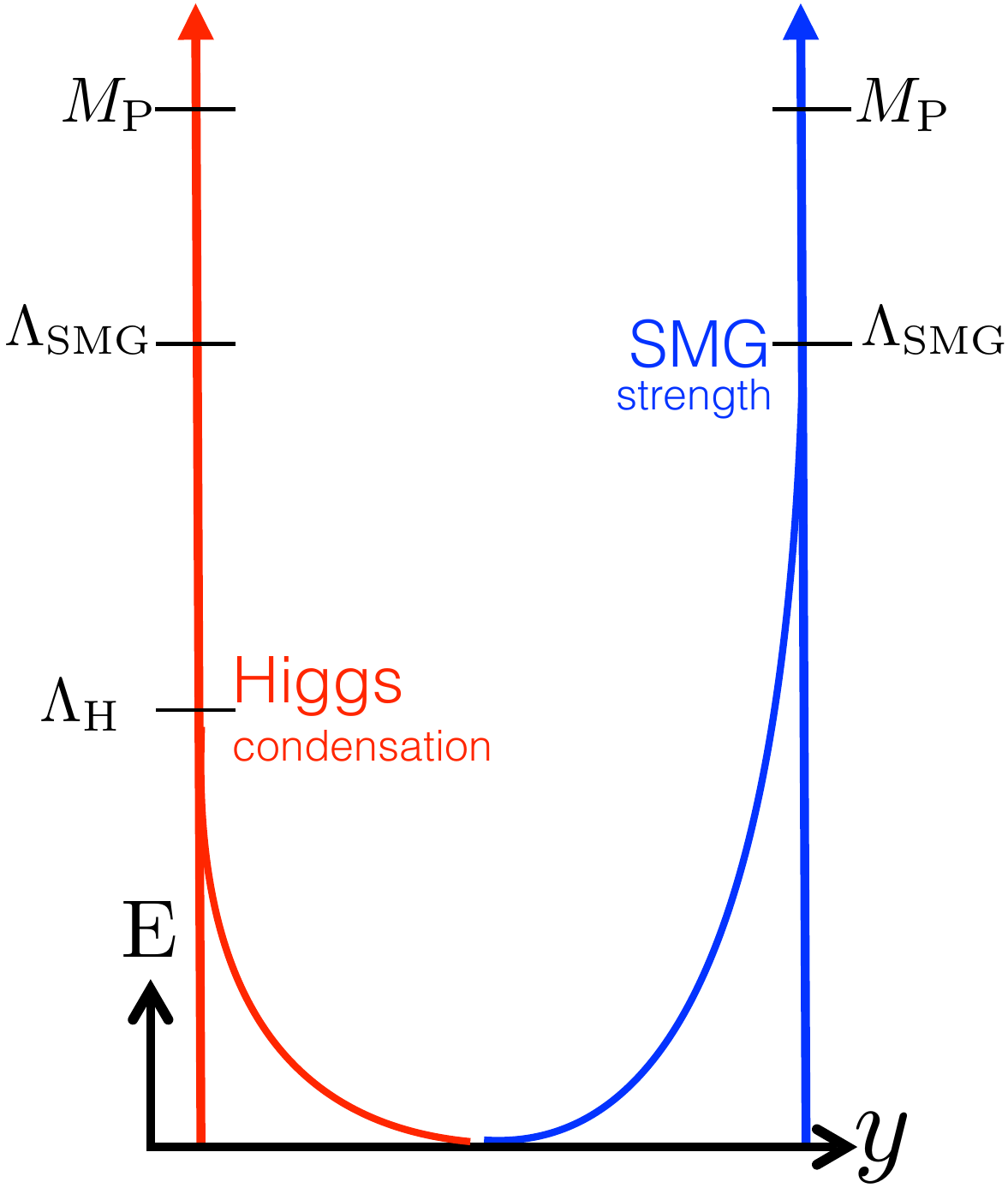} 
\caption{(a) The relations between the energy scale $\rE$ and the CT or P solution. 
%
%
Since the energy scale $\rE \sim  t^{-1}_{\rm period}$ is inverse of the time scale $ t_{\rm period}$,
this shows a time evolution of this quantum universe from the early universe
to the later universe.
(b) The Higgs condensation profile (red curve) becomes dominant at the energy scale $\Lambda_{\rm H}$
but becomes exponentially small when going into the bulk ($+\hat y$ direction).
On the other hand, the SMG strength becomes dominant at a higher energy scale $\Lambda_{\rm SMG}$,
also it becomes exponentially small when going into the bulk ($-\hat y$ direction). 
The horizontal axis labels the bulk direction,
while the vertical axis labels both the energy scale and also schematically the strength of (Higgs condensate or SMG) interaction terms. 
In principle, it is preferred that the Higgs and SMG dynamics do not interfere with each other in any spacetime region 
(i.e., the red and blue curves do not both have nonzero values at the same region). 
However, even if the Higgs condensation and SMG interfere at the same spacetime region, 
as long as the SMG does not generate any mean-field mass as Higgs condensation does,
we can still maintain our solution of the 1+1d CT or P problem
to the low energy (IR below the $\Lambda_{\rm H}$ scale).
}
\label{fig:2}
\end{figure}

We provide another new type of solution to the P problem \eq{eq:CT-problem} via Symmetric Mass Generation (SMG).
A schematic spatial configuration of the solution is shown in \Fig{fig:1} (a).
The prevailing physical mechanisms at different energy scales $\rE$
are shown in \Fig{fig:2} (b).
We summarize the solution by enumerating the setup step by step below. 


\begin{enumerate}[leftmargin=-0mm, label=\textcolor{blue}{\arabic*}., ref={\arabic*}]
\item {\bf 1+1d theory on the boundary of a finite-width 2+1d bulk}:
For the convenience, 
the chiral fermion action \Eq{eq:free} 
can be regarded as the 1+1d boundary theory on ${\cM^2}$ 
of a 2+1d bulk abelian $K$-matrix Chern-Simons theory $S_{\rm bulk}$ on ${\cM^3}$ (here with a rank-4 matrix $K=\diag(1,1,-1,-1)$ and ${\rm I,J} \in \{1,2,3,4\}$)
as an invertible topological field theory (iTFT) but with the real-valued \emph{dynamical} gauge field ``${\rm a}$'' with a compact $\U(1)$ gauge group:
\bea 
\label{eq:CSbulk} 
\int [\cD {\rm a}]\exp(\ii S_{\rm bulk})=
\int [\cD {\rm a}]\exp(\ii \frac{K_{\rm IJ}}{4\pi}\int_{\cM^3}   {\rm a}_{\rm I} \wedge \dd {\rm a}_{\rm J}) = 
\int [\cD {\rm a}]\exp(\ii\frac{K_{\rm IJ}}{4\pi}\int_{\cM^3}  \dd t \dd x  \dd y  \, \varepsilon^{\mu\nu\rho} {\rm a}_{{\rm I},\mu} \partial_\nu {\rm a}_{{\rm J},\rho}).\;\;\;\;\;\;
\eea
This theory is known as four layers of \emph{integer quantum Hall states} in condensed matter.
The 1+1d massless fermion theory \eq{eq:free} on ${\cM^2}$
is a specific Lorentz invariant boundary condition of the 2+1d theory 
\eq{eq:CSbulk} on $\prt{\cM^3}={\cM^2}$.\footnote{The free massless
boundary condition is
${\rm a}_{{\rm I}, t}-K_{\rm IJ'}^{-1}V_{\rm J'J} {\rm a}_{{\rm J}, x}=0$.
We can further choose the velocity matrix $V_{\rm IJ}$ in the Lorentz invariant relativistic system 
as $V_{\rm IJ}=c \delta_{\rm IJ}$ with the speed of light $c=1$.
So for each component ${\rm I}$, we have ${\rm a}_{{\rm I}, t} -K_{\rm II}^{-1}  {\rm a}_{{\rm I}, x}=0$ to give the massless 1+1d theory \eq{eq:free} on the boundary of a 2+1d bulk.
See Appendix \ref{appendix:A} for details.}
The bulk theory \eq{eq:CSbulk} is the familiar abelian Chern-Simons theory with a $\U(1)^4$ gauge group and the apparent gauge transformation
${\rm a}_{\rm I} \to {\rm a}_{\rm I}+\dd \uplambda_{\rm I}$.
This 2+1d $\U(1)^4$ gauge theory also has the extra $\U(1)^4$ global symmetry.
Each of the $\U(1)^4$ global symmetry has its associate $\U(1)$ global symmetry current
that couples to the external background field $A_{\rm I}$ of the \Sec{sec:problem}.
The conserved current of the $\U(1)$ global symmetry is $J_{\rm I}^{\rm Bulk} = \star \frac{1}{2 \pi}\dd {\rm a}_{\rm I}$,
which obeys the current conservation $\dd \star J_{\rm I} = 0$
thanks to $\dd^2=0$ and the Bianchi identity, which holds strictly when ${\cM^3}$ is a closed manifold with no boundary.

We can choose the $\U(1)^4$ global symmetry to be compatible with \Table{table:3450charge},
so we redefine $J_{\al}^{\rm Bulk} = K_{\rm I J} q_{\al,\rm J}  (\star \frac{1}{2 \pi}\dd {\rm a}_{\rm I}) \equiv Q_{\al,\rm I}  (\star \frac{1}{2 \pi}\dd {\rm a}_{\rm I})$ 
where the ${\rm I}$ is summed over and the fixed $\al \in  \{1,2,3,4\}$ labels
the 1st, 2nd, 3rd, or 4th linear independent $\U(1)$ symmetry in \Table{table:3450charge}.
The external background field $A_{\rm I}$ couples to the bulk current via 
$A \wedge \star J_{\al}^{\rm Bulk} \sim Q_{\al,\rm I}  \frac{1}{2 \pi} A_{\rm I} \wedge \dd {\rm a}_{\rm I}$ for the $\U(1)_{\al{\text{-th}}}$ global symmetry.
The bulk-boundary correspondence maps the bulk current coupling term $A \wedge \star J_{\al}^{\rm Bulk}$
to the boundary current coupling term $A \wedge \star J_{\al}$ with the boundary current $J_{\al}$ given in \eq{eq:1st2ndcurrent} and \eq{eq:3rd4thcurrent},
up to a rescaling on the $K$ matrix between two conventions, explained in Appendix \ref{appendix:A} and a footnote.\footnote{Explained in Appendix \ref{appendix:A},
commonly there
are two different choices of 1+1d free fermion lagrangian $\cL_{\Psi,\rm free}$ written as 
$\ii \Psi_{\rm I}^\dagger ( K_{\rm IJ} \partial_t- V_{\rm IJ}\partial_x) \Psi_{\rm J}$
or
$\ii \Psi_{\rm I}^\dagger (\delta_{\rm IJ} \partial_t-K_{\rm IJ'}^{-1}V_{\rm J'J} \partial_x) \Psi_{\rm J}$.
This will affect a factor of $K$ matrix difference for 
the bulk current, written as 
$J_{\al}^{\rm Bulk} = K_{\rm I J} q_{\al,\rm J}  (\star \frac{1}{2 \pi}\dd {\rm a}_{\rm I}) \equiv Q_{\al,\rm I}  (\star \frac{1}{2 \pi}\dd {\rm a}_{\rm I})$
or
$J_{\al}^{\rm Bulk} = q_{\al,\rm I}  (\star \frac{1}{2 \pi}\dd {\rm a}_{\rm I})$.
Similarly for the boundary current $J^\mu_{\alpha}$, it could be written as
$\Psi_{\rm I}^\dagger
{K_{\rm I J} q_{\alpha,\rm J }}
\tau^\mu_{ L/R} 
\Psi_{\rm I} \equiv
\Psi_{\rm I}^\dagger
{ Q_{\alpha,\rm I }}
\tau^\mu_{ L/R} 
\Psi_{\rm I} $
or 
$\Psi_{\rm I}^\dagger  q_{{\al},\rm I} \tau^\mu_{ L/R} \Psi_{\rm I}$.
In \eq{eq:CSbulk} and the paragraphs before this footnote,
we use the first convention.
After this footnote, we switch to the second convention to match the previous  \eq{eq:1st2ndcurrent} and \eq{eq:3rd4thcurrent}.\\
It is transparent to see from the first convention, that the partition function depends on the external background field $A$
as $Z[A]=
\int [\cD {\rm a}]
\exp(\ii 
\frac{K_{\rm IJ}}{4\pi}\int_{\cM^3}   {\rm a}_{\rm I} \wedge \dd {\rm a}_{\rm J}
+\ii
Q_{\al,\rm I}  \frac{1}{2 \pi} A_{\rm I} \wedge \dd {\rm a}_{\rm I} )
=
\exp(\frac{ - \ii }{4 \pi} Q_{\al,\rm I} {K_{\rm IJ}^{-1}} Q_{\al,\rm J} A_{\rm I} \wedge \dd {\rm A}_{\rm I})
=
\exp(\frac{ - \ii }{4 \pi} q_{\al,\rm I} {K_{\rm IJ}^{}} q_{\al,\rm J} A_{\rm I} \wedge \dd {\rm A}_{\rm J})
\equiv 
\exp(\frac{ - \ii }{2} \sigma_{\rm H} A_{\rm I} \wedge \dd {\rm A}_{\rm J})$,
where the Hall conductance is $\sigma_{\rm H}
\equiv \frac{Q_{\al,\rm I} {K_{\rm IJ}^{-1}} Q_{\al,\rm J}}{2 \pi}
\equiv \frac{q_{\al,\rm I} {K_{\rm IJ}^{}} q_{\al,\rm J} }{2 \pi}
$. This $Z[A]$ is also an iTFT but now written in the real-valued external \emph{non-dynamical} background field $A$.
}

{If $\U(1)_{\al{\text{-th}}}$ is treated as a global symmetry, 
its 1-connection $q_{\al,\rm I}  A_{\rm I}$ is just a non-dynamical background probe gauge field.}
If instead we promote the non-dynamical $q_{\al,\rm I}  A_{\rm I}$ to the dynamical $q_{\al,\rm I}  a_{\rm I}$,
then the path integral's current coupling factor has to be adjusted to:
\bea
\text{Bulk: }&&
\exp(\ii   \frac{q_{\al,\rm I}}{2 \pi} \int A_{\rm I} \wedge \dd {\rm a}_{\rm I}) 
\mapsto
\int[\cD a] \exp(\ii   \frac{q_{\al,\rm I}}{2 \pi} \int a_{\rm I} \wedge \dd {\rm a}_{\rm I}). \quad \cr
\text{Boundary: }&&
\exp(\ii   \frac{1}{2 \pi} 
\int \dd t \dd x \; A_{\rm I, \mu}  
(\Psi_{\rm I}^\dagger  q_{{\al},\rm I} \tau^\mu_{ L/R} \Psi_{\rm I})) 
\mapsto
\int[\cD a] \exp(\ii   \frac{1}{2 \pi} 
\int \dd t \dd x \; a_{\rm I, \mu}  
(\Psi_{\rm I}^\dagger  q_{{\al},\rm I} \tau^\mu_{ L/R} \Psi_{\rm I})).
\eea
Due to 
the mixed anomaly between 
$\U(1)_{\rm 1st} \times \U(1)_{\rm 2nd}$ and $\U(1)_{\rm 3rd} \times \U(1)_{\rm 4th}$,
we will only gauge $\U(1)_{\rm 1st} \times \U(1)_{\rm 2nd}$
to get a consistent 1+1d theory. 
These mixed anomalies can be captured by Laughlin's style of
the flux insertion thought experiment \cite{Laughlin1981PRB}, namely threading a magnetic flux through 
the hole of the annulus or cylinder strip to induce the electric field and the boundary-bulk anomalous current transport 
\cite{Wang2013ytaJW1307.7480, SantosWang1310.8291}.\footnote{{After gaining an analytic understanding, 
later we will demonstrate Laughlin's style of thought experiment together with the SMG solution to the 1+1d CT or P problem 
by physical pictures in the Conclusion \Sec{sec:Conclusion}.}}

In fact, if we choose the 2+1d bulk has only a finite width (say, along the $y$ direction so $- w \leq y  \leq w$), then the whole theory can be effectively regarded as a 1+1d system \cite{Wang2013ytaJW1307.7480, Wang2018ugfJW1807.05998, ZengZhuWangYou3450SMG2202.12355}.
We can just choose the chiral fermion $\Psi$ on one edge on $y=-w$, there are two possible scenarios on the other edge on $y=w$:
(1) In one scenario, we choose a symmetric gapped boundary condition on the other edge $y=w$; 
then we shrink the size of $w$ while lifting the mirror edge energy gapped spectrum from $y=w$ to a higher energy. 
(2) In another scenario, we have the mirror fermion $\Psi'$ as the fermion doubling \cite{NielsenNinomiya1981hkPLB} on the other edge $y=w$;
then we can further introduce the SMG to fully gap the mirror fermion at some energy scale 
\cite{Wang2013ytaJW1307.7480, Wang2018ugfJW1807.05998, ZengZhuWangYou3450SMG2202.12355}.
To resolve the 1+1d CT or P problem of the original chiral fermion theory, we will start with this second scenario.

\item {\bf Fermion doubling --- chiral fermion and mirror fermion spectra}:
\label{Sol-Remark:3}
The 1+1d chiral fermion theory can be defined nonperturbatively (for example, on a UV regularization such as a lattice 
\cite{Wang2013ytaJW1307.7480, Wang2018ugfJW1807.05998, ZengZhuWangYou3450SMG2202.12355}).
But n\"aively, the 1+1d chiral fermion theory suffers from the Nielsen-Ninomiya (NN) fermion-doubling problem \cite{NielsenNinomiya1981hkPLB} ---
There are chiral fermions and mirror fermions on two domain walls separated from each other from a finite-width 2+1d bulk.
The chiral and mirror fermions localized on the two domain walls are related to the lattice domain wall fermion construction \cite{Kaplan1992A-method}.
The mirror fermion sector is the generalized parity ($\cP \equiv \rP \rR$) and 
the generalized time-reversal ($\cT \equiv \rT \rR$) partner of the chiral fermion sector (see the detail in Remark \ref{Sol-Remark:3}).
We can use the subindices $L$ and $R$ to denote the chirality of Weyl fermions.
In the chiral fermion theory side, the four Weyl fermions 
$\Psi_{\rm I} \equiv \Big(\begin{smallmatrix}
\psi_{L,3}\\
\psi_{L,4}\\
\psi_{R,5}\\
\psi_{R,0}
\end{smallmatrix}\Big)_{\rm I}$ 
carry U(1) charges: $3_L$, $4_L$, and $5_R$, $0_R$.
In the mirror fermion theory side, the four Weyl fermions 
$\Psi'_{\rm I} \equiv \Big(\begin{smallmatrix}
\psi'_{R,3}\\
\psi'_{R,4}\\
\psi'_{L,5}\\
\psi'_{L,0}
\end{smallmatrix}\Big)_{\rm I}$
carry U(1) charges: $3_R$, $4_R$, and $5_L$, $0_L$.
On the chiral fermion side, we have the free fermion action \Eq{eq:free}.
Similarly, on the mirror fermion side, we have the free mirror fermion action (due to the fermion doubling):
 \bea \label{eq:free-mirror}
 S_{\Psi',\rm free}=\int  \dd t \dd x  \big(
 \ii \psi'^\dagger_{R,3} (\partial_t+\partial_x) \psi_{R,3}
 + \ii\psi'^\dagger_{R,4} (\partial_t+\partial_x) \psi_{R,4}
+ \ii \psi'^\dagger_{L,5} (\partial_t-\partial_x) \psi_{L,5}
+\ii\psi'^\dagger_{L,0} (\partial_t-\partial_x) \psi_{L,0}\big). \quad
\eea

\item {\bf Discrete R, C, $\cP=$ PR and $\cT=$ TR symmetry transformations}: 
\label{Sol-Remark:3}
{Here we follow the systematic analysis on the C-P-T symmetries together with the fermion parity $\Z_2^{\rm F}$ in \cite{WangCPT2109.15320}.}
In particular, we generalize the 
previous C-P-T transformation in \eq{eq:CPT}
to include the full theory with two domain walls.
We focus on the field configurations on the two domain walls at $y=-w$ and $y=+w$.
Other than the chiral fermion $\Psi$ at $y=-w$  and the mirror fermion $\Psi'$ at $y=+w$,
there can also be additional bosonic fields $\Phi$ at $y=-w$ and $\Phi'$ at $y=+w$,
also the same set of gauge field $A$ at $y=-w$ and $y=+w$ that can leak to the finite-width bulk. 

Let us spell out the precise symmetry transformations for the \emph{unitary} reflection R, charge conjugation C, generalized parity $\cP= \rP \rR$, 
and the \emph{anti-unitary} generalized time reversal $\cT = \rT \rR$ of this system with chiral and mirror fermions (\Fig{fig:1})
in terms of the active transformations on the multiplet of fermionic fields, bosonic fields, and gauge field: 
\bea
\label{eq:R}
\Z_2^{\rm R} &:& 
\Psi(t,x,-w) 
\equiv \Big(\begin{smallmatrix}
\psi_{L,3}(t,x)\\
\psi_{L,4}(t,x)\\
\psi_{R,5}(t,x)\\
\psi_{R,0}(t,x)
\end{smallmatrix}\Big) 
\mapsto \Psi'(t,x,w)
\equiv
 \Big(\begin{smallmatrix}
\psi'_{R,3}(t,x)\\
\psi'_{R,4}(t,x)\\
\psi'_{L,5}(t,x)\\
\psi'_{L,0}(t,x)
\end{smallmatrix}\Big). \quad 
\Phi(t,x,-w) 
\mapsto 
\Phi'(t,x,w).
\;
\left\{
\begin{array}{lcr}
A(t,x,-w)
&\mapsto&
A(t,x,w).\\
F(t,x,-w)
&\mapsto& 
F(t,x,w).
\end{array}
\right.
\quad\quad
\\
\label{eq:C}
\Z_2^{\rm C} &:& \Psi(t,x,-w)  
\mapsto (-)^L\Psi^*(t,x,-w)
\equiv
 \Big(\begin{smallmatrix}
-\psi^*_{L,3}(t,x)\\
-\psi^*_{L,4}(t,x)\\
\psi^*_{R,5}(t,x)\\
\psi^*_{R,0}(t,x)
\end{smallmatrix}\Big). \quad 
\Phi(t,x,-w) 
\mapsto 
\Phi^*(t,x,-w). \;
\left\{
\begin{array}{lcr}
A(t,x,-w)
&\mapsto&
-A(t,x,-w).\\
F(t,x,-w)
&\mapsto& 
-F(t,x,-w).
\end{array}
\right.
\quad
\\
\label{eq:P}
\Z_2^{\cP} &:& \Psi(t,x,-w)   
\mapsto \Psi'(t,-x,w)
\equiv
 \Big(\begin{smallmatrix}
\psi'_{R,3}(t,-x)\\
\psi'_{R,4}(t,-x)\\
\psi'_{L,5}(t,-x)\\
\psi'_{L,0}(t,-x)
\end{smallmatrix}\Big). \quad 
\Phi(t,x,-w) 
\mapsto 
\Phi'(t,-x,w).
\quad
\left\{
\begin{array}{lcr}
A_0(t,x,-w)
&\mapsto& 
A_0(t,-x,w).\\
A_1(t,x,-w)
&\mapsto& 
-A_1(t,-x,w).\\
F_{01}(t,x,-w)
&\mapsto&
-F_{01}(t,-x,w).
\end{array}
\right.
\quad\quad
 \\
\label{eq:T}
\Z_2^{\cT} &:& \Psi(t,x,-w)   
\mapsto \Psi'(-t,x,w)
\equiv
 \Big(\begin{smallmatrix}
\psi'_{R,3}(-t,x)\\
\psi'_{R,4}(-t,x)\\
\psi'_{L,5}(-t,x)\\
\psi'_{L,0}(-t,x)
\end{smallmatrix}\Big).
\quad 
\Phi(t,x,-w) 
\mapsto 
\Phi'(-t,x,w).
\quad
\left\{
\begin{array}{lcr}
A_0(t,x,-w)
&\mapsto&
A_0(-t,x,w).\\
A_1(t,x,-w)
&\mapsto& 
-A_1(-t,x,w).\\
F_{01}(t,x,-w)
&\mapsto&
F_{01}(-t,x,w).\\
\end{array}
\right.
\quad\quad
\eea 
The precise $\Phi$ content will soon become clear when we build the full model including interactions --- 
later $\Phi$ will stand for the multiplet of scalars including 
the Higgs or disorder scalars $(\phi_H, \phi_1, \phi_2)$ on the domain wall $y=-w$,
and similarly  $\Phi'$ stands for $(\phi'_H, \phi'_1, \phi'_2)$ on the wall $y=w$.
In the above, we write the symmetry transformations of $\Psi$ and $\Phi$;
{similarly, we have analogous transformations for $\Psi'$ and $\Phi'$.}

Below let us check what are the symmetries of the chiral fermion theory,
without or with the mirror sector, and without or with the $\theta$ term:

$\bullet$ 
The 1+1d U(1) symmetric free massless chiral fermion 3-4-5-0 theory alone is only C symmetric, but it violates the reflection R, and violates the ordinary P and T symmetries (thus being chiral \cite{Wang2013ytaJW1307.7480}).
The generic $\theta$ term
$\frac{\theta}{2 \pi} \int F$ violates C, CT, and P symmetries,
but preserves T symmetry.
So the 1+1d chiral fermion theory together with a generic $\theta$ term
violates all R, C, P, and T symmetries. 

$\bullet$ 
The parent theory with free massless chiral and mirror fermions (above the energy $\rE > \Lambda_{\SMG}$)
is not only C symmetric but also $\cP$ and $\cT$ symmetric, however not R symmetric. 
Notice that the full parent theory
has the generalized $\cP$ and $\cT$ symmetries 
that contain not merely the ordinary P and T symmetries but also the R symmetry. 
Namely, we have already defined the ordinary PR $\equiv \cP$ as the generalized $\cP$ in \eq{eq:P},
and the ordinary TR $\equiv \cT$ as the generalized $\cT$ in \eq{eq:T}.

\item {\bf Theta term in the parent theory with chiral and mirror fermions}: 
\label{Sol-Remark:4}
The parent theory can include a generic $\theta$ term with the field strength $F=\dd A$ for some $\U(1)$ internal symmetry group (e.g., in our \Table{table:3450charge},
this $\U(1)$ can be chosen as either $\U(1)_{\rm 1st}$ or $\U(1)_{\rm 2nd}$)
on two domain walls as
\bea \label{eq:thetaF-theta'F}
\frac{\theta}{2 \pi} \int_{\cM^2} F(t,x,-w) + \frac{\theta'}{2 \pi} \int_{\cM'^2} F(t,x,w).
\eea
One domain wall has its 2d spacetime ${\cM^2}$, the other has its 2d spacetime ${\cM'^2}$. 
These 2d spacetime can be generally curved manifolds, 
where the ${\cM^2} \sqcup {{\overline{\cM'}^2}}= \partial \cM^3$ as the two boundaries of a 2+1d bulk manifold $\cM^3$,
with the overline ${\overline{\cM'}^2}$ implying the orientation reversal ${\cM'}^2$. 
For the specific configuration in \Fig{fig:1} (a),
we can choose ${\cM^2}={\cM^1_{\text{space}}} \times {\cM^1_{\text{time}}}$ and ${\cM'^2}={\cM'^1_{\text{space}}} \times {\cM^1_{\text{time}}}$,
where the 1d time ${\cM^1_{\text{time}}}$ is shared 
and the ${\cM^1_{\text{space}}} \sqcup {\overline{\cM}^1_{\text{space}}}$ is the boundary of a 2-dimensional strip or a cylinder.
The generic $\theta$ and $\theta'$ vacua violate
the R, C, $\cP$, and $\cT$ symmetries, if we treat the abelian gauge field 
on both sides of domain walls distinctly.

However, the gauge fields on both sides are the same gauge field,
so we could even combine the effects into
a combined theta term:
\bea \label{eq:theta-theta'}
\frac{\theta+\theta'}{2 \pi} \int F(t,x).
\eea
In this case, we have to consider a finite-width strip as \Fig{fig:1} (a) 
so we set the electric field strength $F(t,x,-w)=F(t,x,w)$
while ${\cM^2}$ is identified as ${\cM'^2}$ in the spacetime integration range.
The $F_{01}(t,x,-w)=F_{01}(t,x,w)=E_1$ 
identified as the same electric field
on two sides of the strip 
can be precisely derived from 
Laughlin's flux insertion thought experiment \cite{Laughlin1981PRB}.

\item {\bf Impose the $\Z_2^{\rm P R} \equiv \Z_2^{\cP}$ parity-reflection symmetry on the theta term}: 
\label{Sol-Remark:5} 
If we impose the PR symmetry
on the parent theory with chiral and mirror fermions 
and with the ${\theta+\theta'}$ term in \eq{eq:theta-theta'}
(above the energy $\rE > \Lambda_{\SMG}$),
then the PR symmetry in \eq{eq:P} demands that the $\theta$ term on the domain wall on ${\cM^2}$
maps to the $\theta'$ term on the domain wall on ${\cM'^2}$
\bea
\frac{\theta}{2 \pi} \int_{\cM^2} F(t,x,-w) \mapsto 
\frac{\theta}{2 \pi} \int_{\cM'^2} - F(t,-x,w) =\frac{-\theta}{2 \pi} \int_{\cM'^2} F(t,x,w).
\eea
Note that the right-hand side is the $\frac{\theta'}{2 \pi} \int_{\cM'^2} F(t,x,w)$ term on ${\cM'^2}$. 
Thus, the PR symmetry for the full parent theory 
implies that ${\theta'} =- \theta$, 
\bea \label{eq:theta=0}
\theta + \theta' =0 
\quad 
\Rightarrow 
\quad \frac{\theta+\theta'}{2 \pi} \int F(t,x) =0.
\eea
Namely \eq{eq:thetaF-theta'F} and \eq{eq:theta-theta'} vanishes. 
Note that (1) the chiral fermions on ${\cM^2}$ and the mirror fermions on ${\cM'^2}$ have the opposite chirality,
and (2) their chiral symmetries (with the opposite chiralities on the two domain walls) 
coupled to the same U(1) gauge field, the chiral U(1) symmetry transformation will rotate $\theta \mapsto \theta + \upalpha$
and $\theta' \mapsto \theta' - \upalpha$ oppositely, but keeps the
$\theta + \theta' \mapsto \theta + \theta'$ invariant.
Because the above two reasons,
this means that the PR $=\cP$ symmetry
at the parent theory solves the zero theta angle problem at a high energy,
since $\theta + \theta' =0$ and the chiral transformation with an appropriate $\upalpha$
allows us to choose both $\theta=0$ and $\theta'=0$.

\item 
{\bf  Solution to the 1+1d CT or P problem by the imposed $\Z_2^{\rm P R} \equiv \Z_2^{\cP}$ symmetry at UV and the SMG at intermediate energy}:
\label{rk:Z2R}

We proceed our solution to the 1+1d CT or P problem by imposing the
$\Z_2^{\rm P R} \equiv \Z_2^{\cP}$ symmetry further on the fully interacting parent theory,
not just on the free fermions and the $\theta+\theta'$ term,
but also on the interaction terms including the Yukawa-Higgs terms and the SMG multi-fermion terms that we will introduce.

The $\Z_2^{\rm R}$ reflection symmetry along the $y=0$ axis can map the edge theory on one side $y = -w$ 
to the mirror edge theory on the other side $y = w$. The $\Z_2^{\rm R}$ symmetry transformation sends
$y \mapsto -y$ in the passive coordinate transformation, or sends the fields in the active transformation.
Due to the chirality of the chiral fermion at $y = -w$ is opposite to the mirror fermion at $y = w$, 
even the full free massless chiral and mirror fermion theory does not preserve the $\Z_2^{\rm R}$ symmetry.

However, there is also a 1+1d parity symmetry $\Z_2^{\rm P}$ in \eq{eq:CPT-psi} 
that can flip the chirality of chiral fermions.
So we can consider a combined parity-reflection symmetry $\Z_2^{\rm PR}$.
{The $\Z_2^{\rm PR}$'s passive transformation on the coordinates $x \mapsto -x,  \; y \mapsto -y$ is obvious, 
but we shall follow the active transformation \eq{eq:P} on the fields in this work.}
The full free massless chiral and mirror fermion theory does preserve the $\Z_2^{\rm PR}$ symmetry.
In fact, the $\Z_2^{\rm PR}$ symmetry is the $\pi$-rotation in the $x$-$y$ plane with respect to the 2+1d theory.
But here the concerned $\Z_2^{\rm PR}$ symmetry only needs to map between the two 1+1d domain walls.
 
$\bullet$ $\rE > \Lambda_{\SMG}$: We hypothesize that a $\Z_2^{\rm PR}$ reflection symmetry is preserved at some deep ultraviolet (UV) high energy 
full parent theory including interaction terms
above some energy scale 
$\rE > \Lambda_{\SMG}$, while the SMG dominates below $\Lambda_{\SMG}$.
So a full parent field theory with an action 
\bea
{\Z_2^{\rm PR}}: S_{\rm UV}=\int \dd t \dd x \big( 
\cL_{\rm UV}(t, x, -w ) + \cL_{\rm UV}(t, x,w)
\big) \;  {\mapsto} \;  S_{\rm UV}
\eea 
with its Lagrangian density $\cL_{\rm UV}(x,y,t)$ on $y = -w$ and $y = w$
at UV respects a $\Z_2^{\rm PR}$ symmetry.

This implies that the $\Z_2^{\rm PR}$ symmetry is preserved at some UV scale, for the fermion doubling, 
all field contents, the kinetic, theta, and interaction terms in the Lagrangian at $y=-w$ and $y=w$, 
at least kinematically.

$\bullet$ $\rE \geq  M_{\rm P}$ while $M_{\rm P} \gg \Lambda_{\SMG}$: 
If the quantum gravity effect is involved at some Planck energy scale $M_{\rm P}$,
then preferably the symmetries (including $\Z_2^{\rm PR}$) can be either dynamically gauged or explicitly broken.

$\bullet$ $\Lambda_{H} <E < \Lambda_{\SMG}$: 
Below the SMG energy scale $\Lambda_{\SMG}$ 
and above the Higgs condensation energy scale $\Lambda_{H}$, 
the mirror fermion sector shall be fully gapped by the dominant SMG effect.

$\bullet$ $\rE < \Lambda_{H}$: 
Below the Higgs condensation energy scale $\Lambda_{H}$, the condensed Higgs gives the symmetry-breaking mass to the chiral fermion sector.

Since the energy scale $\rE \sim  t^{-1}_{\rm period}$ is inverse proportional to the time scale $ t_{\rm period}$,
we can also regard \Fig{fig:2} (b) as the time evolution of this quantum universe from the beginning (higher energy $\rE$ and small $t$ at UV as the early universe)
to the late time (lower energy $\rE$ and large $t$ at IR as the late universe).

\item {\bf Yukawa-Higgs term for the fermion doubling}:
\label{Sol-Remark:YH}
To respect the $\Z_2^{\rm PR}$ for a parent theory at the energy $\rE > \Lambda_{\SMG}$,
we introduce the same kind of YH term on the chiral and mirror fermion sectors:
\bea
\label{eq:YH-Chiral}
  S_{{\text{YH-}}\Psi}
& \equiv& 
 \int  \dd t \dd x  \;
(\phi_H^3 \e^{\ii \vartheta_{3,0} }\psi_{L,3}^\dagger \psi_{R,0} + 
\phi_H^\dagger \e^{\ii \vartheta_{4,5} } \psi_{L,4}^\dagger \psi_{R,5}
+\text{h.c.} 
+ |D_\mu \phi_H|^2
- \mu_H |\phi_H|^2 - \lambda_H |\phi_H|^4 ).\cr
\label{eq:YH-Mirror}
 S_{{\text{YH-}}\Psi'}
& \equiv&
 \int  \dd t \dd x \; 
(\phi'^3_H \e^{\ii \vartheta_{3,0} } {\psi'^\dagger_{R,3}} \psi'_{L,0} + 
\phi'^\dagger_H \e^{\ii \vartheta_{4,5} } \psi'^\dagger_{R,4} \psi'_{L,5}
+\text{h.c.} 
+ |D_\mu \phi_H|^2
- \mu'_H |\phi'_H|^2 - \lambda'_H |\phi'_H|^4 ). 
\eea
Note that $\e^{\ii \vartheta_{3,0}}$ and $\e^{\ii \vartheta_{4,5}}$ have the same complex values 
in the $S_{{\text{YH-}}\Psi}$ and $S_{{\text{YH-}}\Psi'}$ to respect the $\Z_2^{\rm PR}$ symmetry.
There are two possible choices of Higgs profiles.
(1) We can choose different Higgs fields $\phi_H$ and $\phi_H'$ for the chiral and mirror sectors.
(2) Alternatively, we may even choose the same Higgs field on the chiral and mirror sectors (so the Higgs 
$\phi_H = \phi_H'$ can propagate from either of two 1+1d domain walls to the 2+1d finite-width bulk).

The Yukawa-Higgs mass matrices from \eq{eq:YH-Chiral}
for the chiral sector 
and
for the mirror sector
are respectively:
\bea \label{eq:YH-matrix}
(\psi_{L,3}^\dagger,  \psi_{L,4}^\dagger) 
M_{\rm H} 
\big(\begin{smallmatrix} 
\psi_{R,0}\\
\psi_{R,5}
\end{smallmatrix} \big) + {\rm h.c.}
&\equiv&
(\psi_{L,3}^\dagger,  \psi_{L,4}^\dagger) 
\begin{pmatrix} 
\phi_H^3 \e^{\ii \vartheta_{3,0} } & 0\\
0 & \phi_H^\dagger \e^{\ii \vartheta_{4,5} } 
\end{pmatrix} 
\big(\begin{smallmatrix} 
\psi_{R,0}\\
\psi_{R,5}
\end{smallmatrix} \big)+ {\rm h.c.}
\cr
(\psi'^\dagger_{L,3},  \psi'^\dagger_{L,4}) 
M'^{\dagger}_{\rm H} 
\big(\begin{smallmatrix} 
\psi'_{R,0}\\
\psi'_{R,5}
\end{smallmatrix} \big)+ {\rm h.c.}
&\equiv&
(\psi'^\dagger_{L,3},  \psi'^\dagger_{L,4}) 
\begin{pmatrix} 
\phi'^{\dagger 3}_H
\e^{-\ii \vartheta_{3,0} } & 0\\
0 & \phi'_H \e^{-\ii \vartheta_{4,5} } 
\end{pmatrix} 
\big(\begin{smallmatrix} 
\psi'_{R,0}\\
\psi'_{R,5}
\end{smallmatrix} \big)+ {\rm h.c.}
\eea
We shall set $\phi_H$ and $\phi'_H$ to be real-valued,
and the complex phases ($\vartheta_{3,0}$ and $\vartheta_{4,5}$)
capture the full complexity.

%

\item {\bf SMG by multi-fermion interactions}:
We follow the 1+1d SMG in \eq{eq:multi-fermion-SMG} by the multi-fermion interactions 
\cite{Wang2013ytaJW1307.7480, Wang2018ugfJW1807.05998, ZengZhuWangYou3450SMG2202.12355, Tong2104.03997}
that satisfy a mathematically rigorous gapping condition \cite{Haldane1995Stability, KapustinSaulina1008.0654KS, Wang2015Boundary, Levin1301.7355} (that defines a topological gapped boundary condition of the 2+1d bulk Chern-Simons theory).
To respect the $\Z_2^{\rm PR}$ symmetry for a parent theory at the energy $\rE > \Lambda_{\SMG}$,
we introduce the same multi-fermion interactions for the chiral and mirror fermion sectors:
\bea
\hspace{-6mm}
\label{eq:multi-fermion}
g S_{\text{multi-}\Psi} &\equiv& 
 \int  \dd t \dd x \; \big(
 {g}_{1}   ({\psi}_{L,3}) 
(  {\psi}_{L,4}^\dagger)^2
( {\psi}_{R,5}  )(  {\psi}_{R,0})^2 
+  {g}_{2}     ( {\psi}_{L,3} )^2
({\psi}_{L,4})
( {\psi}_{R,5}^\dagger )^2
(  {\psi}_{R,0})+\text{h.c.} \big).\cr
\hspace{-6mm}
g' S_{\text{multi-}\Psi'} &\equiv&  \int  \dd t \dd x \; \big(
 {g}'_{\rI}   ({\psi}'_{R,3}) 
(  \psi'^\dagger_{R,4})^2
( \psi'_{L,5}  )(  \psi'_{L,0} )^2
+  {g}'_{2}     ( \psi'_{R,3})^2
(\psi'_{R,4})
( \psi'^\dagger _{L,5})^2
(  \psi'_{L,0})+\text{h.c.} \big).
\eea
The multi-fermion interactions are derived based on the bosonization method \cite{Wang2013ytaJW1307.7480, Wang2018ugfJW1807.05998}.  
Since the fermions are Grassmann numbers, all higher power (larger than 1) of 
any fermion needs to be point-splitted.
These terms preserve the Lorentz spacetime Spin group, and the internal $\U(1)_{\rm 1st}$ and  $\U(1)_{\rm 2nd}$ symmetries.
The coupling $g$ and $g'$ control the coupling ${g}_{1}$,  ${g}_{2}$, $ {g}'_{1}$,  and ${g}'_{2}$
in a dynamical manner --- in the sense that $g$ and $g'$ can depend on the time or energy scales (or the length or momentum scales)
illustrated in \Fig{fig:2}.

\item {\bf SMG by disorder-scalar interactions}: 
Follow \eq{eq:disorder-SMG}, we can rewrite the multi-fermion interactions.
Integrating out the disorder scalar $\int [\cD \phi_1][\cD \phi_1'][\cD \phi_2][\cD \phi_2']$
in the interaction term 
below \cite{ZengZhuWangYou3450SMG2202.12355}
derives the multi-fermion interactions \eq{eq:multi-fermion}: 
\bea
\label{eq:disorder}
\hspace{-6mm}
&&g S_{\text{disorder}} \equiv \int  \dd t \dd x \;  \Big(
\phi_1^2 \psi_{L,3} \psi_{R,5}
+\phi_1^\dagger \psi_{L,4}^\dagger \psi_{R,0}
+\phi_2^\dagger \psi_{L,3}\psi_{R,5}^\dagger 
+\phi_2^2 \psi_{L,4} \psi_{R,0}
+\text{h.c.}
+\frac{1}{{\tilde{g}_1}} \phi_1^\dagger\phi_1
+\frac{1}{{\tilde{g}_2}} \phi_2^\dagger\phi_2\Big)
.\cr
&&g' S_{\text{disorder}'} \equiv \int  \dd t \dd x  \;  \Big(
\phi'^2_1 \psi'_{R,3} \psi'_{L,5}
+\phi'^\dagger_1 \psi'^\dagger_{R,4} \psi'_{L,0}
+\phi'^\dagger_2 \psi'_{R,3}\psi'^\dagger _{L,5}
+\phi'^2_2 \psi'_{R,4} \psi'_{L,0}
+\text{h.c.}
+\frac{1}{{\tilde{g}_1}} \phi'^\dagger_1\phi'_1
+\frac{1}{{\tilde{g}_2}} \phi'^\dagger_2\phi'_2\Big)
.\quad\quad
\eea
We do not have to introduce explicit kinetic terms for these scalar fields ($\phi_1, \phi_1', \phi_2, \phi_2'$).
We also shall include any random configuration of these scalar fields in the spacetime when doing the path integral $\int [\cD \phi_1][\cD \phi_1'][\cD \phi_2][\cD \phi_2']$.
The coupling is related by $g_\alpha \sim {\tilde{g}_\alpha}^2$.

\item {\bf $\bar\theta$ and $\bar\theta'$ terms that include the theta terms and complex phases of mass matrices}: 
Now we aim to show that in the full UV parent theory,
the analogous $\bar\theta$ and $\bar\theta'$ terms like \eq{eq:bar-theta} vanish once the $\Z_2^{\rm PR}$ is imposed.

There could be theta terms for both  $\U(1)_{\rm 1st} \times \U(1)_{\rm 2nd}$.
Denote the chiral sector's as $\theta_1$ and $\theta_2$,
and the mirror sector's as $\theta_1'$ and $\theta_2'$.
The $\Z_2^{\rm PR}$ symmetry at some higher energy imposes that \eq{eq:theta=0}, so
$\theta_1 + \theta_1' =0$ and $\theta_2 + \theta_2' =0$.

In the Higgs condensed low energy theory, breaking $\U(1)_{\rm 1st} \times \U(1)_{\rm 2nd}$ to $\U(1)_{\rm vector}$,
we can also pay attention to the  $\Z_2^{\rm PR}$ symmetry on the unbroken $\U(1)_{\rm vector}$'s theta term of the chiral and mirror sector: 
\bea \label{eq:thetavv'=0}
\theta_v + \theta_v' =0.
\eea
Follow \eq{eq:bar-theta-} and \eq{eq:YH-Mirror}, we have
${\bar{\theta}_{\rm v}} \equiv {\theta_{\rm v}}- {\bf q}_{{\rm v}}^{\intercal} \cdot \arg (\diag M) = {\theta_{\rm v}}-\vartheta_{3,0} - 3 \vartheta_{4,5}$
and 
${\bar{\theta}'_{\rm v}} \equiv {\theta'_{\rm v}}- {\bf q}_{{\rm v}}^{\intercal} \cdot \arg (\diag M'^\dagger) = {\theta'_{\rm v}} + \vartheta_{3,0} + 3 \vartheta_{4,5}$.
Notice that the overall
\bea \label{eq:barthetavv'}
{\bar{\theta}_{\rm v}} + 
{\bar{\theta}'_{\rm v}} \equiv 
\big( {\theta_{\rm v}}- {\bf q}_{{\rm v}}^{\intercal} \cdot \arg (\diag M) \big)
+
\big( {\theta'_{\rm v}} - {\bf q}_{{\rm v}}^{\intercal} \cdot \arg (\diag M'^\dagger) \big)
={{\theta}_{\rm v}} + 
{{\theta}'_{\rm v}},
\eea
the total mass matrix contributions to the complex phases cancel out,
because (1) the chiral fermion mass $M$ and mirror fermion mass $M'$
have the opposite pairing on the chiral fermions, and (2) the magnitude of phases (i.e., here $\vartheta_{3,0}$ and $\vartheta_{4,5}$)
are imposed by the $\Z_2^{\rm PR}$ symmetry.

So from \eq{eq:thetavv'=0} and \eq{eq:barthetavv'},  we derive that 
${\bar{\theta}_{\rm v}} +  {\bar{\theta}'_{\rm v}}=0$.
Just like Remark \ref{Sol-Remark:5},
because the ${\bar{\theta}_{\rm v}} $ and ${\bar{\theta}'_{\rm v}}$ couple to the same $\U(1)_{\rm vector}$,
we can rotate ${\bar{\theta}_{\rm v}}\mapsto {\bar{\theta}_{\rm v}} + \delta \upalpha =0$ and ${\bar{\theta}'_{\rm v}} \mapsto {\bar{\theta}'_{\rm v}} - \delta \upalpha =0$
by the same symmetry transformation by those anomalous symmetries that have mixed anomalies with the $\U(1)_{\rm vector}$ symmetry,  with an amount $\delta \upalpha$.
This means effectively we have 
\bea
{\bar{\theta}_{\rm v}} +  {\bar{\theta}'_{\rm v}}=0
 \; \Rightarrow  \;
\bar{\theta}_v = \bar{\theta}_v' =0
\eea 
at UV, thanks to the imposed $\Z_2^{\rm PR}$ symmetry.

\item {\bf UV solution to the 1+1d CT or P problem}:
The full parent theory at UV has the path integral:
\bea \label{eq:Z-full}
 \int  [\cD \psi][\cD \bar\psi]
[\cD \psi'][\cD \bar\psi']
 \e^{\ii (S_{\rm bulk}+
S_{\Psi,\rm free} +  S_{\Psi',\rm free} 
+S_{{\text{YH-}}\Psi}
+S_{{\text{YH-}}\Psi'}
+g S_{\text{multi-}\Psi}
+g' S_{\text{multi-}\Psi'}
)} 
 \e^{\ii \int_{\cM^2} \frac{1}{2 \pi} 
(\theta_{\rm v} + \theta'_{\rm v})    
F^{\rm v} 
}.
\eea
In earlier remarks, we have shown that as long as the $\Z_2^{\rm PR}$ symmetry is respected for this full UV theory
above some energy scale $\rE > \Lambda_{\SMG}$, then $\bar{\theta}_v = \bar{\theta}_v' =0$ solves the 1+1d CT or P problem.

$\bullet$ The multi-fermion interactions are irrelevant operators at IR, but becomes non-perturbatively important at UV.
Thus, the multi-fermion interactions are \emph{dangerously irrelevant operators} that can drive the phase transitions when the coupling strength becomes large
above some critical strength \cite{Wang2013ytaJW1307.7480, Wang2018ugfJW1807.05998, ZengZhuWangYou3450SMG2202.12355}. 
This is consistent with the fact that we set the $\Lambda_{\SMG}$ at a higher UV energy scale (\Fig{fig:2}).
The multi-fermion interactions are not renormalizable, however it is fine to include non-renormalizable terms in the effective field theory formulation.
Moreover, the multi-fermion interactions can be regularized on the UV scale $\Lambda_{\SMG}$ 
(e.g., on a lattice \cite{Wang2013ytaJW1307.7480, Wang2018ugfJW1807.05998, ZengZhuWangYou3450SMG2202.12355} or by a short distance cutoff in quantum gravity).
 
$\bullet$ 1+1d abelian gauge theory exhibits confinement at IR, while shows asymptotic freedom at UV \cite{Schwinger1962PR2}. 
Namely, the $\U(1)$ gauge theory becomes weak coupling at UV.
Thus, conceptually, we may regard $A$ as background gauge field at UV;
we regard it as a dynamical gauge field $a$ only after going below the SMG scale, 
only then 
$$\int [\cD a_{\rm v}] \exp(\ii \int  (\frac{1}{2 e^2} F^{\rm v} \wedge \star F^{\rm v}) )$$ 
in the path integral becomes significant.
In contrast, in the limit when $A$ is treated as a background gauge field,
the SMG gapping the mirror fermion sector while 
keeping the chiral fermion sector gapless is indeed successfully numerically verified in \cite{ZengZhuWangYou3450SMG2202.12355}.

\item {\bf Flow to IR solution to the 1+1d CT or P problem}: We also need to demonstrate that $\bar{\theta}_v = \bar{\theta}_v' =0$ at UV under 
the renormalization group (RG) flows to
below the Higgs condensation $\Lambda_{\rm H}$ scale, the model still keeps 
$\bar{\theta}_v \simeq \bar{\theta}_v' \simeq 0$ at IR.
Let us list down and then rule out probable dangers that can ruin the zero or small IR theta angles.

$\bullet$ If there is a mass matrix $M=M_{\rm H} + g M_{\rm SMG}$ mixing between the Higgs condensation mass $M_{\rm H}$ and SMG mass $g M_{\rm SMG}$ at IR, 
then the theta angle contribution from \eq{eq:barthetavv'} gives:\footnote{Precisely
we should look at 
this form 
${\bf q}_{{\rm v}}^{\intercal} \cdot \arg (\diag M)+{\bf q}_{{\rm v}}^{\intercal} \cdot \arg (\diag M'^\dagger) $ 
in \eq{eq:barthetavv'},
but here we give the argument in
$\arg (\cred{\det} (M  M'^\dagger))$ 
for simplicity.
}
\bea \label{eq:IR-phase}
 \arg (\cred{\det} 
 \begin{pmatrix} 
M & 0 \\
0 &  M'^\dagger
\end{pmatrix}) 
=
 \arg (\cred{\det} (M  M'^\dagger))
 &=&
  \arg (\cred{\det} ( (M_{\rm H} + g M_{\rm SMG})  (M'^\dagger_{\rm H} + g' M'^\dagger_{\rm SMG})))\cr
   &=&
  \arg (\cred{\det} ( M_{\rm H}M'^\dagger_{\rm H} + g g'  M_{\rm SMG}M'^\dagger_{\rm SMG} +  g M_{\rm SMG} M'^\dagger_{\rm H} + g' M_{\rm H} M'^\dagger_{\rm SMG})).
\eea
The ${\arg} (\cred{\det} (M_{\rm H}M'^\dagger_{\rm H} + g g'  M_{\rm SMG}M'^\dagger_{\rm SMG}))=0$ alone can possibly hold, 
because the complex phase contributions from $M_{\rm H}$ and ${M'}_{\rm H}^\dagger$ may cancel with each other, 
and also $M_{\rm SMG}$ and ${M'}_{\rm SMG}^\dagger$ may cancel with each other,
when the product matrices are due to the same mechanisms.
But the mixing term $\arg (\cred{\det}  ( g M_{\rm SMG} M'^\dagger_{\rm H} + g' M_{\rm H} M'^\dagger_{\rm SMG} ) )$ can have a nonzero angle,
because we choose the strength in \Fig{fig:2} (b) as
$\frac{|M'^\dagger_{\rm H}|}{|M_{\rm H}|} \to 0 \ll 1$ 
and
$\frac{| g'  M'^\dagger_{\rm SMG}  |}{| g M_{\rm SMG} |} \to \infty \gg 1$,
so
$ \frac{ |g M_{\rm SMG} M'^\dagger_{\rm H}|}{ |g' M_{\rm H} M'^\dagger_{\rm SMG}|} \to 0 \ll 1$.
So the complex phase contribution of \eq{eq:IR-phase} mainly weighs in
$\arg (\cred{\det}  (g' M_{\rm H} M'^\dagger_{\rm SMG} ) )$ when the mass gap strength profile is the \Fig{fig:2} (b) type,
while the total complex phase \eq{eq:IR-phase} is still pending to be computed.
In the next remark, we argue how 
\eq{eq:IR-phase} with $\arg (\cred{\det}  (M  M'^\dagger)) \simeq 0$ can solve the CT or P problem at IR.

\item {\bf Non-Mean-Field SMG Conditions to solve the CT or P problem at IR}:
Yukawa-Higgs mass matrices in \eq{eq:YH-Chiral} and \eq{eq:YH-matrix} are the mean-field mass, so $\< M_{\rm H} \>$ and $\< M'_{\rm H} \>$ are generally non-zeros.
In comparison, 
we show that as long as the SMG mass in \eq{eq:multi-fermion} does not contribute any \emph{mean-field quadratic mass matrix} form, 
then $\<M_{\rm SMG} \>=0$ and $ \< M'^\dagger_{\rm SMG}  \>=0$, which gives zero IR complex phase in \eq{eq:IR-phase}.
We list down the conditions for 
$\<M_{\rm SMG}\>=\<M'^\dagger_{\rm SMG}\>=0$ for the inserted operators \eq{eq:multi-fermion} for the mirror fermion $\Psi'$ must obey that:
\bea \label{eq:0MF-mirror}
&&\CO'_{\iota \upsilon}  \psi'_{R,\iota} \psi'_{L,\upsilon}  +{\rm h.c.} \text{ requires } \<\CO'_{\iota \upsilon}\> =0,\cr
&&\tilde{\CO}'_{\iota \upsilon}  \psi'^\dagger_{R,\iota} \psi'_{L,\upsilon}  +{\rm h.c.} \text{ requires } \<\tilde{\CO}'_{\iota \upsilon}\> =0,
\eea
where $\psi'_{R,\iota}$ represents $\U(1)_{\rm 1st}$ charge 3 or charge 4 Weyl fermions,
while $\psi'_{L,\upsilon}$ represents $\U(1)_{\rm 1st}$ charge 5 or charge 0 Weyl fermions.
Similar conditions hold for the inserted operators \eq{eq:multi-fermion} for the chiral fermion $\Psi$:
\bea \label{eq:0MF-chiral}
&&\CO_{\iota \upsilon}  \psi_{L,\iota} \psi_{R,\upsilon}  +{\rm h.c.} \text{ requires } \<\CO_{\iota \upsilon}\> =0,\cr
&&\tilde{\CO}_{\iota \upsilon}  \psi^\dagger_{L,\iota} \psi_{R,\upsilon}  +{\rm h.c.} \text{ requires } \<\tilde{\CO}_{\iota \upsilon}\> =0.
\eea
We know that the above conditions in \eq{eq:0MF-mirror} and in \eq{eq:0MF-chiral} are indeed true for the SMG interactions \eq{eq:multi-fermion} ---
because if any of these expectation values $\<\CO'_{\iota \upsilon}\>, \<\tilde{\CO}'_{\iota \upsilon}\>, \<\CO_{\iota \upsilon}\>,$ and $ \<\tilde{\CO}_{\iota \upsilon}\>$
are nonzero, they would break the $\U(1)_{\rm 1st}$ and $\U(1)_{\rm 2nd}$ symmetries of \Table{table:3450charge}. 
But the SMG exactly demands that those anomaly-free symmetries must be preserved, so
$\<\CO'_{\iota \upsilon}\> =  \<\tilde{\CO}'_{\iota \upsilon}\> =  \<\CO_{\iota \upsilon}\> = \<\tilde{\CO}_{\iota \upsilon}\> =0$.

The $\e^{\ii (S_{\text{multi-}\Psi}
+S_{\text{multi-}\Psi'})}$
can also be rewritten as the disorder scalar
$\int [\cD \phi_1][\cD \phi_1'][\cD \phi_2][\cD \phi_2']$ $\e^{\ii (g S_{\text{disorder}}+g ' S_{\text{disorder}'})}$ 
formulation in \eq{eq:disorder}. In that case, the above no-mean-field-mass conditions \eq{eq:0MF-mirror} and \eq{eq:0MF-chiral} become
\bea
&&
\< \phi'_1 \> =\< \phi'_2 \> =
\< \phi'^2_1 \> =\< \phi'^2_2 \> = 0, \text{ and }
\< \phi_1 \> =\< \phi_2 \> =
\< \phi^2_1 \> =\< \phi^2_2 \> = 0.
\eea
These above conditions guarantee that
$\<M_{\rm SMG}\> =\<M'^\dagger_{\rm SMG}\>=0$. 
If we take the mean-field interpretation of \eq{eq:IR-phase},
then 
\begin{multline}
\arg (\cred{\det} (\< M \> \< M'^\dagger\>))
 =
\arg (\cred{\det} ( \< M_{\rm H} + g M_{\rm SMG}\>  \< M'^\dagger_{\rm H} + g' M'^\dagger_{\rm SMG} \> ))\\
=
\arg (\cred{\det} ( \< M_{\rm H}\>  \<M'^\dagger_{\rm H}\>  
+ g g'  \< M_{\rm SMG}\>  \< M'^\dagger_{\rm SMG} \> 
+  g \< M_{\rm SMG}\> \< M'^\dagger_{\rm H} \> 
+ g' \< M_{\rm H}\> \< M'^\dagger_{\rm SMG}\> ))
.
\end{multline}
 At IR, only the Higgs condensate develops at the chiral fermion side so $\<M_{\rm H}\>\vert_{\rm IR} \neq 0$ but $\<M'_{\rm H}\> \vert_{\rm IR} =0$,
 we expect that our $\arg (\cred{\det}  (\< M\> \<  M'^\dagger\> )) \vert_{\rm IR}=\arg (\cred{\det}  ( \<M_{\rm H}\> \< M'^\dagger_{\rm H}\>)) \vert_{\rm IR} 
\approx 0$ strictly. 

In summary, our solution has a smaller or zero $\bar{\theta}$ at IR, 
even better than the higher-loop calculation arguments given in \cite{BabuMohapatra1989, BabuMohapatra1989rbPRD, BarrChangSenjanovic1991qxPRL}.
Because \Refe{BabuMohapatra1989, BabuMohapatra1989rbPRD, BarrChangSenjanovic1991qxPRL} require only the Higgs mechanism, 
thus their Higgs mixing on both the chiral and mirror sectors can generate complex phases at the higher-order quantum corrections 
beyond the tree-level semiclassical analysis discussed above.
Here instead we have two mechanisms in our solution: Higgs mechanism dominates on the chiral sector and the SMG dominates on the mirror sector, 
and there is no mean-field mass matrix mixing to generate quantum corrections to $\bar{\theta} \approx 0$.

{There is another reason that the IR correction to our UV solution $\bar{\theta} = 0$ is extremely small. 
Because the SMG multi-fermion interaction and disorder scalar interaction are highly irrelevant operators at IR, 
the lower the energy, the weaker effects are these interactions on the IR correction of $\bar{\theta}$.}

\item {\bf Low-energy dynamics}:
Here we comment on the IR dynamics of the theory with $\bar{\theta}_v = \bar{\theta}_v' =0$.

$\bullet$ Without $\U(1)$ gauge fields, the multi-fermion interaction drives from the gapless phase to the gapped phase via the Berezinskii-Kosterlitz-Thouless (BKT) universality class transition \cite{ZengZhuWangYou3450SMG2202.12355}. In terms of the mirror fermion theory in \eq{eq:multi-fermion},
this means that $|g' |< |g_c'|$ for the gapless phase, 
and $|g'| >  |g_c' |$ for the SMG phase, for some nonperturbative critical strength \cite{ZengZhuWangYou3450SMG2202.12355}.

$\bullet$ With $\U(1)$ gauge fields,
in the perturbative field theory analysis around the free field Gaussian fixed point,
the gauge coupling $[e]=1$ has an energy scaling dimension 1, 
the Anderson-Higgs mass of fermion $[m_{\Psi}]=1$
(proportional to some power of the Higgs condensate $\< \phi_H \>$)
and the Higgs quadratic mass coefficient $[\mu_H]=1$ in \eq{eq:YH-Chiral}
 all have an energy scaling dimension 1.
We can consider the phase diagram parametrized by $\frac{m_{\Psi}^2}{e^2}$ and also 
$\frac{\mu_H}{e^2}$ similar to the analysis in \cite{Tong2104.03997}.

$\bullet$ With $\U(1)$ gauge fields, but suppose we remove the chiral fermions out of the theory (e.g. $|\frac{m_{\Psi}^2}{e^2}| \gg 1$):
In the 1+1d U(1) gauge theory phase diagram,
the $\frac{\mu_H}{e^2} \gg 1$ gives the Coulomb phase,
and $\frac{\mu_H}{e^2} \ll -1$ gives the Higgs phase.\footnote{More
precisely, in our model with \Table{table:3450charge}, 
the $\frac{\mu_H}{e^2} \gg 1$ gives the Coulomb phase for both $\U(1)_{\rm 1st}$ and $\U(1)_{\rm 2nd}$.
The $\frac{\mu_H}{e^2} \ll -1$ Higgs down $\U(1)_{\rm 1st}$ and $\U(1)_{\rm 2nd}$ to the Coulomb phase of $\U(1)_{\rm vector}$.}
But for 1+1d U(1) gauge theory with $\bar{\theta}_v = \bar{\theta}_v' =0$, 
the Coulomb phase and Higgs phase all are confined phases \cite{CallanDashenGross1977}. 
How to understand this confinement? 
In the pure electrodynamics Coulomb phase, the U(1) electric field $F_{01}=E_1$ has a constant profile solution in the spacetime under the equation of motion. 
The U(1) electric field goes along the only available spatial direction $x$ in 1+1d, 
and the energy stored in two U(1) unit charged $\pm e$ particles with a distance $l$ apart
 is $\frac{1}{2} e^2 l$. This is a linear potential 
proportional to the stretched distance $l$ for the confinement.
In the Higgs phase, the low energy theory is a gas of vortices and anti-vortices, but it also shows the confinement.\\
 
 $\bullet$ With $\U(1)$ gauge fields, also we include the chiral fermions into the theory in the bare massless limit $m_{\Psi}= 0$ or $\frac{m_{\Psi}^2}{e^2} \to 0$:
 This also means that no Higgs condensate $\< \phi_H \>=0$,
or the Higgs scalar becomes massive and decouples $\frac{\mu_H}{e^2} \gg 1$.
The 1+1d $\U(1)$ gauge theory is in the Coulomb phase which is a confined phase.\\[-2mm]

We are able to regard one term of multi-fermion interactions \eq{eq:multi-fermion} as the gauge singlet
Dirac fermion mass term of $\U(1)_{\rm 1st}$ for $q_1=(3,4,5,0)$;
namely, ${g}_{2}  \big(( {\psi}_{L,3} )^2
({\psi}_{L,4})
( {\psi}_{R,5}^\dagger )^2 \big)
(  {\psi}_{R,0})+\text{h.c.} \sim 
{g}_{2}
{\psi}_{L,0}^\dagger
{\psi}_{R,0}
+\text{h.c.}$ and 
${g}'_{2}   \big(  ( \psi'_{R,3})^2
(\psi'_{R,4})
( \psi'^\dagger _{L,5})^2\big)
(  \psi'_{L,0})+\text{h.c.}
\sim
{g}'_{2}
 \psi'^\dagger_{R,0}
\psi'_{L,0}+\text{h.c.}$ 
The ${\psi}_{L,0}$ and ${\psi}'_{R,0}$ may be regarded as the gauge singlet composite fermion 
due to the $\U(1)_{\rm 1st}$ confinement \cite{Tong2104.03997}.
\\[-2mm]

We are also able to regard 
the other term of multi-fermion interactions as the gauge singlet
of another Dirac fermion mass term of $\U(1)_{\rm 2nd}$ for $q_2=(0,5,4,3)$;  
namely,
$ {g}_{1}   ({\psi}_{L,3}) 
(  {\psi}_{L,4}^\dagger)^2
( {\psi}_{R,5}  )(  {\psi}_{R,0})^2 +\text{h.c.} \sim  
{g}_{1}
{\psi}_{L,3}
{\psi}_{R,3}^\dagger
+\text{h.c.}$
and 
$ {g}'_{1}   ({\psi}'_{R,3}) 
(  \psi'^\dagger_{R,4})^2
( \psi'_{L,5}  )(  \psi'_{L,0} )^2 +\text{h.c.}
 \sim  
{g}'_{1}
{\psi}'_{R,3}
{\psi}'^\dagger_{L,3}
+\text{h.c.}
$
The ${\psi}_{R,3}$ and ${\psi}'_{L,3}$ may be regarded as the gauge singlet composite fermion due to the $\U(1)_{\rm 2nd}$ confinement.
\\[-2mm]
 
However, there are no gauge singlet mean-field mass term deformations to express all multi-fermion interactions to preserve both $\U(1)_{\rm 1st}$ and $\U(1)_{\rm 2nd}$.
Thus, this suggests that the multi-fermion SMG interactions \eq{eq:multi-fermion} by preserving all anomaly-free U(1) (both $\U(1)_{\rm 1st}$ and $\U(1)_{\rm 2nd}$)
are beyond the mean-field descriptions even at IR.
Similarly, the disorder scalar interactions \eq{eq:disorder} are also beyond the mean-field descriptions.\\

$\bullet$ With $\U(1)$ gauge fields, also we include the chiral fermions paired with nonzero mass $m_{\Psi} \neq 0$ via
the Higgs condensate $\< \phi_H \> \neq 0$:
The 1+1d  gauge theory is Higgs down to $\U(1)_{\rm vec}$ but still in a confined phase.
The fermion theory becomes vector $q_{\rm v}= (1, 3,3,1)$ charged under the $\U(1)_{\rm vec}$ gauge field.
The fermions pair up with nonzero vector mass 
$\< \phi_H^3\> \psi_{L,3}^\dagger \psi_{R,0} +  \< \phi_H^\dagger \> \psi_{L,4}^\dagger \psi_{R,5} +\text{h.c.}$ preserving the $\U(1)_{\rm vec}$.
The low energy property of this resulting theory is controlled by the energy scales of 
the fermion mass $m_{\Psi}$ and the $\U(1)_{\rm vec}$ confinement energy gap.

\end{enumerate}



\section{Conclusion} 
\label{sec:Conclusion}

The 1+1d toy model formulated in this work is certainly far from a realistic model for the 3+1d Standard Model world.
However, inspired by many previous milestone works \cite{GrossNeveu1974, CallanDashenGross1977} studied 1+1d toy models as simplified models of the 3+1d world,
we believe that the phenomena demonstrated in the 1+1d toy model indicate some important features 
that we also anticipate in the more realistic 3+1d model to solve the Strong CP problem \cite{StrongCPtoappear-2212.14036}.


\subsection{Massless Fermion vs Mean-Field Ordered Mass vs Interacting Non-Mean-Field Disordered Mass Fermion}
\label{sec:transition}

\begin{figure}[!h] 
\centering
 \includegraphics[width=6.5in]{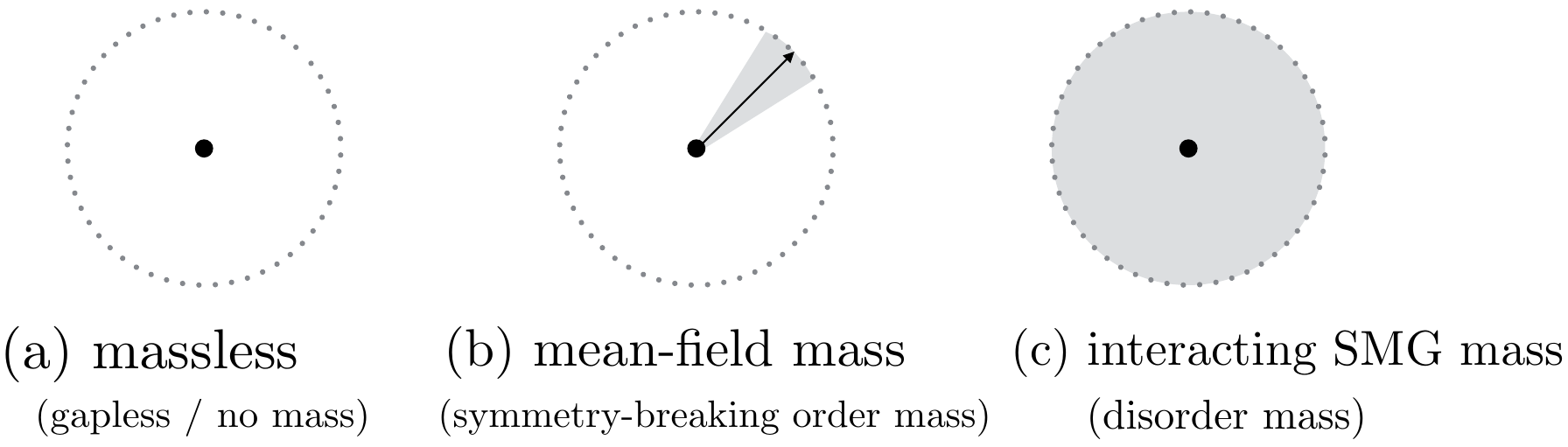}
\caption{(a) The massless fermion is drawn as no mass field (such as the Higgs scalar $\phi_H$).
The dotted circle means the target space of the mass field. The origin means the mass field ordering is zero (such as $\phi_H=0$). 
(b) The mean-field or symmetry-breaking ordered mass is shown as the mass field ordered in a fixed-oriented arrow direction in the target space,
while small continuous fluctuations in the gray wedge region around the arrow direction give Goldstone modes or axions. 
In 1+1d, this superfluid-type solution is unavailable, but the algebraic superfluid solution is available.
Since the $\bar{\theta}$ and the complex phase of the mass field are related, we may also interpret this picture as the oriented/ordered $\bar{\theta}$-clock.
(c) The interacting SMG or disordered mass gives mass even without symmetry-breaking within an anomaly-free symmetry.
The mass field is disordered so its arrow is not oriented in a fixed direction in the target space. 
The disordered arrow in the target space is shown as the full gray sphere in the whole target space.
Since the $\bar{\theta}$ and the complex phase of the mass field are related, we may also interpret this picture as the disoriented/disordered $\bar{\theta}$-clock.
}
\label{fig:mass-field}
\end{figure}

We can cross-compare the massless fermion solution of 't Hooft \cite{tHooft1976ripPRL} (\Fig{fig:mass-field} (a)), 
the Peccei-Quinn symmetry-breaking mass solution with axions 
\cite{PecceiQuinn1977hhPRL,PecceiQuinn1977urPRD, Weinberg1977ma1978PRL, Wilczek1977pjPRL} (\Fig{fig:mass-field} (b)),
and \Sec{sec:Another-Solution}'s interacting SMG mass solution (\Fig{fig:mass-field} (c)), 
schematically shown in \Fig{fig:mass-field}.

\noindent
$\bullet$ {\bf The massless fermion solution}  \cite{tHooft1976ripPRL} (\Fig{fig:mass-field} (a))
says that if any of the quark $\psi$ (say up quark) has no mean-field nor Higgs mass,
then we can do the chiral transformation on this quark $\psi$ alone to rotate the $\theta$ away but without gaining a complex phase 
in the mean-field mass matrix (since there is no mean-field mass for $\psi$).
This sets $\bar{\theta}=0$ thus solving the 1+1d CT or P problem (or Strong CP problem in 3+1d).\\

\noindent
$\bullet$ {\bf The mean-field symmetry-breaking ordered mass and Peccei-Quinn solution} (\Fig{fig:mass-field} (b)):
The arbitrariness of $\bar\theta$ can be relaxed by the symmetry breaking dynamics to $\bar\theta \simeq 0$.
In the weakly gauge limit or global symmetry limit of $G$, 
the Peccei-Quinn solution can be regarded as a symmetry breaking solution that also uses the mean-field Yukawa-Higgs symmetry breaking mass term.
The fluctuation around a fixed $\bar\theta = 0$ gives a low-energy Goldstone mode or an axion.
However, this superfluid solution is unavailable in 1+1d.
{There are no spontaneous continuous symmetry breaking, no truly long-range order, 
and no Goldstone modes in 1+1d \cite{Coleman1973Goldstone, MerminWagner1966PhysRevLett, Hohenberg1967PhysRev}. 
So we do not have a Peccei-Quinn solution \cite{PecceiQuinn1977hhPRL,PecceiQuinn1977urPRD, Weinberg1977ma1978PRL, Wilczek1977pjPRL},  
 no corresponding pseudo Goldstone modes or nearly massless axion in 1+1d.
But we have the quasi-long-range-order algebraic-superfluid solution instead in 1+1d.
 }\\

\noindent
$\bullet$ {\bf Interacting SMG or disordered mass} (\Fig{fig:mass-field} (c)): 
In contrast, \Sec{sec:Another-Solution} solution says that if some of the fermions are anomaly-free under a $G$-symmetry,
then they can gain the mass via SMG to preserve $G$. 
The SMG mass can be viewed as an interacting mass from multi-fermion interactions, or the disordered mass from disorder interactions.
But they have no mean-field (nor Higgs-induced) mass. Therefore, we can also do the chiral transformation on any fermion $\psi$ that has only SMG-induced mass
to rotate the $\theta$ away but without gaining a complex phase in the mass matrix, since there is no mean-field mass for $\psi$.
This also sets $\bar{\theta}=0$ thus also solving the 1+1d CT or P problem.\\

\begin{figure}[!h] 
\centering
 \includegraphics[width=6.8in]{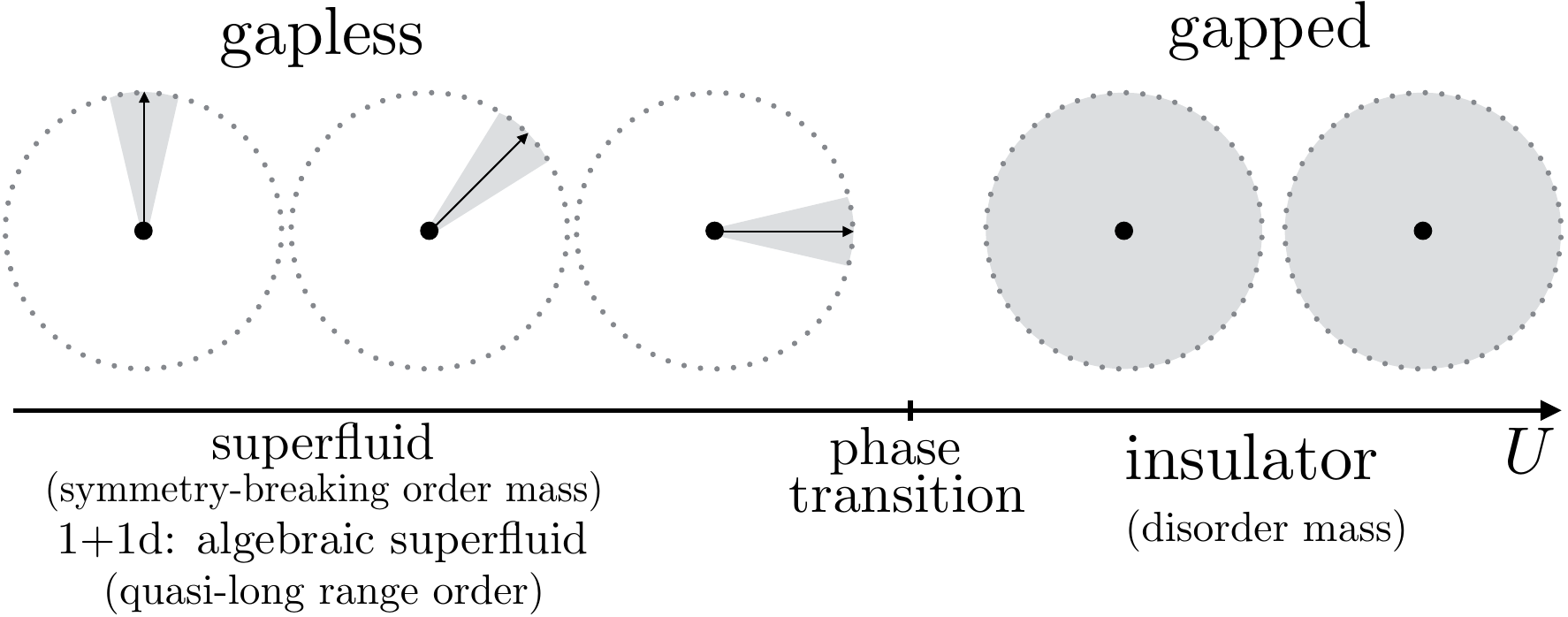}
\caption{The superfluid to insulator transition as an analogy of our Peccei-Quinn ordered phase solution to the SMG disordered phase solution.
The superfluid here has to be an algebraic superfluid in 1+1d. The algebraic superfluid in 1+1d
has an algebraic long-range order with a decay tail $\< \CO(x_i) \CO(x_j) \> \sim  |x_i-x_j|^{-\alpha}$ instead of the truly 
long-range order with $\< \CO(x_i) \CO(x_j) \> \sim$ constant at a large distance $ |x_i-x_j|$.
The truly superfluid to insulator transition analogy is applicable in spacetime dimensions higher than 1+1d in the weakly gauge or global symmetry $G$ limit.
Several thick black origins indicate different site locations in spacetime. 
On the ordered superfluid side, the fluctuation around the ordered mean-field value
gives rise to analogous (nearly) gapless Goldstone modes.
On the disordered insulator side, the excitations are fully gapped above the energy gap.
The superfluid-insulator transition can be realized in the Bose Hubbard model \cite{Fisher1989zza} or in the abelian duality of QFT \cite{Deligne1999qpQFTIAS}} 
\label{fig:superfluid-insulator}
\end{figure}

The transition from the mean-field mass to the interacting SMG mass
is analogous to the phase transition between the ordered phase and the disordered phase
--- for example, the famous superfluid-to-insulator type of phase transition is shown in \Fig{fig:superfluid-insulator}.
The superfluid-insulator transition can be realized on a UV regularized lattice system such as the Bose Hubbard toy model  
\cite{Fisher1989zza} with the following Hamiltonian:
\bea \label{eq:bose-hubbard}
\hat{H} = - {\rm t}_{\rm hop} \sum_{\<i,j \>} (b_i^\dagger b_j + {\rm h.c.}) +U \sum_i (b_i^\dagger b_i)(b_i^\dagger b_i-1)
= - {\rm t}_{\rm hop} \sum_{\<i,j\>} (b_i^\dagger b_j + {\rm h.c.}) +U \sum_i (n_i)(n_i-1). 
\eea
The $b_i$ and $b_i^\dagger$ are bosonic annihilation/creation operators on the site $i$. 
In \eq{eq:bose-hubbard}, the first term is the kinetic term of the bosons with the hopping constant ${\rm t}_{\rm hop}$ of the nearest neighbor ${\<i,j\>}$
between neighbor sites $i $ and $j$.  The second term is the onsite repulsion with a potential strength $U$ penalizing the boson number 
$n_i \equiv b_i^\dagger b_i$ when $n_i$ is different from $0$ or $1$.
{In our theory, we \emph{do} require a modification on this Bose Hubbard model to include fermions and $\frac{\theta}{2 \pi} \int F$ term, 
which we will leave the detail for future work \cite{DisorderTheta}.}
Nonetheless, we can still find this bosonic toy model capturing the basic analogous physics of the modified Peccei-Quinn solution and the SMG solution as follows.
We write $b_i \equiv \sqrt{n_j} e^{\ii \theta_j} \equiv \sqrt{\<n_j\> +\delta n_j} e^{\ii \theta_j}$,
with the boson number $n_j$ contains the mean-field $\<n_j\> $ and its fluctuation $\delta n_j$. 
It is easy to show that the standard commutator
$[b_i , b_j^\dagger]=  \delta_{ij}$ implies that 
$[n_i, b_j] = [b_i^\dagger b_i, b_j] =- \delta_{ij}$, so the particle number $n_i$ and the bosonic complex phase field $\theta_j$
are canonical conjugate operators with respect to each other with the commutator $[n_i, \theta_j]=\ii \delta_{ij}$.
We can derive the low energy theory of \eq{eq:bose-hubbard} in terms of $\theta$ in the Euclidean partition function with Euclidean time $\tau$:
\bea
\label{eq:Ztheta}
Z &\sim& 
\int [D \theta][D n] \exp( \int \dd \tau \dd x (\ii (\prt_\tau \theta) n + {\rm t}_{\rm hop}  (\prt_x \theta)^2  + U n^2))
\sim
\int [D \theta] \exp( \int \dd \tau \dd x (  \frac{1}{U} (\prt_\tau \theta)^2+ {\rm t}_{\rm hop}  (\prt_x \theta)^2  ))\cr
&\sim& \int [D \theta] \exp(\sum_{\<i,j \>} R \cos(\theta_i - \theta_j)).
\eea
In the last line, we change to the spacetime lattice, such that $R \propto \sqrt{\frac{{\rm t}_{\rm hop}}{U}}$.
So intuitively,\\ 
$\bullet$ When ${\frac{{\rm t}_{\rm hop}}{U}} \gg 1$, the model is in the \emph{algebraic superfluid phase} with an algebraic quasi-long-range order
$\<\exp(\ii \theta_i) \exp(-\ii \theta_j) \> \sim  |x_i-x_j|^{-\alpha}$. This is analogous to the Peccei-Quinn solution with nearly gapless $\theta$.\\
$\bullet$ When ${\frac{{\rm t}_{\rm hop}}{U}} \ll 1$ instead,
the model is in the \emph{insulator phase} which $\theta_i$ is disorder thus in short-range order
with an unoriented $\theta_i$ clock with $\<\exp(\ii \theta_i)\> \simeq 0$,
as in \Fig{fig:mass-field} (c). This is analogous to our SMG solution with a gapped spectrum.

Apply the techniques like the abelian duality \cite{Deligne1999qpQFTIAS} to rewrite the \eq{eq:Ztheta} in terms of the boson number $n$, we obtain
\bea
\label{eq:Zn}
Z &\sim& 
\int [D n] \exp(\sum_{\<i,j \>} \frac{1}{R} ( n_i - n_j)^2 )
\sim
\int [D n] \exp( \int \dd \tau \dd x ( \frac{1}{R}  (\prt_\mu n)^2  - g \cos(2 \pi  n))).
\eea
In the first form of \eq{eq:Zn}, the $n \in \Z$ is discrete in a quantized integer.
In the second form of \eq{eq:Zn}, the $n \in \R$ is continuous, pinned down by the energetic penalty of the cosine term.
Intuitively,\\ 
$\bullet$ When ${\frac{{\rm t}_{\rm hop}}{U}} \gg 1$ thus with a small $1/R$ and a relatively large $g$, we get the low energy action
$\int \dd \tau \dd x ( \frac{1}{R}  (\prt_\mu n)^2  - g \frac{1}{2}(2 \pi  n)^2 + \dots)$ such that $n$ has a gapped spectrum
with an exponential decay correlation function $\< n_i n_j \>$. The  $n$ is disorder thus in short-range order.
But we have the $\theta$ algebraic quasi-long-range order in the \emph{algebraic superfluid phase}.
\\
$\bullet$ When ${\frac{{\rm t}_{\rm hop}}{U}} \ll 1$, 
 thus with a large $1/R$ and a relatively small $g$, the energy penalizes the $(\prt_\mu n)^2$ fluctuation so $(n_i - n_j)$ is small.
Hence $n_i$ is long-range order with its \emph{discrete symmetry spontaneously broken}
and $\< n_i n_j \> \sim$ constant, but with no Goldstone modes.
Again this is consistent with the insulator phase with $\theta_i$ disorder thus short-range order,
with an unoriented $\theta_i$ clock with $\<\exp(\ii \theta_i)\> \simeq 0$.
The case of including fermions and $\frac{\theta}{2 \pi} \int F$ term in this SMG insulator phase deserves further future investigation \cite{DisorderTheta}.

\subsection{Thought Experiment on Symmetric Gapped Mirror Fermion}
\label{sec:thought-exp}

\begin{figure}[!h] 
\centering
(a) \includegraphics[height=1.6in]{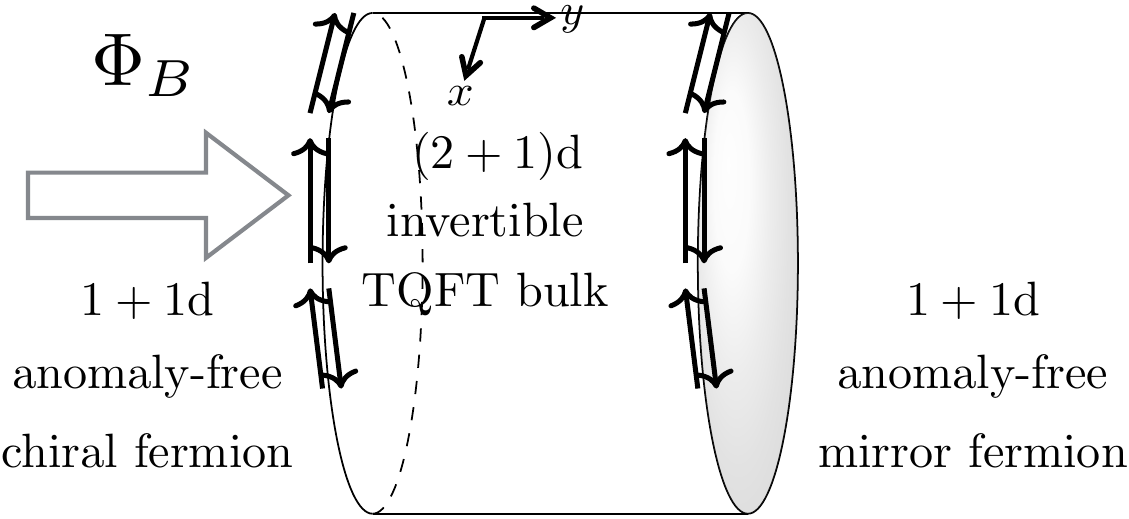}
(b) \includegraphics[height=1.6in]{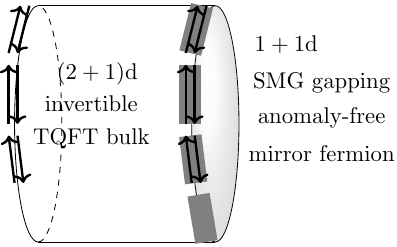}
\caption{(a) The time-dependent magnetic flux from an extra-dimension (here 2+1d)
threading through the hole 
of the cylinder strip can induce the 1+1d electric field (say of $\U(1)_{\al{\text{-th}}}$) along the $x$ direction, which follows from 
$\frac{\dd \Phi_{B,y}}{\dd t}= - \oint E_x \cdot \dd l$ and the Bianchi identity of the field strength $\dd F =0$.
The electric field can further cause current transport (say of $\U(1)_{\bt{\text{-th}}}$) from one edge of chiral fermion to the other edge of mirror fermion,
if and only if the anomaly-free coefficient $q^{\intercal}_\al K^{} q_\bt$ and the Hall conductance $\sigma_{xy}=\frac{e^2}{2 \pi} q^{\intercal}_\al K^{} q_\bt$ is nonzero.
(b).
Because the 1+1d fermion is anomaly-free within the symmetry $G$, 
the 2+1d bulk invertible TQFT is a trivial vacuum and thus is detachable. The 1+1d mirror fermion is also SMG gappable without symmetry-breaking in $G$.}
\label{fig:flux}
\end{figure}

Let us summarize \Sec{sec:Solution}'s
{solution to the 1+1d CT or P problem via SMG on the mirror fermions}
in terms of Laughlin's style of the flux insertion thought experiment \cite{Laughlin1981PRB} (see also \cite{Wang2013ytaJW1307.7480, SantosWang1310.8291})
in \Fig{fig:flux}.

\noindent
$\bullet$ $\rE > \Lambda_{\SMG}$: We hypothesize that a $\Z_2^{\rm PR}$ reflection symmetry mapping between the chiral and mirror fermion
is preserved at some deep ultraviolet (UV) high energy 
full parent theory including interaction terms
above some energy scale. The $\theta$ in the chiral sector and the $\theta'=-\theta$ in the mirror sector are canceled. 
The $\bar{\theta} = \bar{\theta}' = \theta = \theta' =0$ restricts the induced electric field to be zero, because the expectation value of the electric field 
$\< F_{01} \>={e^2}{n - \frac{\theta}{2 \pi}} + \dots$ with $n \in \Z$ and $\dots$ are the contribution from the matter contents (e.g., bosonized phase fields of fermions). 
The restricted theta also restricts Laughlin flux insertion types of the transport shown in \Fig{fig:flux} (a).

\noindent
$\bullet$ $\Lambda_{H} <E < \Lambda_{\SMG}$: 
Below the SMG energy scale $\Lambda_{\SMG}$ 
but above the Higgs condensation energy scale $\Lambda_{H}$, 
the mirror fermion sector shall be fully gapped out by the SMG, shown in \Fig{fig:flux} (b).

\noindent
$\bullet$ $\rE < \Lambda_{H}$: 
Below the Higgs condensation energy scale $\Lambda_{H}$, the condensed Higgs gives the symmetry-breaking mass to the chiral fermion sector.\\


\noindent
\emph{Acknowledgments} ---
JW thanks Yuta Hamada, Daniel Jafferis, Matthew Reece, Yi-Zhuang You, and Meng Zheng for the encouraging conversations. 
JW especially thanks Yuta Hamada for the discussions 
on applying the Symmetric Mass Generation \cite{StrongCPtoappear-2212.14036,HamadaWang2209.15244} to beyond-the-Standard-Model problems, 
also sincerely thanks Yi-Zhuang You on the discussion of \Sec{sec:transition} and
the order to disorder transition \cite{DisorderTheta}.
After the manuscript
submission, JW is grateful to receive helpful comments from Matthew Strassler and Edward Witten.
{JW appreciate the generous feedback from many participants (e.g., Isabel Garcia Garcia, Sungwoo Hong, and Seth Koren, etc.)
of Harvard CMSA Phase Transitions and Topological Defects in the Early Universe workshop (August 2-5, 2022)
and of Simons Center for Geometry and Physics's
Generalized Global Symmetries, Quantum Field Theory, and Geometry at Stony Brook University (September 19-23, 2022).}
JW is supported by Harvard University CMSA.

\appendix

\section{Fermion to Boson Field Theory}
\label{appendix:A}
\subsection{Fermion}

The original free massless fermion theory action is
\bea \label{eq:fermion-action}
S_{\Psi,\rm free}
=\int  \dd t \dd x \cL_{\Psi,\rm free}
=\int  \dd t \dd x \ii \Psi_{\rm I}^\dagger (\delta_{\rm IJ} \partial_t-K_{\rm IJ'}^{-1}V_{\rm J'J} \partial_x) \Psi_{\rm J} 
=\int  \dd t \dd x \ii  \Psi_{\rm I}^\dagger (\partial_t-K_{\rm II}^{-1}  \partial_x) \Psi_{\rm I}.
\eea
The velocity matrix $V_{\rm IJ}$ in the Lorentz invariant relativistic system 
can be chosen as $V_{\rm IJ}=c \delta_{\rm IJ}$, here we set the speed of light $c=1$.
Given the $\U(1)$ charge vector $q$ (like in \Table{table:3450charge}),
the $\U(1)$ symmetry transformation acts on $\psi_I$ as:
\bea \label{eq:fermion-symmetry}
\Psi_{\rm I}\mapsto \exp(\ii q_{\rm I}\vartheta )\Psi_{\rm I}, \quad \vartheta \in [0, 2 \pi) \cong \U(1).
\eea
For example, for $q =q_1$ as the first charge vector $\U(1)_{\rm 1st}$ 
in \Table{table:3450charge}, then $q_{\rm I} =q_{1, \rm I} = (3,4,5,0)$ where the $\rm I$ specifies the component of the $q$ vector.
The symmetry transformation on the 0d point-like \emph{\bf charged object} $\Psi_{\rm I}$ implies 
that there is a corresponding co-dimension-1 \emph{\bf charge operator} $U$ also known as the \emph{\bf symmetry generator}.
The symmetry generator $U$ is also a topological operator meaning that the deformation cannot change the symmetry measurement ---
as long as the topological configuration (i.e. the link between the charged object and charge operator) is not altered under the deformation \cite{Gaiotto2014kfa1412.5148}. 

More precisely for this specific 1+1d example, the co-dimension-1 charge operator $U$ is 1d,
\bea \label{eq:U-charge-operator}
U &=& 
\exp(-\ii \theta \int_{} \dd x' J^0(x') ) 
\equiv \exp(- \ii \theta 
 \int_{} \dd x'  
 \Psi_{\rm I}^\dagger
{q_{\rm I } }
\Psi_{\rm I})
.
\\
\label{eq:fermion-shift}
U \Psi_{\rm I} U^{-1} &=& \exp(\ii q_{\rm I}\vartheta )\Psi_{\rm I}, \quad \vartheta \in [0, 2 \pi) \cong \U(1).
\eea
More generally, 
$J^\mu \equiv
 \Psi_{\rm I}^\dagger
{q_{\rm I}}
\tau^\mu_{ L/R} 
\Psi_{\rm I}$ with $\tau^\mu_{ L/R}$ 
given in the main text \eq{eq:tau-mu}.
In the above, the first line has the repeated indices summed over (namely $\sum_{\rm I}$);
the second line has the fixed index true for each ${\rm I}$.
To derive \eq{eq:fermion-shift}, we use the fact that 
the $\Psi_{\rm I}$ and the conjugate momentum
$\frac{\delta {\cL_{\Psi,\rm free}}}{\delta (\partial_t \Psi_{\rm I} )} =\ii   \Psi_{\rm I}^\dagger$
satisfies the canonical quantization relation
\bea \label{eq:commutation}
\{\Psi_{\rm I}(x), \ii   \Psi_{\rm J}^\dagger(x')\} =
  \ii \delta_{\rm IJ}  \delta(x - x') \Rightarrow
\{\Psi_{\rm I}(x),  \Psi_{\rm J}^\dagger(x')\} =
\delta_{\rm IJ} \delta(x-x').
\eea
The anti-commutator for each 1+1d one-component Weyl fermion $\Psi_{\rm I}$ is in agreement with 
the anti-commutator of the 1+1d two-component Dirac fermion 
$\Psi_{\rm D}
=\Big(\begin{smallmatrix}
\psi_{L}\\
\psi_{R}
\end{smallmatrix}\Big)$, which says
$\{\Psi_{\rm D, a}(x),  \Psi_{\rm D, b}^\dagger(x')\} =
\delta(x-x') \delta_{\rm a b}$
where ${\rm a, b}$ are the 2-component indices.

\subsection{Bosonization}

It is helpful to compare with the bosonization language \cite{Coleman1974PRD, Mandelstam1975PRD} of this 1+1d chiral fermion theory.
For the convenience to compare with a familiar generic $K$-matrix formulation of a bosonized theory,
we rewrite the above fermionic action \eq{eq:fermion-action} 
by rescaling with the unimodular $K$ matrix 
as:
\bea \label{eq:fermion-action-2}
\int  \dd t \dd x \cL_{\Psi,\rm free}
=\int  \dd t \dd x \ii \Psi_{\rm I}^\dagger ( K_{\rm IJ} \partial_t- V_{\rm IJ}\partial_x) \Psi_{\rm J}. 
\eea
Define ${ Q_{\rm I}} \equiv {K_{\rm I J} q_{\rm J}}$, 
the co-dimension-1 charge operator $U$ 
in \eq{eq:U-charge-operator}
is modified to
\bea \label{eq:U-charge-operator}
U &=& 
\exp(-\ii \theta \int_{} \dd x' J^0(x') ) 
\equiv \exp(- \ii \theta 
 \int_{} \dd x'  
 \Psi_{\rm I}^\dagger
{K_{\rm I J} q_{\rm J}}
\Psi_{\rm I})
\equiv
\exp(-\ii \theta 
 \int_{} \dd x'
\Psi_{\rm I}^\dagger
Q_{\rm I}
\Psi_{\rm I}).
\\
\label{eq:fermion-shift-2}
U \Psi_{\rm I} U^{-1} &=& \exp(\ii q_{\rm I}\vartheta )\Psi_{\rm I}, \quad \vartheta \in [0, 2 \pi) \cong \U(1).
\eea
In the above, the first line has the repeated indices summed over (namely $\sum_{\rm I,J}$ and $\sum_{\rm I}$);
the second line has the fixed index true for each ${\rm I}$.
To derive \eq{eq:fermion-shift-2}, we also use the fact that 
the $\Psi_{\rm I}$ and the conjugate momentum
$\frac{\delta {\cL_{\Psi,\rm free}}}{\delta (\partial_t \Psi_{\rm I} )} =\ii  K_{\rm IJ}  \Psi_{\rm J}^\dagger$
satisfies the canonical quantization relation
\bea \label{eq:commutation-2}
\{\Psi_{\rm I}(x), \ii  K_{\rm IJ}  \Psi_{\rm J}^\dagger(x')\} =
  \ii \delta(x - x') \Rightarrow
\{\Psi_{\rm I}(x),  \Psi_{\rm J}^\dagger(x')\} =
 K^{-1}_{\rm IJ} \delta(x-x').
\eea
Because the action \eq{eq:fermion-action} is modified to \eq{eq:fermion-action-2},
the commutation relation is modified from \eq{eq:commutation} to \eq{eq:commutation-2}. 

This free massless chiral fermion theory action \eq{eq:fermion-action-2}
can be bosonized to a multiplet compact chiral boson $\varphi_{\rm I}$ theory \cite{FloreaniniJackiw1987PRL}, 
here each $\varphi_{\rm I} \in [0, 2 \pi)$, with a bosonized action:
\bea
 \label{eq:boson-action}
S_{\varphi, \rm free}
=\int  \dd t \dd x \cL_{\varphi,\rm free}
= \frac{1}{4\pi} \int \dd t \dd x ( K_{\rm IJ} \partial_t \varphi_{\rm I} \partial_x \varphi_{\rm J} 
-V_{\rm IJ}\partial_x \varphi_{\rm I}   \partial_x \varphi_{\rm J} ).
\eea
Again we choose the velocity matrix $V_{\rm IJ}$
as $V_{\rm IJ}=c \delta_{\rm IJ}$ with the speed of light $c=1$.
Although most of the discussions here are true for any general symmetric $K$ matrix, 
we consider in particular the diagonal $K_{\rm IJ}= K_{\rm II} \delta_{\rm IJ}$ for this 1+1d chiral fermion theory,
such that $K_{\rm II} = +1$ for the left-moving mode and 
$K_{\rm II} = -1$ for the right-moving mode. Here $K_{\rm II}$ is the ${\rm I}$-th diagonal element \emph{without} summing over ${\rm I}$.

Each complex chiral Weyl fermionic field $\Psi_{\rm I}$ is bosonized to the vertex operator of a compact chiral bosonic field
\bea \label{eq:fermion-boson}
{\Psi_{\rm I} \equiv : \exp( \ii K_{\rm II}\varphi_{\rm I} ): =: \exp(\pm \ii \varphi_{\rm I} ):}
\eea
with the proper normal ordering $::$ that moves the annihilation operators to the right and the creation operators to the left.
The $\pm$ sign in the exponent of \eq{eq:fermion-boson} depends on the chirality of fermions (e.g., left-moving $\psi_L$ for $K_{\rm II}=+1$, right-moving $\psi_R$ for $K_{\rm II}=-1$). 
{Under the \eq{eq:fermion-boson} map, the multi-fermion interaction \eq{eq:multi-fermion} such as $g S_{\text{multi-}\Psi}$ and $g' S_{\text{multi-}\Psi'}$ on the chiral and mirror sectors
become:
\bea \label{eq:interaction-cosine-term}
\tilde{g} S_{\cos{\text-}\varphi}&\equiv&\int  \dd t \dd x \; \big(  \tilde{g}_{1} \cos(  {\varphi}_{L,3} - 2 {\varphi}_{L,4}  + \varphi_{R,5} + 2\varphi_{R,0}  )
+
 \tilde{g}_{2} \cos(  2{\varphi}_{L,3} + {\varphi}_{L,4}  -2 \varphi_{R,5} + \varphi_{R,0}  )
\big),
\cr
\tilde{g}' S_{\cos{\text-}\varphi'}&\equiv&\int  \dd t \dd x \; \big(  \tilde{g}'_{1} \cos(  {\varphi}'_{R,3} - 2 {\varphi}'_{R,4}  + \varphi'_{L,5} + 2\varphi'_{L,0}  )
+
 \tilde{g}'_{2} \cos(  2{\varphi}'_{R,3} + {\varphi}'_{R,4}  -2 \varphi'_{L,5} + \varphi'_{L,0}  )
\big)
.
\eea
}

The U(1) symmetry transformation on the fermion $\psi_{\rm I}$ in \eq{eq:fermion-symmetry}
becomes the following U(1) symmetry transformation on the compact bosonized field $\varphi_{\rm I}$:
\bea
\varphi_{\rm I} \mapsto \varphi_{\rm I}  + q_{\rm I} \vartheta, \quad \varphi_{\rm I}, \vartheta \in [0, 2 \pi) \cong \U(1), \quad  q_{\rm I} \in \Z.
\eea
The co-dimension-1 charge operator $U$ as a 1d topological operator in the bosonized field is,
\bea
U &=& 
 \exp(-\ii \theta \int_{} \dd x' J^0(x') ) 
 \equiv
\exp(\ii \theta \frac{K_{\rm I' J'} q_{\rm J'}}{2 \pi} \int_{} \dd x' \prt_{x'} \varphi_{\rm I'}  ) 
\equiv \exp(\ii \theta \frac{ Q_{\rm I'}}{2 \pi} 
\int_{} \dd x' \prt_{x'} \varphi_{\rm I'} ).\\
\label{eq:boson-varphi-shift}
U \varphi_{\rm I} U^{-1} &=&
\varphi_{\rm I} - \ii \theta \frac{K_{\rm I' J'} q_{\rm J'}}{2 \pi} 
\int_{} \dd x' 
 [ \varphi_{\rm I}, \prt_{x'} \varphi_{\rm I'}  ]
=\varphi_{\rm I}  + q_{\rm I}\vartheta. 
\eea
Again we define ${ Q_{\rm I}} \equiv {K_{\rm I J} q_{\rm J}}$. 
{If we consider the case that the 1d space is extended along the real line $x' \in \mathbb{R} =(-\infty, \infty)$,
then the integration range is $\int_{- \infty}^\infty$. 
Alternatively for the 1d circular space $x' \in [0, 2 \pi R)$ of the radius size $R$, 
the integration range needs to be appropriately modified to $\int_{0}^{2 \pi R}$ if we choose the reference point at $x'=0$.}
In \eq{eq:boson-varphi-shift}, we also use the fact that 
the $\varphi_{\rm I}$ and the conjugate momentum
$\Pi_{\rm I} \equiv \frac{\delta {\cL_{\varphi,\rm free}}}{\delta (\partial_t \varphi_{\rm I} )} =\frac{1}{4\pi} K_{\rm IJ} \partial_x \varphi_{\rm J}$
satisfies the canonical quantization relation
\bea
[\varphi_{\rm I}(x), \Pi_{\rm J}(x')] =
\frac{1}{2} \ii  \delta_{\rm IJ} \delta(x - x') \Rightarrow
[\varphi_{\rm I}(x),  \partial_{x'} \varphi_{\rm J}(x')] =
\ii  2 \pi K^{-1}_{\rm IJ} \delta(x-x').
\eea
Here the commutator gives a $\frac{1}{2}$ factor because we have each $\varphi_{\rm I}$ mode as a 1+1d chiral boson.

More generally, we can place the 1d charge operator $U$ along any 1d curve in the 1+1d spacetime,
while $U$ is obtained by the exponential of integrating $J^\mu$ current along this 1d curve. 
There is a correspondence between the operators from fermionic $\Psi$ and bosonic $\varphi$ fields:
\bea
J^\mu &\equiv& 
 \Psi_{\rm I'}^\dagger
{K_{\rm I' J'} q_{\rm J'}}
\tau^\mu_{ L/R} 
\Psi_{\rm I'}
=
-\frac{ 1}{2 \pi} {K_{\rm I' J'} q_{\rm J'}} \epsilon^{\mu \nu}
\prt_{\nu} \varphi_{\rm I'}
= -\frac{ 1}{2 \pi} {Q_{\rm I' }} \epsilon^{\mu \nu}
\prt_{\nu} \varphi_{\rm I'}. \\
J &\equiv&  J^\mu \dd x_{\mu}. 
\quad\quad\quad
U \equiv
\exp(- \ii \theta   {\int (\dd x'_{\mu})^{\perp}   J^\mu}). 
\eea
{Here we denote $(\dd x'_{\mu})^{\perp}$ as the co-dimension volume form of the current direction $x'_{\mu}$.
For example, for $\mu=0$, the density $J^0$ has a corresponding volume $(\dd x'_{0})^{\perp}= \dd x$ along the $\mu=1$ spatial direction
that we integrate over.}

\subsection{Light Sector vs Dark Sector: 
Anomaly-free Condition vs Topological Gapping Condition}

\begin{figure}[!h] 
\centering
{{(a)}} \includegraphics[height=2.8in]{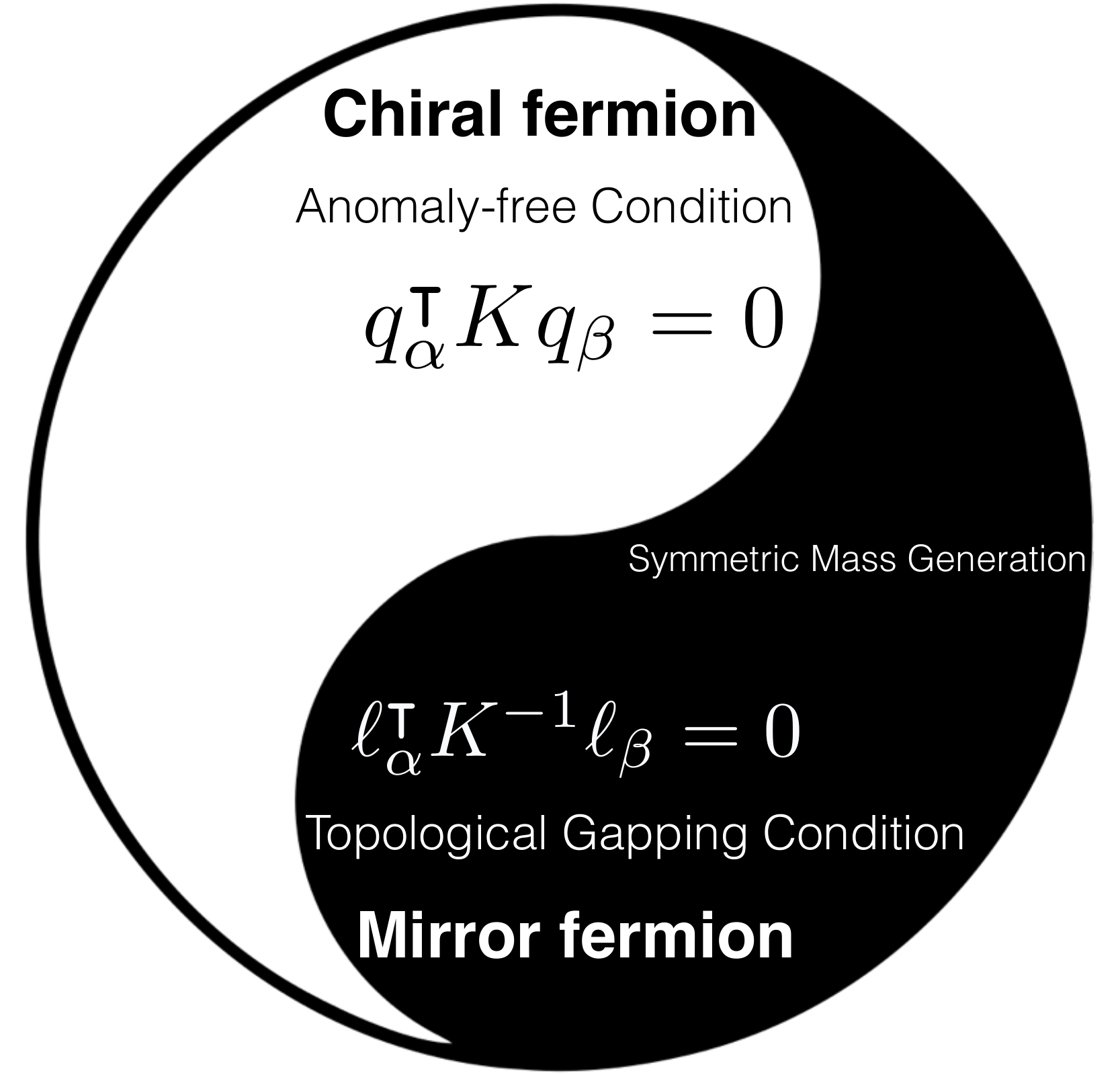} \quad\quad\quad\quad \quad\quad\\ 
{{(b)}} \includegraphics[height=1.3in]{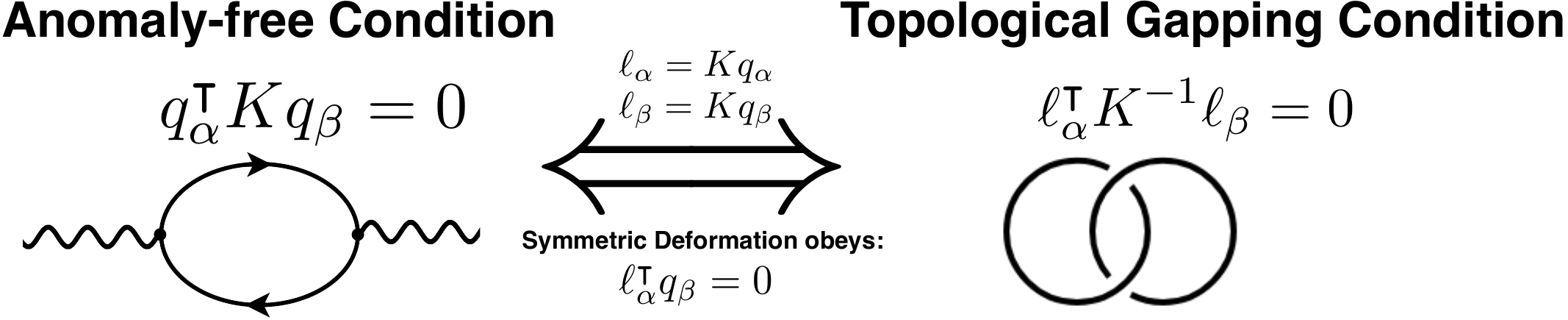}  
\caption{(a) The $\mathbb{light}$ sector is the chiral fermion sector, which is also light nearly massless/gapless.
The {\bf dark} sector is the mirror fermion sector, which is also heavy and gapped.
(b) The anomaly-free condition
\eq{eq:1+1d-anomaly}'s
$q^{\intercal}_\al K^{} q_\bt=0$
and the topological gapping condition
\eq{eq:1+1d-gap}'s
$\ell^{\intercal}_\al K^{-1} \ell_\bt$ can be mapped to each other.
It is proven in \cite{Wang2013ytaJW1307.7480} (also reviewed in \cite{WangYou2204.14271})
that one condition holds if and only if the other condition holds.
Thus, the anomaly-free condition in the chiral fermion sector guarantees 
the design of the topological gapping condition in the mirror fermion sector.
}
\label{fig:light-dark}
\end{figure}

The bosonized interaction cosine terms \eq{eq:interaction-cosine-term} 
(and their fermionized multi-fermion interaction counterpart \eq{eq:multi-fermion}) in the {\bf dark} sector (gapped mirror sector)
are carefully designed to satisfy the {\bf topological gapping condition} \cite{Haldane1995Stability, KapustinSaulina1008.0654KS, Wang2015Boundary, Levin1301.7355},
which are tightly constrained by the  $\mathbb{anomaly}$-$\mathbb{free}$ $\mathbb{condition}$ in the $\mathbb{light}$ sector (nearly gapless chiral sector).
In fact, it is proven in \cite{Wang2013ytaJW1307.7480} (also reviewed in \cite{WangYou2204.14271})
that the anomaly-free condition can be mapped to the topological gapping condition; thus these conditions are equivalent (if and only if) or dual to each other, schematically
shown in \Fig{fig:light-dark}.

To explain \Fig{fig:light-dark} and its use in our 1+1d toy model of $N$-left moving and $N$-right moving Weyl fermions with
zero central charge $c_-=0$ and with an internal symmetry $\U(N)_L \times \U(N)_R  \supset \U(1)^{2N}$, 
we recall that
the anomaly-free condition for the $\U(1)$ charges $q_\al$ and $q_\bt$ is
 the anomaly coefficient \eq{eq:U1anomaly-free} vanishes:
\bea \label{eq:1+1d-anomaly}
q^{\intercal}_\al K^{} q_\bt \equiv \sum_{\rm I,J} q_{\al,\rm I} K^{}_{\rm IJ} q_{\bt,\rm J}=0,
\eea 
where $K$ can be a general rank-$2N$ symmetric bilinear integer-valued matrix (with $2N \times 2N$ integer entries), and ${\rm I,J} \in \{1,\dots, 2N\}$. 
Note at most ${\al , \bt} \in \{1,\dots, N\}$, because  the largest anomaly-free subgroup of $\U(1)^{2N}$ is $\U(1)^{N}$.
In particular, for the 3-4-5-0 model, we take $\al, \bt \in \{1,2\}$ for $\U(1)_{\rm 1st} \times \U(1)_{\rm 2nd}$ in \Table{table:3450charge}, for $N=2$ and $2N=4$.

On the other hand, the topological gapping condition \cite{Haldane1995Stability, KapustinSaulina1008.0654KS, Wang2015Boundary, Levin1301.7355} 
demands to find a set of $N$ cosine terms, 
$$\sum_{\alpha=1}^N g_{\alpha}\cos(\ell_{\alpha,{\rm I} }\varphi_{\rm I}),$$
to gap the theory.
To design the cosine terms preserving the $\U(1)$ symmetry
$\varphi_{\rm I} \to \varphi_{\rm I} +  {q_{\beta,{\rm I}} \upalpha}$, we require that
\bea \label{eq:1+1d-symm}
\ell_\alpha^\intercal q_{\beta} \equiv \sum_{\rm I} \ell_{\alpha, {\rm I}} q_{\beta,{\rm I}} =0.
\eea
The large $g$ coupling limit of the cosine term causes the condensation of the vertex operator $\<\exp(\ii \ell_{\alpha,{\rm I} }\varphi_{\rm I}) \>\neq 0$ in the 1+1d.
The 1+1d gapping condition can be regarded as the topological boundary condition specified by the line operator $A$ in a 2+1d bulk of $K$ matrix Chern-Simons theory.
The topological boundary condition is the Dirichlet boundary condition 
$$\sum_{\rm I} \ell_{\alpha, {\rm I}} A_{\rm I} \big\vert_{\cM^2}=0$$
of line operator with $\alpha \in \{1,\dots, N\}$ and ${\rm I,J} \in \{1,\dots, 2N\}$. 
There are $N$ different Dirichlet boundary conditions among $\alpha \in \{1,\dots, N\}$,
and they must be compatible such that the link invariant of any linked loop in the 2+1d bulk
must give a trivial phase:
$$
\text{$\exp(\ii \pi \ell^{\intercal}_\al K^{-1} \ell_\al) =1$ and
$\exp(\ii 2\pi \ell^{\intercal}_\al K^{-1} \ell_\bt) =1$.}
$$
This phase 
can be computed as the ratio of path integral $Z[\text{Hopf link};S^3]/Z[S^3]$ \cite{Putrov2016qdo1612.09298PWY},
with two line operators labeled by $\ell_\al$ and $\ell_\bt$ linked as a Hopf link say in a 3-sphere $S^3$.
Thus the topological boundary condition is now related to the trivial Berry phase or statistical phase between the bulk lines,
which projects to the boundary constraint:
\bea \label{eq:1+1d-gap}
\ell^{\intercal}_\al K^{-1} \ell_\bt \equiv \sum_{\rm I,J} \ell_{\al,\rm I} K^{-1}_{\rm IJ} \ell_{\bt,\rm J}=0.
\eea
We may call the symmetry condition \eq{eq:1+1d-symm} and the topological gapping condition \eq{eq:1+1d-gap} together as the SMG gapping condition.
The 
boundary condition 
$\sum_{\rm I} \ell_{\alpha, {\rm I}} A_{\rm I} \big\vert_{\cM^2}=0$
implies that a certain set of anyons live on the open endpoint of the Wilson line 
$\exp(\ii \int \sum_{\rm I} \ell_{\alpha, {\rm I}} A_{\rm I})$  can end on the 1+1d boundary.
Together with \eq{eq:1+1d-gap}, these indicate physically that a set of anyons from a 2+1d bulk can condense on the 1+1d boundary in a compatible manner 
to fully gap the boundary --- this is known as the gapped boundary defined by the compatible anyon condensation.

To explain that the SMG gapping condition 
holds implies that the anomaly-free condition also holds,
\Refe{Wang2013ytaJW1307.7480} shows that given the set of
$\ell_\alpha$ satisfying \eq{eq:1+1d-symm} and \eq{eq:1+1d-gap}, 
one can find the set of $q_{\alpha}$ simultaneously satisfying \eq{eq:1+1d-anomaly}.
This is true because that given $\ell_\alpha$, {we can choose $q_{\alpha} =K^{-1} \ell_\alpha$.} 

To explain that the anomaly-free condition 
holds implies that the SMG gapping condition also holds,
\Refe{Wang2013ytaJW1307.7480} shows that given the set of
$q_\alpha$ satisfying \eqref{eq:1+1d-anomaly}, 
one can find the set of $l_{\alpha}$ simultaneously satisfying \eq{eq:1+1d-symm} and \eq{eq:1+1d-gap}.
This is true because that given $q_\alpha$, {we can choose $\ell_{\alpha}=K q_\alpha$.} 

{In \Table{table:3450charge}, we choose the two mutual anomaly-free $\U(1)_{\rm 1st} \times \U(1)_{\rm 2nd}$ charge
$q_{{1}} =( 3 , 4 , 5 , 0)$ and  
$q_{{2}} =( 0,  5, 4, 3)$.
So the two corresponding gapping terms can be $\ell_{1}=K q_1= q_3=( 3 , 4 , -5 , 0)$
and $\ell_{2}=K q_2= q_4=( 0,  5, -4, -3)$.
Note that they are also the other mutual anomaly-free $\U(1)_{\rm 3rd} \times \U(1)_{\rm 4th}$ charge $q_3$ and $q_4$.
In \Table{table:3450charge}, we choose a different linear combination,
we redefine
 $\frac{1}{3} (\ell_1 -2 \ell_2) =(1, -2, 1, 2)$ as $\ell_{1}$
and redefine $\frac{1}{3} (2 \ell_1 - \ell_2) = ( 2, 1, -2,1)$ as $\ell_{2}$, 
to minimize the scaling dimensions of the cosine terms. 
These data are compatible with the conditions of \eq{eq:1+1d-anomaly}, \eq{eq:1+1d-symm},  and \eq{eq:1+1d-gap}.
The underlying structure is also known as the Narain lattice of the compact boson theory \cite{Narain:1986am}.
}

This concludes the demonstration of the map ``$\Leftrightarrow$'' (namely, if and only if conditions) in \Fig{fig:light-dark}
\cite{Wang2013ytaJW1307.7480, ZengZhuWangYou3450SMG2202.12355}.
Because this proof works for a large class of rank-$2N$ symmetric bilinear integer-valued $K$ matrix,
we can generalize the 1+1d 3-4-5-0 toy model to a general 1+1d U(1)$^N$ symmetry fermion or gauge theory.
This means that the 1+1d CT or P solution provided in this work 
can be straightforwardly generalized to all 1+1d U(1)$^N$-anomaly-free models.


\bibliography{BSM-CP-2d.bib}

\begin{thebibliography}{72}%
\makeatletter
\providecommand \@ifxundefined [1]{%
 \@ifx{#1\undefined}
}%
\providecommand \@ifnum [1]{%
 \ifnum #1\expandafter \@firstoftwo
 \else \expandafter \@secondoftwo
 \fi
}%
\providecommand \@ifx [1]{%
 \ifx #1\expandafter \@firstoftwo
 \else \expandafter \@secondoftwo
 \fi
}%
\providecommand \natexlab [1]{#1}%
\providecommand \enquote  [1]{``#1''}%
\providecommand \bibnamefont  [1]{#1}%
\providecommand \bibfnamefont [1]{#1}%
\providecommand \citenamefont [1]{#1}%
\providecommand \href@noop [0]{\@secondoftwo}%
\providecommand \href [0]{\begingroup \@sanitize@url \@href}%
\providecommand \@href[1]{\@@startlink{#1}\@@href}%
\providecommand \@@href[1]{\endgroup#1\@@endlink}%
\providecommand \@sanitize@url [0]{\catcode `\\12\catcode `\$12\catcode
  `\&12\catcode `\#12\catcode `\^12\catcode `\_12\catcode `\%12\relax}%
\providecommand \@@startlink[1]{}%
\providecommand \@@endlink[0]{}%
\providecommand \url  [0]{\begingroup\@sanitize@url \@url }%
\providecommand \@url [1]{\endgroup\@href {#1}{\urlprefix }}%
\providecommand \urlprefix  [0]{URL }%
\providecommand \Eprint [0]{\href }%
\providecommand \doibase [0]{http://dx.doi.org/}%
\providecommand \selectlanguage [0]{\@gobble}%
\providecommand \bibinfo  [0]{\@secondoftwo}%
\providecommand \bibfield  [0]{\@secondoftwo}%
\providecommand \translation [1]{[#1]}%
\providecommand \BibitemOpen [0]{}%
\providecommand \bibitemStop [0]{}%
\providecommand \bibitemNoStop [0]{.\EOS\space}%
\providecommand \EOS [0]{\spacefactor3000\relax}%
\providecommand \BibitemShut  [1]{\csname bibitem#1\endcsname}%
\let\auto@bib@innerbib\@empty
\bibitem [{\citenamefont {Smith}\ \emph {et~al.}(1957)\citenamefont {Smith},
  \citenamefont {Purcell},\ and\ \citenamefont {Ramsey}}]{Smith1957ht}%
  \BibitemOpen
  \bibfield  {author} {\bibinfo {author} {\bibfnamefont {J.~H.}\ \bibnamefont
  {Smith}}, \bibinfo {author} {\bibfnamefont {E.~M.}\ \bibnamefont {Purcell}},
  \ and\ \bibinfo {author} {\bibfnamefont {N.~F.}\ \bibnamefont {Ramsey}},\
  }\bibfield  {title} {\enquote {\bibinfo {title} {{Experimental limit to the
  electric dipole moment of the neutron}},}\ }\href {\doibase
  10.1103/PhysRev.108.120} {\bibfield  {journal} {\bibinfo  {journal} {Phys.
  Rev.}\ }\textbf {\bibinfo {volume} {108}},\ \bibinfo {pages} {120--122}
  (\bibinfo {year} {1957})}\BibitemShut {NoStop}%
\bibitem [{\citenamefont {Baker}\ \emph {et~al.}(2006)\citenamefont {Baker}
  \emph {et~al.}}]{Bakerhepex0602020}%
  \BibitemOpen
  \bibfield  {author} {\bibinfo {author} {\bibfnamefont {C.~A.}\ \bibnamefont
  {Baker}} \emph {et~al.},\ }\bibfield  {title} {\enquote {\bibinfo {title}
  {{An Improved experimental limit on the electric dipole moment of the
  neutron}},}\ }\href {\doibase 10.1103/PhysRevLett.97.131801} {\bibfield
  {journal} {\bibinfo  {journal} {Phys. Rev. Lett.}\ }\textbf {\bibinfo
  {volume} {97}},\ \bibinfo {pages} {131801} (\bibinfo {year} {2006})},\
  \Eprint {http://arxiv.org/abs/hep-ex/0602020} {arXiv:hep-ex/0602020}
  \BibitemShut {NoStop}%
\bibitem [{\citenamefont {Abel}\ \emph {et~al.}(2020)\citenamefont {Abel} \emph
  {et~al.}}]{Abel2020PRL2001.11966}%
  \BibitemOpen
  \bibfield  {author} {\bibinfo {author} {\bibfnamefont {C.}~\bibnamefont
  {Abel}} \emph {et~al.},\ }\bibfield  {title} {\enquote {\bibinfo {title}
  {{Measurement of the Permanent Electric Dipole Moment of the Neutron}},}\
  }\href {\doibase 10.1103/PhysRevLett.124.081803} {\bibfield  {journal}
  {\bibinfo  {journal} {Phys. Rev. Lett.}\ }\textbf {\bibinfo {volume} {124}},\
  \bibinfo {pages} {081803} (\bibinfo {year} {2020})},\ \Eprint
  {http://arxiv.org/abs/2001.11966} {arXiv:2001.11966 [hep-ex]} \BibitemShut
  {NoStop}%
\bibitem [{\citenamefont {Dine}(2000)}]{Dine2000TASI0011376}%
  \BibitemOpen
  \bibfield  {author} {\bibinfo {author} {\bibfnamefont {Michael}\ \bibnamefont
  {Dine}},\ }\bibfield  {title} {\enquote {\bibinfo {title} {{TASI lectures on
  the strong CP problem}},}\ }in\ \href@noop {} {\emph {\bibinfo {booktitle}
  {{Theoretical Advanced Study Institute in Elementary Particle Physics (TASI
  2000): Flavor Physics for the Millennium}}}}\ (\bibinfo {year} {2000})\ pp.\
  \bibinfo {pages} {349--369},\ \Eprint {http://arxiv.org/abs/hep-ph/0011376}
  {arXiv:hep-ph/0011376} \BibitemShut {NoStop}%
\bibitem [{\citenamefont {Hook}(2019)}]{Hook2018TASI1812.02669}%
  \BibitemOpen
  \bibfield  {author} {\bibinfo {author} {\bibfnamefont {Anson}\ \bibnamefont
  {Hook}},\ }\bibfield  {title} {\enquote {\bibinfo {title} {{TASI Lectures on
  the Strong CP Problem and Axions}},}\ }\href@noop {} {\bibfield  {journal}
  {\bibinfo  {journal} {PoS}\ }\textbf {\bibinfo {volume} {TASI2018}},\
  \bibinfo {pages} {004} (\bibinfo {year} {2019})},\ \Eprint
  {http://arxiv.org/abs/1812.02669} {arXiv:1812.02669 [hep-ph]} \BibitemShut
  {NoStop}%
\bibitem [{\citenamefont {Belavin}\ \emph {et~al.}(1975)\citenamefont
  {Belavin}, \citenamefont {Polyakov}, \citenamefont {Schwartz},\ and\
  \citenamefont {Tyupkin}}]{BelavinBPST1975}%
  \BibitemOpen
  \bibfield  {author} {\bibinfo {author} {\bibfnamefont {A.A.}\ \bibnamefont
  {Belavin}}, \bibinfo {author} {\bibfnamefont {Alexander~M.}\ \bibnamefont
  {Polyakov}}, \bibinfo {author} {\bibfnamefont {A.S.}\ \bibnamefont
  {Schwartz}}, \ and\ \bibinfo {author} {\bibfnamefont {Yu.S.}\ \bibnamefont
  {Tyupkin}},\ }\bibfield  {title} {\enquote {\bibinfo {title} {{Pseudoparticle
  Solutions of the Yang-Mills Equations}},}\ }\href {\doibase
  10.1016/0370-2693(75)90163-X} {\bibfield  {journal} {\bibinfo  {journal}
  {Phys. Lett. B}\ }\textbf {\bibinfo {volume} {59}},\ \bibinfo {pages}
  {85--87} (\bibinfo {year} {1975})}\BibitemShut {NoStop}%
\bibitem [{\citenamefont {'t~Hooft}(1976{\natexlab{a}})}]{tHooft1976instanton}%
  \BibitemOpen
  \bibfield  {author} {\bibinfo {author} {\bibfnamefont {Gerard}\ \bibnamefont
  {'t~Hooft}},\ }\bibfield  {title} {\enquote {\bibinfo {title} {{Computation
  of the Quantum Effects Due to a Four-Dimensional Pseudoparticle}},}\ }\href
  {\doibase 10.1103/PhysRevD.14.3432} {\bibfield  {journal} {\bibinfo
  {journal} {Phys. Rev. D}\ }\textbf {\bibinfo {volume} {14}},\ \bibinfo
  {pages} {3432--3450} (\bibinfo {year} {1976}{\natexlab{a}})},\ \bibinfo
  {note} {[Erratum: Phys.Rev.D 18, 2199 (1978)]}\BibitemShut {NoStop}%
\bibitem [{\citenamefont {'t~Hooft}(1980)}]{tHooft1979ratanomaly}%
  \BibitemOpen
  \bibfield  {author} {\bibinfo {author} {\bibfnamefont {Gerard}\ \bibnamefont
  {'t~Hooft}},\ }\bibfield  {title} {\enquote {\bibinfo {title} {{Naturalness,
  chiral symmetry, and spontaneous chiral symmetry breaking}},}\ }\bibfield
  {booktitle} {\emph {\bibinfo {booktitle} {{Recent Developments in Gauge
  Theories. Proceedings, Nato Advanced Study Institute, Cargese, France, August
  26 - September 8, 1979}}},\ }\href {\doibase 10.1007/978-1-4684-7571-5_9}
  {\bibfield  {journal} {\bibinfo  {journal} {NATO Sci. Ser. B}\ }\textbf
  {\bibinfo {volume} {59}},\ \bibinfo {pages} {135--157} (\bibinfo {year}
  {1980})}\BibitemShut {NoStop}%
\bibitem [{\citenamefont {'t~Hooft}(1976{\natexlab{b}})}]{tHooft1976ripPRL}%
  \BibitemOpen
  \bibfield  {author} {\bibinfo {author} {\bibfnamefont {Gerard}\ \bibnamefont
  {'t~Hooft}},\ }\bibfield  {title} {\enquote {\bibinfo {title} {{Symmetry
  Breaking Through Bell-Jackiw Anomalies}},}\ }\href {\doibase
  10.1103/PhysRevLett.37.8} {\bibfield  {journal} {\bibinfo  {journal} {Phys.
  Rev. Lett.}\ }\textbf {\bibinfo {volume} {37}},\ \bibinfo {pages} {8--11}
  (\bibinfo {year} {1976}{\natexlab{b}})}\BibitemShut {NoStop}%
\bibitem [{\citenamefont {Peccei}\ and\ \citenamefont
  {Quinn}(1977{\natexlab{a}})}]{PecceiQuinn1977hhPRL}%
  \BibitemOpen
  \bibfield  {author} {\bibinfo {author} {\bibfnamefont {R.~D.}\ \bibnamefont
  {Peccei}}\ and\ \bibinfo {author} {\bibfnamefont {Helen~R.}\ \bibnamefont
  {Quinn}},\ }\bibfield  {title} {\enquote {\bibinfo {title} {{CP Conservation
  in the Presence of Instantons}},}\ }\href {\doibase
  10.1103/PhysRevLett.38.1440} {\bibfield  {journal} {\bibinfo  {journal}
  {Phys. Rev. Lett.}\ }\textbf {\bibinfo {volume} {38}},\ \bibinfo {pages}
  {1440--1443} (\bibinfo {year} {1977}{\natexlab{a}})}\BibitemShut {NoStop}%
\bibitem [{\citenamefont {Peccei}\ and\ \citenamefont
  {Quinn}(1977{\natexlab{b}})}]{PecceiQuinn1977urPRD}%
  \BibitemOpen
  \bibfield  {author} {\bibinfo {author} {\bibfnamefont {R.~D.}\ \bibnamefont
  {Peccei}}\ and\ \bibinfo {author} {\bibfnamefont {Helen~R.}\ \bibnamefont
  {Quinn}},\ }\bibfield  {title} {\enquote {\bibinfo {title} {{Constraints
  Imposed by CP Conservation in the Presence of Instantons}},}\ }\href
  {\doibase 10.1103/PhysRevD.16.1791} {\bibfield  {journal} {\bibinfo
  {journal} {Phys. Rev. D}\ }\textbf {\bibinfo {volume} {16}},\ \bibinfo
  {pages} {1791--1797} (\bibinfo {year} {1977}{\natexlab{b}})}\BibitemShut
  {NoStop}%
\bibitem [{\citenamefont {Weinberg}(1978)}]{Weinberg1977ma1978PRL}%
  \BibitemOpen
  \bibfield  {author} {\bibinfo {author} {\bibfnamefont {Steven}\ \bibnamefont
  {Weinberg}},\ }\bibfield  {title} {\enquote {\bibinfo {title} {{A New Light
  Boson?}}}\ }\href {\doibase 10.1103/PhysRevLett.40.223} {\bibfield  {journal}
  {\bibinfo  {journal} {Phys. Rev. Lett.}\ }\textbf {\bibinfo {volume} {40}},\
  \bibinfo {pages} {223--226} (\bibinfo {year} {1978})}\BibitemShut {NoStop}%
\bibitem [{\citenamefont {Wilczek}(1978)}]{Wilczek1977pjPRL}%
  \BibitemOpen
  \bibfield  {author} {\bibinfo {author} {\bibfnamefont {Frank}\ \bibnamefont
  {Wilczek}},\ }\bibfield  {title} {\enquote {\bibinfo {title} {{Problem of
  Strong $P$ and $T$ Invariance in the Presence of Instantons}},}\ }\href
  {\doibase 10.1103/PhysRevLett.40.279} {\bibfield  {journal} {\bibinfo
  {journal} {Phys. Rev. Lett.}\ }\textbf {\bibinfo {volume} {40}},\ \bibinfo
  {pages} {279--282} (\bibinfo {year} {1978})}\BibitemShut {NoStop}%
\bibitem [{\citenamefont {Nelson}(1984)}]{Nelson1983zb1984PLB}%
  \BibitemOpen
  \bibfield  {author} {\bibinfo {author} {\bibfnamefont {Ann~E.}\ \bibnamefont
  {Nelson}},\ }\bibfield  {title} {\enquote {\bibinfo {title} {{Naturally Weak
  CP Violation}},}\ }\href {\doibase 10.1016/0370-2693(84)92025-2} {\bibfield
  {journal} {\bibinfo  {journal} {Phys. Lett. B}\ }\textbf {\bibinfo {volume}
  {136}},\ \bibinfo {pages} {387--391} (\bibinfo {year} {1984})}\BibitemShut
  {NoStop}%
\bibitem [{\citenamefont {Barr}(1984)}]{Barr1984qx1984PRL}%
  \BibitemOpen
  \bibfield  {author} {\bibinfo {author} {\bibfnamefont {Stephen~M.}\
  \bibnamefont {Barr}},\ }\bibfield  {title} {\enquote {\bibinfo {title}
  {{Solving the Strong CP Problem Without the Peccei-Quinn Symmetry}},}\ }\href
  {\doibase 10.1103/PhysRevLett.53.329} {\bibfield  {journal} {\bibinfo
  {journal} {Phys. Rev. Lett.}\ }\textbf {\bibinfo {volume} {53}},\ \bibinfo
  {pages} {329} (\bibinfo {year} {1984})}\BibitemShut {NoStop}%
\bibitem [{\citenamefont {Babu}\ and\ \citenamefont
  {Mohapatra}(1989)}]{BabuMohapatra1989}%
  \BibitemOpen
  \bibfield  {author} {\bibinfo {author} {\bibfnamefont {K.~S.}\ \bibnamefont
  {Babu}}\ and\ \bibinfo {author} {\bibfnamefont {Rabindra~N.}\ \bibnamefont
  {Mohapatra}},\ }\bibfield  {title} {\enquote {\bibinfo {title} {{{CP}
  Violation in Seesaw Models of Quark Masses}},}\ }\href {\doibase
  10.1103/PhysRevLett.62.1079} {\bibfield  {journal} {\bibinfo  {journal}
  {Phys. Rev. Lett.}\ }\textbf {\bibinfo {volume} {62}},\ \bibinfo {pages}
  {1079} (\bibinfo {year} {1989})}\BibitemShut {NoStop}%
\bibitem [{\citenamefont {Babu}\ and\ \citenamefont
  {Mohapatra}(1990)}]{BabuMohapatra1989rbPRD}%
  \BibitemOpen
  \bibfield  {author} {\bibinfo {author} {\bibfnamefont {K.~S.}\ \bibnamefont
  {Babu}}\ and\ \bibinfo {author} {\bibfnamefont {Rabindra~N.}\ \bibnamefont
  {Mohapatra}},\ }\bibfield  {title} {\enquote {\bibinfo {title} {{A Solution
  to the Strong {CP} Problem Without an Axion}},}\ }\href {\doibase
  10.1103/PhysRevD.41.1286} {\bibfield  {journal} {\bibinfo  {journal} {Phys.
  Rev. D}\ }\textbf {\bibinfo {volume} {41}},\ \bibinfo {pages} {1286}
  (\bibinfo {year} {1990})}\BibitemShut {NoStop}%
\bibitem [{\citenamefont {Barr}\ \emph {et~al.}(1991)\citenamefont {Barr},
  \citenamefont {Chang},\ and\ \citenamefont
  {Senjanovic}}]{BarrChangSenjanovic1991qxPRL}%
  \BibitemOpen
  \bibfield  {author} {\bibinfo {author} {\bibfnamefont {Stephen~M.}\
  \bibnamefont {Barr}}, \bibinfo {author} {\bibfnamefont {D.}~\bibnamefont
  {Chang}}, \ and\ \bibinfo {author} {\bibfnamefont {G.}~\bibnamefont
  {Senjanovic}},\ }\bibfield  {title} {\enquote {\bibinfo {title} {{Strong CP
  problem and parity}},}\ }\href {\doibase 10.1103/PhysRevLett.67.2765}
  {\bibfield  {journal} {\bibinfo  {journal} {Phys. Rev. Lett.}\ }\textbf
  {\bibinfo {volume} {67}},\ \bibinfo {pages} {2765--2768} (\bibinfo {year}
  {1991})}\BibitemShut {NoStop}%
\bibitem [{\citenamefont {Wang}\ and\ \citenamefont
  {You}(2022{\natexlab{a}})}]{WangYou2204.14271}%
  \BibitemOpen
  \bibfield  {author} {\bibinfo {author} {\bibfnamefont {Juven}\ \bibnamefont
  {Wang}}\ and\ \bibinfo {author} {\bibfnamefont {Yi-Zhuang}\ \bibnamefont
  {You}},\ }\bibfield  {title} {\enquote {\bibinfo {title} {{Symmetric Mass
  Generation}},}\ }\href {\doibase 10.3390/sym14071475} {\bibfield  {journal}
  {\bibinfo  {journal} {Symmetry}\ }\textbf {\bibinfo {volume} {14}} (\bibinfo
  {year} {2022}{\natexlab{a}}),\ 10.3390/sym14071475},\ \Eprint
  {http://arxiv.org/abs/2204.14271} {arXiv:2204.14271 [cond-mat.str-el]}
  \BibitemShut {NoStop}%
\bibitem [{\citenamefont {Schwinger}(1962)}]{Schwinger1962PR2}%
  \BibitemOpen
  \bibfield  {author} {\bibinfo {author} {\bibfnamefont {Julian~S.}\
  \bibnamefont {Schwinger}},\ }\bibfield  {title} {\enquote {\bibinfo {title}
  {{Gauge Invariance and Mass. 2.}}}\ }\href {\doibase
  10.1103/PhysRev.128.2425} {\bibfield  {journal} {\bibinfo  {journal} {Phys.
  Rev.}\ }\textbf {\bibinfo {volume} {128}},\ \bibinfo {pages} {2425--2429}
  (\bibinfo {year} {1962})}\BibitemShut {NoStop}%
\bibitem [{\citenamefont
  {Wang}(2022{\natexlab{a}})}]{StrongCPtoappear-2212.14036}%
  \BibitemOpen
  \bibfield  {author} {\bibinfo {author} {\bibfnamefont {Juven}\ \bibnamefont
  {Wang}},\ }\bibfield  {title} {\enquote {\bibinfo {title} {{Strong CP Problem
  and Symmetric Interacting Mass Solution}},}\ }\href@noop {} {\  (\bibinfo
  {year} {2022}{\natexlab{a}})},\ \Eprint {http://arxiv.org/abs/2212.14036}
  {arXiv:2212.14036 [hep-ph]} \BibitemShut {NoStop}%
\bibitem [{\citenamefont {Wang}(2022{\natexlab{b}})}]{WangCPT2109.15320}%
  \BibitemOpen
  \bibfield  {author} {\bibinfo {author} {\bibfnamefont {Juven}\ \bibnamefont
  {Wang}},\ }\bibfield  {title} {\enquote {\bibinfo {title} {{C-P-T
  fractionalization}},}\ }\href {\doibase 10.1103/PhysRevD.106.105009}
  {\bibfield  {journal} {\bibinfo  {journal} {Phys. Rev. D}\ }\textbf {\bibinfo
  {volume} {106}},\ \bibinfo {pages} {105009} (\bibinfo {year}
  {2022}{\natexlab{b}})},\ \Eprint {http://arxiv.org/abs/2109.15320}
  {arXiv:2109.15320 [hep-th]} \BibitemShut {NoStop}%
\bibitem [{Dis()}]{DisorderTheta}%
  \BibitemOpen
  \bibfield  {title} {\enquote {\bibinfo {title} {{Order to Disorder Theta
  Vacua}},}\ }\href@noop {} {\bibinfo  {journal} {to appear}\ }\BibitemShut
  {NoStop}%
\bibitem [{\citenamefont {Fisher}\ \emph {et~al.}(1989)\citenamefont {Fisher},
  \citenamefont {Weichman}, \citenamefont {Grinstein},\ and\ \citenamefont
  {Fisher}}]{Fisher1989zza}%
  \BibitemOpen
\bibfield  {journal} {  }\bibfield  {author} {\bibinfo {author} {\bibfnamefont
  {Matthew P.~A.}\ \bibnamefont {Fisher}}, \bibinfo {author} {\bibfnamefont
  {Peter~B.}\ \bibnamefont {Weichman}}, \bibinfo {author} {\bibfnamefont
  {G.}~\bibnamefont {Grinstein}}, \ and\ \bibinfo {author} {\bibfnamefont
  {Daniel~S.}\ \bibnamefont {Fisher}},\ }\bibfield  {title} {\enquote {\bibinfo
  {title} {{Boson localization and the superfluid-insulator transition}},}\
  }\href {\doibase 10.1103/PhysRevB.40.546} {\bibfield  {journal} {\bibinfo
  {journal} {Phys. Rev. B}\ }\textbf {\bibinfo {volume} {40}},\ \bibinfo
  {pages} {546--570} (\bibinfo {year} {1989})}\BibitemShut {NoStop}%
\bibitem [{\citenamefont {Adler}(1969)}]{Adler1969gkABJ}%
  \BibitemOpen
  \bibfield  {author} {\bibinfo {author} {\bibfnamefont {Stephen~L.}\
  \bibnamefont {Adler}},\ }\bibfield  {title} {\enquote {\bibinfo {title}
  {{Axial vector vertex in spinor electrodynamics}},}\ }\href {\doibase
  10.1103/PhysRev.177.2426} {\bibfield  {journal} {\bibinfo  {journal} {Phys.
  Rev.}\ }\textbf {\bibinfo {volume} {177}},\ \bibinfo {pages} {2426--2438}
  (\bibinfo {year} {1969})}\BibitemShut {NoStop}%
\bibitem [{\citenamefont {Bell}\ and\ \citenamefont
  {Jackiw}(1969)}]{Bell1969tsABJ}%
  \BibitemOpen
  \bibfield  {author} {\bibinfo {author} {\bibfnamefont {J.~S.}\ \bibnamefont
  {Bell}}\ and\ \bibinfo {author} {\bibfnamefont {R.}~\bibnamefont {Jackiw}},\
  }\bibfield  {title} {\enquote {\bibinfo {title} {{A PCAC puzzle: $\pi^0 \to
  \gamma \gamma$ in the $\sigma$ model}},}\ }\href {\doibase
  10.1007/BF02823296} {\bibfield  {journal} {\bibinfo  {journal} {Nuovo Cim.}\
  }\textbf {\bibinfo {volume} {A60}},\ \bibinfo {pages} {47--61} (\bibinfo
  {year} {1969})}\BibitemShut {NoStop}%
\bibitem [{\citenamefont {Nielsen}\ and\ \citenamefont
  {Ninomiya}(1981)}]{NielsenNinomiya1981hkPLB}%
  \BibitemOpen
  \bibfield  {author} {\bibinfo {author} {\bibfnamefont {Holger~Bech}\
  \bibnamefont {Nielsen}}\ and\ \bibinfo {author} {\bibfnamefont
  {M.}~\bibnamefont {Ninomiya}},\ }\bibfield  {title} {\enquote {\bibinfo
  {title} {{No Go Theorem for Regularizing Chiral Fermions}},}\ }\href
  {\doibase 10.1016/0370-2693(81)91026-1} {\bibfield  {journal} {\bibinfo
  {journal} {Phys. Lett. B}\ }\textbf {\bibinfo {volume} {105}},\ \bibinfo
  {pages} {219--223} (\bibinfo {year} {1981})}\BibitemShut {NoStop}%
\bibitem [{\citenamefont {Craig}\ \emph {et~al.}(2021)\citenamefont {Craig},
  \citenamefont {Garcia~Garcia}, \citenamefont {Koszegi},\ and\ \citenamefont
  {McCune}}]{CraigGarciaKoszegiMcCune2012.13416}%
  \BibitemOpen
  \bibfield  {author} {\bibinfo {author} {\bibfnamefont {Nathaniel}\
  \bibnamefont {Craig}}, \bibinfo {author} {\bibfnamefont {Isabel}\
  \bibnamefont {Garcia~Garcia}}, \bibinfo {author} {\bibfnamefont {Giacomo}\
  \bibnamefont {Koszegi}}, \ and\ \bibinfo {author} {\bibfnamefont {Amara}\
  \bibnamefont {McCune}},\ }\bibfield  {title} {\enquote {\bibinfo {title} {{P
  not PQ}},}\ }\href {\doibase 10.1007/JHEP09(2021)130} {\bibfield  {journal}
  {\bibinfo  {journal} {JHEP}\ }\textbf {\bibinfo {volume} {09}},\ \bibinfo
  {pages} {130} (\bibinfo {year} {2021})},\ \Eprint
  {http://arxiv.org/abs/2012.13416} {arXiv:2012.13416 [hep-ph]} \BibitemShut
  {NoStop}%
\bibitem [{\citenamefont {Eichten}\ and\ \citenamefont
  {Preskill}(1986)}]{Eichten1985ftPreskill1986}%
  \BibitemOpen
  \bibfield  {author} {\bibinfo {author} {\bibfnamefont {Estia}\ \bibnamefont
  {Eichten}}\ and\ \bibinfo {author} {\bibfnamefont {John}\ \bibnamefont
  {Preskill}},\ }\bibfield  {title} {\enquote {\bibinfo {title} {{Chiral Gauge
  Theories on the Lattice}},}\ }\href {\doibase 10.1016/0550-3213(86)90207-5}
  {\bibfield  {journal} {\bibinfo  {journal} {Nucl. Phys.}\ }\textbf {\bibinfo
  {volume} {B268}},\ \bibinfo {pages} {179--208} (\bibinfo {year}
  {1986})}\BibitemShut {NoStop}%
\bibitem [{\citenamefont {Wang}\ and\ \citenamefont
  {Wen}(2013)}]{Wang2013ytaJW1307.7480}%
  \BibitemOpen
  \bibfield  {author} {\bibinfo {author} {\bibfnamefont {Juven}\ \bibnamefont
  {Wang}}\ and\ \bibinfo {author} {\bibfnamefont {Xiao-Gang}\ \bibnamefont
  {Wen}},\ }\bibfield  {title} {\enquote {\bibinfo {title} {{Non-Perturbative
  Regularization of 1+1D Anomaly-Free Chiral Fermions and Bosons: On the
  equivalence of anomaly matching conditions and boundary gapping rules}},}\
  }\href@noop {} {\  (\bibinfo {year} {2013})},\ \Eprint
  {http://arxiv.org/abs/1307.7480} {arXiv:1307.7480 [hep-lat]} \BibitemShut
  {NoStop}%
\bibitem [{\citenamefont {Wang}\ and\ \citenamefont
  {Wen}(2019)}]{Wang2018ugfJW1807.05998}%
  \BibitemOpen
  \bibfield  {author} {\bibinfo {author} {\bibfnamefont {Juven}\ \bibnamefont
  {Wang}}\ and\ \bibinfo {author} {\bibfnamefont {Xiao-Gang}\ \bibnamefont
  {Wen}},\ }\bibfield  {title} {\enquote {\bibinfo {title} {{A Solution to the
  1+1D Gauged Chiral Fermion Problem}},}\ }\href {\doibase
  10.1103/PhysRevD.99.111501} {\bibfield  {journal} {\bibinfo  {journal} {Phys.
  Rev.}\ }\textbf {\bibinfo {volume} {D99}},\ \bibinfo {pages} {111501}
  (\bibinfo {year} {2019})},\ \Eprint {http://arxiv.org/abs/1807.05998}
  {arXiv:1807.05998 [hep-lat]} \BibitemShut {NoStop}%
\bibitem [{\citenamefont {Zeng}\ \emph {et~al.}(2022)\citenamefont {Zeng},
  \citenamefont {Zhu}, \citenamefont {Wang},\ and\ \citenamefont
  {You}}]{ZengZhuWangYou3450SMG2202.12355}%
  \BibitemOpen
  \bibfield  {author} {\bibinfo {author} {\bibfnamefont {Meng}\ \bibnamefont
  {Zeng}}, \bibinfo {author} {\bibfnamefont {Zheng}\ \bibnamefont {Zhu}},
  \bibinfo {author} {\bibfnamefont {Juven}\ \bibnamefont {Wang}}, \ and\
  \bibinfo {author} {\bibfnamefont {Yi-Zhuang}\ \bibnamefont {You}},\
  }\bibfield  {title} {\enquote {\bibinfo {title} {{Symmetric Mass Generation
  in the 1+1 Dimensional Chiral Fermion 3-4-5-0 Model}},}\ }\href {\doibase
  10.1103/PhysRevLett.128.185301} {\bibfield  {journal} {\bibinfo  {journal}
  {Phys. Rev. Lett.}\ }\textbf {\bibinfo {volume} {128}},\ \bibinfo {pages}
  {185301} (\bibinfo {year} {2022})},\ \Eprint
  {http://arxiv.org/abs/2202.12355} {arXiv:2202.12355 [cond-mat.str-el]}
  \BibitemShut {NoStop}%
\bibitem [{\citenamefont {Fidkowski}\ and\ \citenamefont
  {Kitaev}(2011)}]{FidkowskifSPT1}%
  \BibitemOpen
  \bibfield  {author} {\bibinfo {author} {\bibfnamefont {Lukasz}\ \bibnamefont
  {Fidkowski}}\ and\ \bibinfo {author} {\bibfnamefont {Alexei}\ \bibnamefont
  {Kitaev}},\ }\bibfield  {title} {\enquote {\bibinfo {title} {Topological
  phases of fermions in one dimension},}\ }\href {\doibase
  10.1103/PhysRevB.83.075103} {\bibfield  {journal} {\bibinfo  {journal} {Phys.
  Rev. B}\ }\textbf {\bibinfo {volume} {83}},\ \bibinfo {pages} {075103}
  (\bibinfo {year} {2011})}\BibitemShut {NoStop}%
\bibitem [{\citenamefont {Fidkowski}\ and\ \citenamefont
  {Kitaev}(2010)}]{FidkowskifSPT2}%
  \BibitemOpen
  \bibfield  {author} {\bibinfo {author} {\bibfnamefont {Lukasz}\ \bibnamefont
  {Fidkowski}}\ and\ \bibinfo {author} {\bibfnamefont {Alexei}\ \bibnamefont
  {Kitaev}},\ }\bibfield  {title} {\enquote {\bibinfo {title} {Effects of
  interactions on the topological classification of free fermion systems},}\
  }\href {\doibase 10.1103/PhysRevB.81.134509} {\bibfield  {journal} {\bibinfo
  {journal} {Phys. Rev. B}\ }\textbf {\bibinfo {volume} {81}},\ \bibinfo
  {pages} {134509} (\bibinfo {year} {2010})}\BibitemShut {NoStop}%
\bibitem [{\citenamefont {Wen}(2013)}]{Wen2013ppa1305.1045}%
  \BibitemOpen
  \bibfield  {author} {\bibinfo {author} {\bibfnamefont {Xiao-Gang}\
  \bibnamefont {Wen}},\ }\bibfield  {title} {\enquote {\bibinfo {title} {{A
  lattice non-perturbative definition of an SO(10) chiral gauge theory and its
  induced standard model}},}\ }\href {\doibase 10.1088/0256-307X/30/11/111101}
  {\bibfield  {journal} {\bibinfo  {journal} {Chin. Phys. Lett.}\ }\textbf
  {\bibinfo {volume} {30}},\ \bibinfo {pages} {111101} (\bibinfo {year}
  {2013})},\ \Eprint {http://arxiv.org/abs/1305.1045} {arXiv:1305.1045
  [hep-lat]} \BibitemShut {NoStop}%
\bibitem [{\citenamefont {Wang}\ and\ \citenamefont
  {Wen}(2020)}]{WangWen2018cai1809.11171}%
  \BibitemOpen
  \bibfield  {author} {\bibinfo {author} {\bibfnamefont {Juven}\ \bibnamefont
  {Wang}}\ and\ \bibinfo {author} {\bibfnamefont {Xiao-Gang}\ \bibnamefont
  {Wen}},\ }\bibfield  {title} {\enquote {\bibinfo {title} {{A Non-Perturbative
  Definition of the Standard Models}},}\ }\href {\doibase
  10.1103/PhysRevResearch.2.023356} {\bibfield  {journal} {\bibinfo  {journal}
  {Phys. Rev. Res.}\ }\textbf {\bibinfo {volume} {2}},\ \bibinfo {pages}
  {023356} (\bibinfo {year} {2020})},\ \Eprint
  {http://arxiv.org/abs/1809.11171} {arXiv:1809.11171 [hep-th]} \BibitemShut
  {NoStop}%
\bibitem [{\citenamefont {You}\ \emph {et~al.}(2014)\citenamefont {You},
  \citenamefont {BenTov},\ and\ \citenamefont
  {Xu}}]{You2014oaaYouBenTovXu1402.4151}%
  \BibitemOpen
  \bibfield  {author} {\bibinfo {author} {\bibfnamefont {Yizhuang}\
  \bibnamefont {You}}, \bibinfo {author} {\bibfnamefont {Yoni}\ \bibnamefont
  {BenTov}}, \ and\ \bibinfo {author} {\bibfnamefont {Cenke}\ \bibnamefont
  {Xu}},\ }\bibfield  {title} {\enquote {\bibinfo {title} {{Interacting
  Topological Superconductors and possible Origin of $16n$ Chiral Fermions in
  the Standard Model}},}\ }\href@noop {} {\  (\bibinfo {year} {2014})},\
  \Eprint {http://arxiv.org/abs/1402.4151} {arXiv:1402.4151 [cond-mat.str-el]}
  \BibitemShut {NoStop}%
\bibitem [{\citenamefont {You}\ and\ \citenamefont {Xu}(2015)}]{YX14124784}%
  \BibitemOpen
  \bibfield  {author} {\bibinfo {author} {\bibfnamefont {Yi-Zhuang}\
  \bibnamefont {You}}\ and\ \bibinfo {author} {\bibfnamefont {Cenke}\
  \bibnamefont {Xu}},\ }\bibfield  {title} {\enquote {\bibinfo {title}
  {Interacting topological insulator and emergent grand unified theory},}\
  }\href {\doibase 10.1103/physrevb.91.125147} {\bibfield  {journal} {\bibinfo
  {journal} {Phys. Rev. B}\ }\textbf {\bibinfo {volume} {91}},\ \bibinfo
  {pages} {125147} (\bibinfo {year} {2015})},\ \Eprint
  {http://arxiv.org/abs/1412.4784} {arXiv:1412.4784} \BibitemShut {NoStop}%
\bibitem [{\citenamefont {BenTov}(2015)}]{BenTov2014eea1412.0154}%
  \BibitemOpen
  \bibfield  {author} {\bibinfo {author} {\bibfnamefont {Yoni}\ \bibnamefont
  {BenTov}},\ }\bibfield  {title} {\enquote {\bibinfo {title} {{Fermion masses
  without symmetry breaking in two spacetime dimensions}},}\ }\href {\doibase
  10.1007/JHEP07(2015)034} {\bibfield  {journal} {\bibinfo  {journal} {JHEP}\
  }\textbf {\bibinfo {volume} {07}},\ \bibinfo {pages} {034} (\bibinfo {year}
  {2015})},\ \Eprint {http://arxiv.org/abs/1412.0154} {arXiv:1412.0154
  [cond-mat.str-el]} \BibitemShut {NoStop}%
\bibitem [{\citenamefont {BenTov}\ and\ \citenamefont
  {Zee}(2016)}]{BenTov2015graZee1505.04312}%
  \BibitemOpen
  \bibfield  {author} {\bibinfo {author} {\bibfnamefont {Yoni}\ \bibnamefont
  {BenTov}}\ and\ \bibinfo {author} {\bibfnamefont {A.}~\bibnamefont {Zee}},\
  }\bibfield  {title} {\enquote {\bibinfo {title} {{Origin of families and
  $SO(18)$ grand unification}},}\ }\href {\doibase 10.1103/PhysRevD.93.065036}
  {\bibfield  {journal} {\bibinfo  {journal} {Phys. Rev.}\ }\textbf {\bibinfo
  {volume} {D93}},\ \bibinfo {pages} {065036} (\bibinfo {year} {2016})},\
  \Eprint {http://arxiv.org/abs/1505.04312} {arXiv:1505.04312 [hep-th]}
  \BibitemShut {NoStop}%
\bibitem [{\citenamefont {Seiberg}(2019)}]{NSeiberg-Strings-2019-talk}%
  \BibitemOpen
  \bibfield  {author} {\bibinfo {author} {\bibfnamefont {Nathan}\ \bibnamefont
  {Seiberg}},\ }\bibfield  {title} {\enquote {\bibinfo {title} {{Thoughts About
  Quantum Field Theory}},}\ }\href@noop {} {\bibfield  {journal} {\bibinfo
  {journal} {(Talk at Strings 2019)}\ } (\bibinfo {year} {2019})}\BibitemShut
  {NoStop}%
\bibitem [{\citenamefont {Wang}(2020{\natexlab{a}})}]{JW2008.06499}%
  \BibitemOpen
  \bibfield  {author} {\bibinfo {author} {\bibfnamefont {Juven}\ \bibnamefont
  {Wang}},\ }\bibfield  {title} {\enquote {\bibinfo {title} {{Anomaly and
  Cobordism Constraints Beyond Grand Unification: Energy Hierarchy}},}\
  }\href@noop {} {\  (\bibinfo {year} {2020}{\natexlab{a}})},\ \Eprint
  {http://arxiv.org/abs/2008.06499} {arXiv:2008.06499 [hep-th]} \BibitemShut
  {NoStop}%
\bibitem [{\citenamefont {Razamat}\ and\ \citenamefont
  {Tong}(2021)}]{RazamatTong2009.05037}%
  \BibitemOpen
  \bibfield  {author} {\bibinfo {author} {\bibfnamefont {Shlomo~S.}\
  \bibnamefont {Razamat}}\ and\ \bibinfo {author} {\bibfnamefont {David}\
  \bibnamefont {Tong}},\ }\bibfield  {title} {\enquote {\bibinfo {title}
  {{Gapped Chiral Fermions}},}\ }\href {\doibase 10.1103/PhysRevX.11.011063}
  {\bibfield  {journal} {\bibinfo  {journal} {Phys. Rev. X}\ }\textbf {\bibinfo
  {volume} {11}},\ \bibinfo {pages} {011063} (\bibinfo {year} {2021})},\
  \Eprint {http://arxiv.org/abs/2009.05037} {arXiv:2009.05037 [hep-th]}
  \BibitemShut {NoStop}%
\bibitem [{\citenamefont {Tong}(2022)}]{Tong2104.03997}%
  \BibitemOpen
  \bibfield  {author} {\bibinfo {author} {\bibfnamefont {David}\ \bibnamefont
  {Tong}},\ }\bibfield  {title} {\enquote {\bibinfo {title} {{Comments on
  symmetric mass generation in 2d and 4d}},}\ }\href {\doibase
  10.1007/JHEP07(2022)001} {\bibfield  {journal} {\bibinfo  {journal} {JHEP}\
  }\textbf {\bibinfo {volume} {07}},\ \bibinfo {pages} {001} (\bibinfo {year}
  {2022})},\ \Eprint {http://arxiv.org/abs/2104.03997} {arXiv:2104.03997
  [hep-th]} \BibitemShut {NoStop}%
\bibitem [{\citenamefont {Wang}\ and\ \citenamefont
  {You}(2022{\natexlab{b}})}]{Wang2106.16248}%
  \BibitemOpen
  \bibfield  {author} {\bibinfo {author} {\bibfnamefont {Juven}\ \bibnamefont
  {Wang}}\ and\ \bibinfo {author} {\bibfnamefont {Yi-Zhuang}\ \bibnamefont
  {You}},\ }\bibfield  {title} {\enquote {\bibinfo {title} {{Gauge enhanced
  quantum criticality beyond the standard model}},}\ }\href {\doibase
  10.1103/PhysRevD.106.025013} {\bibfield  {journal} {\bibinfo  {journal}
  {Phys. Rev. D}\ }\textbf {\bibinfo {volume} {106}},\ \bibinfo {pages}
  {025013} (\bibinfo {year} {2022}{\natexlab{b}})},\ \Eprint
  {http://arxiv.org/abs/2106.16248} {arXiv:2106.16248 [hep-th]} \BibitemShut
  {NoStop}%
\bibitem [{\citenamefont {Freed}\ and\ \citenamefont
  {Hopkins}(2021)}]{2016arXiv160406527F}%
  \BibitemOpen
  \bibfield  {author} {\bibinfo {author} {\bibfnamefont {Daniel~S.}\
  \bibnamefont {Freed}}\ and\ \bibinfo {author} {\bibfnamefont {Michael~J.}\
  \bibnamefont {Hopkins}},\ }\bibfield  {title} {\enquote {\bibinfo {title}
  {{Reflection positivity and invertible topological phases}},}\ }\href
  {\doibase 10.2140/gt.2021.25.1165} {\bibfield  {journal} {\bibinfo  {journal}
  {Geom. Topol.}\ }\textbf {\bibinfo {volume} {25}},\ \bibinfo {pages}
  {1165--1330} (\bibinfo {year} {2021})},\ \Eprint
  {http://arxiv.org/abs/1604.06527} {arXiv:1604.06527 [hep-th]} \BibitemShut
  {NoStop}%
\bibitem [{\citenamefont {Laughlin}(1981)}]{Laughlin1981PRB}%
  \BibitemOpen
  \bibfield  {author} {\bibinfo {author} {\bibfnamefont {R.~B.}\ \bibnamefont
  {Laughlin}},\ }\bibfield  {title} {\enquote {\bibinfo {title} {{Quantized
  Hall conductivity in two-dimensions}},}\ }\href {\doibase
  10.1103/PhysRevB.23.5632} {\bibfield  {journal} {\bibinfo  {journal} {Phys.
  Rev. B}\ }\textbf {\bibinfo {volume} {23}},\ \bibinfo {pages} {5632--5733}
  (\bibinfo {year} {1981})}\BibitemShut {NoStop}%
\bibitem [{\citenamefont {Santos}\ and\ \citenamefont
  {Wang}(2014)}]{SantosWang1310.8291}%
  \BibitemOpen
  \bibfield  {author} {\bibinfo {author} {\bibfnamefont {Luiz~H.}\ \bibnamefont
  {Santos}}\ and\ \bibinfo {author} {\bibfnamefont {Juven}\ \bibnamefont
  {Wang}},\ }\bibfield  {title} {\enquote {\bibinfo {title}
  {{Symmetry-protected many-body Aharonov-Bohm effect}},}\ }\href {\doibase
  10.1103/PhysRevB.89.195122} {\bibfield  {journal} {\bibinfo  {journal} {Phys.
  Rev. B}\ }\textbf {\bibinfo {volume} {89}},\ \bibinfo {pages} {195122}
  (\bibinfo {year} {2014})},\ \Eprint {http://arxiv.org/abs/1310.8291}
  {arXiv:1310.8291 [quant-ph]} \BibitemShut {NoStop}%
\bibitem [{\citenamefont {Kapustin}(2014)}]{Kapustin2014tfa1403.1467}%
  \BibitemOpen
  \bibfield  {author} {\bibinfo {author} {\bibfnamefont {Anton}\ \bibnamefont
  {Kapustin}},\ }\bibfield  {title} {\enquote {\bibinfo {title} {{Symmetry
  Protected Topological Phases, Anomalies, and Cobordisms: Beyond Group
  Cohomology}},}\ }\href@noop {} {\  (\bibinfo {year} {2014})},\ \Eprint
  {http://arxiv.org/abs/1403.1467} {arXiv:1403.1467 [cond-mat.str-el]}
  \BibitemShut {NoStop}%
\bibitem [{\citenamefont {Kapustin}\ \emph {et~al.}(2015)\citenamefont
  {Kapustin}, \citenamefont {Thorngren}, \citenamefont {Turzillo},\ and\
  \citenamefont {Wang}}]{Kapustin1406.7329}%
  \BibitemOpen
  \bibfield  {author} {\bibinfo {author} {\bibfnamefont {Anton}\ \bibnamefont
  {Kapustin}}, \bibinfo {author} {\bibfnamefont {Ryan}\ \bibnamefont
  {Thorngren}}, \bibinfo {author} {\bibfnamefont {Alex}\ \bibnamefont
  {Turzillo}}, \ and\ \bibinfo {author} {\bibfnamefont {Zitao}\ \bibnamefont
  {Wang}},\ }\bibfield  {title} {\enquote {\bibinfo {title} {{Fermionic
  Symmetry Protected Topological Phases and Cobordisms}},}\ }\href {\doibase
  10.1007/JHEP12(2015)052} {\bibfield  {journal} {\bibinfo  {journal} {JHEP}\
  }\textbf {\bibinfo {volume} {12}},\ \bibinfo {pages} {052} (\bibinfo {year}
  {2015})},\ \bibinfo {note} {[JHEP12,052(2015)]},\ \Eprint
  {http://arxiv.org/abs/1406.7329} {arXiv:1406.7329 [cond-mat.str-el]}
  \BibitemShut {NoStop}%
\bibitem [{\citenamefont {Wan}\ and\ \citenamefont
  {Wang}(2019)}]{WanWang2018bns1812.11967}%
  \BibitemOpen
  \bibfield  {author} {\bibinfo {author} {\bibfnamefont {Zheyan}\ \bibnamefont
  {Wan}}\ and\ \bibinfo {author} {\bibfnamefont {Juven}\ \bibnamefont {Wang}},\
  }\bibfield  {title} {\enquote {\bibinfo {title} {{Higher Anomalies, Higher
  Symmetries, and Cobordisms I: Classification of Higher-Symmetry-Protected
  Topological States and Their Boundary Fermionic/Bosonic Anomalies via a
  Generalized Cobordism Theory}},}\ }\href {\doibase
  10.4310/AMSA.2019.v4.n2.a2} {\bibfield  {journal} {\bibinfo  {journal} {Ann.
  Math. Sci. Appl.}\ }\textbf {\bibinfo {volume} {4}},\ \bibinfo {pages}
  {107--311} (\bibinfo {year} {2019})},\ \Eprint
  {http://arxiv.org/abs/1812.11967} {arXiv:1812.11967 [hep-th]} \BibitemShut
  {NoStop}%
\bibitem [{\citenamefont {Wang}(2020{\natexlab{b}})}]{JW2006.16996}%
  \BibitemOpen
  \bibfield  {author} {\bibinfo {author} {\bibfnamefont {Juven}\ \bibnamefont
  {Wang}},\ }\bibfield  {title} {\enquote {\bibinfo {title} {{Anomaly and
  Cobordism Constraints Beyond the Standard Model: Topological Force}},}\
  }\href@noop {} {\  (\bibinfo {year} {2020}{\natexlab{b}})},\ \Eprint
  {http://arxiv.org/abs/2006.16996} {arXiv:2006.16996 [hep-th]} \BibitemShut
  {NoStop}%
\bibitem [{\citenamefont {Wang}(2021)}]{JW2012.15860}%
  \BibitemOpen
  \bibfield  {author} {\bibinfo {author} {\bibfnamefont {Juven}\ \bibnamefont
  {Wang}},\ }\bibfield  {title} {\enquote {\bibinfo {title} {{Ultra
  Unification}},}\ }\href {\doibase 10.1103/PhysRevD.103.105024} {\bibfield
  {journal} {\bibinfo  {journal} {Phys. Rev. D}\ }\textbf {\bibinfo {volume}
  {103}},\ \bibinfo {pages} {105024} (\bibinfo {year} {2021})},\ \Eprint
  {http://arxiv.org/abs/2012.15860} {arXiv:2012.15860 [hep-th]} \BibitemShut
  {NoStop}%
\bibitem [{\citenamefont {Coleman}(1973)}]{Coleman1973Goldstone}%
  \BibitemOpen
  \bibfield  {author} {\bibinfo {author} {\bibfnamefont {Sidney~R.}\
  \bibnamefont {Coleman}},\ }\bibfield  {title} {\enquote {\bibinfo {title}
  {{There are no Goldstone bosons in two-dimensions}},}\ }\href {\doibase
  10.1007/BF01646487} {\bibfield  {journal} {\bibinfo  {journal} {Commun. Math.
  Phys.}\ }\textbf {\bibinfo {volume} {31}},\ \bibinfo {pages} {259--264}
  (\bibinfo {year} {1973})}\BibitemShut {NoStop}%
\bibitem [{\citenamefont {Mermin}\ and\ \citenamefont
  {Wagner}(1966)}]{MerminWagner1966PhysRevLett}%
  \BibitemOpen
  \bibfield  {author} {\bibinfo {author} {\bibfnamefont {N.~D.}\ \bibnamefont
  {Mermin}}\ and\ \bibinfo {author} {\bibfnamefont {H.}~\bibnamefont
  {Wagner}},\ }\bibfield  {title} {\enquote {\bibinfo {title} {{Absence of
  ferromagnetism or antiferromagnetism in one-dimensional or two-dimensional
  isotropic Heisenberg models}},}\ }\href {\doibase
  10.1103/PhysRevLett.17.1133} {\bibfield  {journal} {\bibinfo  {journal}
  {Phys. Rev. Lett.}\ }\textbf {\bibinfo {volume} {17}},\ \bibinfo {pages}
  {1133--1136} (\bibinfo {year} {1966})}\BibitemShut {NoStop}%
\bibitem [{\citenamefont {Hohenberg}(1967)}]{Hohenberg1967PhysRev}%
  \BibitemOpen
  \bibfield  {author} {\bibinfo {author} {\bibfnamefont {P.~C.}\ \bibnamefont
  {Hohenberg}},\ }\bibfield  {title} {\enquote {\bibinfo {title} {{Existence of
  Long-Range Order in One and Two Dimensions}},}\ }\href {\doibase
  10.1103/PhysRev.158.383} {\bibfield  {journal} {\bibinfo  {journal} {Phys.
  Rev.}\ }\textbf {\bibinfo {volume} {158}},\ \bibinfo {pages} {383--386}
  (\bibinfo {year} {1967})}\BibitemShut {NoStop}%
\bibitem [{\citenamefont {{Haldane}}(1995)}]{Haldane1995Stability}%
  \BibitemOpen
  \bibfield  {author} {\bibinfo {author} {\bibfnamefont {F.~D.~M.}\
  \bibnamefont {{Haldane}}},\ }\bibfield  {title} {\enquote {\bibinfo {title}
  {{Stability of Chiral Luttinger Liquids and Abelian Quantum Hall States}},}\
  }\href {\doibase 10.1103/PhysRevLett.74.2090} {\bibfield  {journal} {\bibinfo
   {journal} {Phys. Rev. Lett}\ }\textbf {\bibinfo {volume} {74}},\ \bibinfo
  {pages} {2090--2093} (\bibinfo {year} {1995})},\ \Eprint
  {http://arxiv.org/abs/cond-mat/9501007} {arXiv:cond-mat/9501007 [cond-mat]}
  \BibitemShut {NoStop}%
\bibitem [{\citenamefont {{Kapustin}}\ and\ \citenamefont
  {{Saulina}}(2011)}]{KapustinSaulina1008.0654KS}%
  \BibitemOpen
  \bibfield  {author} {\bibinfo {author} {\bibfnamefont {A.}~\bibnamefont
  {{Kapustin}}}\ and\ \bibinfo {author} {\bibfnamefont {N.}~\bibnamefont
  {{Saulina}}},\ }\bibfield  {title} {\enquote {\bibinfo {title} {{Topological
  boundary conditions in abelian Chern-Simons theory}},}\ }\href {\doibase
  10.1016/j.nuclphysb.2010.12.017} {\bibfield  {journal} {\bibinfo  {journal}
  {Nuclear Physics B}\ }\textbf {\bibinfo {volume} {845}},\ \bibinfo {pages}
  {393--435} (\bibinfo {year} {2011})},\ \Eprint
  {http://arxiv.org/abs/1008.0654} {arXiv:1008.0654 [hep-th]} \BibitemShut
  {NoStop}%
\bibitem [{\citenamefont {{Wang}}\ and\ \citenamefont
  {{Wen}}(2015)}]{Wang2015Boundary}%
  \BibitemOpen
  \bibfield  {author} {\bibinfo {author} {\bibfnamefont {Juven~C.}\
  \bibnamefont {{Wang}}}\ and\ \bibinfo {author} {\bibfnamefont {Xiao-Gang}\
  \bibnamefont {{Wen}}},\ }\bibfield  {title} {\enquote {\bibinfo {title}
  {{Boundary degeneracy of topological order}},}\ }\href {\doibase
  10.1103/PhysRevB.91.125124} {\bibfield  {journal} {\bibinfo  {journal} {Phys.
  Rev. B}\ }\textbf {\bibinfo {volume} {91}},\ \bibinfo {eid} {125124}
  (\bibinfo {year} {2015})},\ \Eprint {http://arxiv.org/abs/1212.4863}
  {arXiv:1212.4863 [cond-mat.str-el]} \BibitemShut {NoStop}%
\bibitem [{\citenamefont {Levin}(2013)}]{Levin1301.7355}%
  \BibitemOpen
  \bibfield  {author} {\bibinfo {author} {\bibfnamefont {Michael}\ \bibnamefont
  {Levin}},\ }\bibfield  {title} {\enquote {\bibinfo {title} {{Protected edge
  modes without symmetry}},}\ }\href {\doibase 10.1103/PhysRevX.3.021009}
  {\bibfield  {journal} {\bibinfo  {journal} {Phys. Rev.}\ }\textbf {\bibinfo
  {volume} {X3}},\ \bibinfo {pages} {021009} (\bibinfo {year} {2013})},\
  \Eprint {http://arxiv.org/abs/1301.7355} {arXiv:1301.7355 [cond-mat.str-el]}
  \BibitemShut {NoStop}%
\bibitem [{\citenamefont {{Chen}}\ \emph {et~al.}(2013)\citenamefont {{Chen}},
  \citenamefont {{Giedt}},\ and\ \citenamefont
  {{Poppitz}}}]{Chen2013Model3451211.6947}%
  \BibitemOpen
  \bibfield  {author} {\bibinfo {author} {\bibfnamefont {Chen}\ \bibnamefont
  {{Chen}}}, \bibinfo {author} {\bibfnamefont {Joel}\ \bibnamefont {{Giedt}}},
  \ and\ \bibinfo {author} {\bibfnamefont {Erich}\ \bibnamefont {{Poppitz}}},\
  }\bibfield  {title} {\enquote {\bibinfo {title} {{On the decoupling of mirror
  fermions}},}\ }\href {\doibase 10.1007/JHEP04(2013)131} {\bibfield  {journal}
  {\bibinfo  {journal} {Journal of High Energy Physics}\ }\textbf {\bibinfo
  {volume} {2013}},\ \bibinfo {eid} {131} (\bibinfo {year} {2013})},\ \Eprint
  {http://arxiv.org/abs/1211.6947} {arXiv:1211.6947 [hep-lat]} \BibitemShut
  {NoStop}%
\bibitem [{\citenamefont {Kaplan}(1992)}]{Kaplan1992A-method}%
  \BibitemOpen
  \bibfield  {author} {\bibinfo {author} {\bibfnamefont {David~B.}\
  \bibnamefont {Kaplan}},\ }\bibfield  {title} {\enquote {\bibinfo {title} {A
  method for simulating chiral fermions on the lattice},}\ }\href {\doibase
  https://doi.org/10.1016/0370-2693(92)91112-M} {\bibfield  {journal} {\bibinfo
   {journal} {Physics Letters B}\ }\textbf {\bibinfo {volume} {288}},\ \bibinfo
  {pages} {342--347} (\bibinfo {year} {1992})}\BibitemShut {NoStop}%
\bibitem [{\citenamefont {Callan}\ \emph {et~al.}(1977)\citenamefont {Callan},
  \citenamefont {Dashen},\ and\ \citenamefont {Gross}}]{CallanDashenGross1977}%
  \BibitemOpen
  \bibfield  {author} {\bibinfo {author} {\bibfnamefont {Curtis~G.}\
  \bibnamefont {Callan}, \bibfnamefont {Jr.}}, \bibinfo {author} {\bibfnamefont
  {Roger~F.}\ \bibnamefont {Dashen}}, \ and\ \bibinfo {author} {\bibfnamefont
  {David~J.}\ \bibnamefont {Gross}},\ }\bibfield  {title} {\enquote {\bibinfo
  {title} {{Instantons and Massless Fermions in Two-Dimensions}},}\ }\href
  {\doibase 10.1103/PhysRevD.16.2526} {\bibfield  {journal} {\bibinfo
  {journal} {Phys. Rev. D}\ }\textbf {\bibinfo {volume} {16}},\ \bibinfo
  {pages} {2526} (\bibinfo {year} {1977})}\BibitemShut {NoStop}%
\bibitem [{\citenamefont {Gross}\ and\ \citenamefont
  {Neveu}(1974)}]{GrossNeveu1974}%
  \BibitemOpen
  \bibfield  {author} {\bibinfo {author} {\bibfnamefont {David~J.}\
  \bibnamefont {Gross}}\ and\ \bibinfo {author} {\bibfnamefont {Andre}\
  \bibnamefont {Neveu}},\ }\bibfield  {title} {\enquote {\bibinfo {title}
  {{Dynamical Symmetry Breaking in Asymptotically Free Field Theories}},}\
  }\href {\doibase 10.1103/PhysRevD.10.3235} {\bibfield  {journal} {\bibinfo
  {journal} {Phys. Rev. D}\ }\textbf {\bibinfo {volume} {10}},\ \bibinfo
  {pages} {3235} (\bibinfo {year} {1974})}\BibitemShut {NoStop}%
\bibitem [{\citenamefont {Deligne}\ \emph {et~al.}(1999)\citenamefont
  {Deligne}, \citenamefont {Etingof}, \citenamefont {Freed}, \citenamefont
  {Jeffrey}, \citenamefont {Kazhdan}, \citenamefont {Morgan}, \citenamefont
  {Morrison},\ and\ \citenamefont {Witten}}]{Deligne1999qpQFTIAS}%
  \BibitemOpen
  \bibinfo {editor} {\bibfnamefont {P.}~\bibnamefont {Deligne}}, \bibinfo
  {editor} {\bibfnamefont {P.}~\bibnamefont {Etingof}}, \bibinfo {editor}
  {\bibfnamefont {D.~S.}\ \bibnamefont {Freed}}, \bibinfo {editor}
  {\bibfnamefont {L.~C.}\ \bibnamefont {Jeffrey}}, \bibinfo {editor}
  {\bibfnamefont {D.}~\bibnamefont {Kazhdan}}, \bibinfo {editor} {\bibfnamefont
  {J.~W.}\ \bibnamefont {Morgan}}, \bibinfo {editor} {\bibfnamefont {D.~R.}\
  \bibnamefont {Morrison}}, \ and\ \bibinfo {editor} {\bibfnamefont {Edward}\
  \bibnamefont {Witten}},\ eds.,\ \href@noop {} {\emph {\bibinfo {title}
  {{Quantum fields and strings: A course for mathematicians. Vol. 1, 2}}}}\
  (\bibinfo {year} {1999})\BibitemShut {NoStop}%
\bibitem [{\citenamefont {Hamada}\ and\ \citenamefont
  {Wang}(2022)}]{HamadaWang2209.15244}%
  \BibitemOpen
  \bibfield  {author} {\bibinfo {author} {\bibfnamefont {Yuta}\ \bibnamefont
  {Hamada}}\ and\ \bibinfo {author} {\bibfnamefont {Juven}\ \bibnamefont
  {Wang}},\ }\bibfield  {title} {\enquote {\bibinfo {title} {{Flavor Hierarchy
  from Smooth Confinement}},}\ }\href@noop {} {\  (\bibinfo {year} {2022})},\
  \Eprint {http://arxiv.org/abs/2209.15244} {arXiv:2209.15244 [hep-ph]}
  \BibitemShut {NoStop}%
\bibitem [{\citenamefont {Gaiotto}\ \emph {et~al.}(2015)\citenamefont
  {Gaiotto}, \citenamefont {Kapustin}, \citenamefont {Seiberg},\ and\
  \citenamefont {Willett}}]{Gaiotto2014kfa1412.5148}%
  \BibitemOpen
  \bibfield  {author} {\bibinfo {author} {\bibfnamefont {Davide}\ \bibnamefont
  {Gaiotto}}, \bibinfo {author} {\bibfnamefont {Anton}\ \bibnamefont
  {Kapustin}}, \bibinfo {author} {\bibfnamefont {Nathan}\ \bibnamefont
  {Seiberg}}, \ and\ \bibinfo {author} {\bibfnamefont {Brian}\ \bibnamefont
  {Willett}},\ }\bibfield  {title} {\enquote {\bibinfo {title} {{Generalized
  Global Symmetries}},}\ }\href {\doibase 10.1007/JHEP02(2015)172} {\bibfield
  {journal} {\bibinfo  {journal} {JHEP}\ }\textbf {\bibinfo {volume} {02}},\
  \bibinfo {pages} {172} (\bibinfo {year} {2015})},\ \Eprint
  {http://arxiv.org/abs/1412.5148} {arXiv:1412.5148 [hep-th]} \BibitemShut
  {NoStop}%
\bibitem [{\citenamefont {Coleman}(1975)}]{Coleman1974PRD}%
  \BibitemOpen
  \bibfield  {author} {\bibinfo {author} {\bibfnamefont {Sidney~R.}\
  \bibnamefont {Coleman}},\ }\bibfield  {title} {\enquote {\bibinfo {title}
  {{The Quantum Sine-Gordon Equation as the Massive Thirring Model}},}\ }\href
  {\doibase 10.1103/PhysRevD.11.2088} {\bibfield  {journal} {\bibinfo
  {journal} {Phys. Rev. D}\ }\textbf {\bibinfo {volume} {11}},\ \bibinfo
  {pages} {2088} (\bibinfo {year} {1975})}\BibitemShut {NoStop}%
\bibitem [{\citenamefont {Mandelstam}(1975)}]{Mandelstam1975PRD}%
  \BibitemOpen
  \bibfield  {author} {\bibinfo {author} {\bibfnamefont {S.}~\bibnamefont
  {Mandelstam}},\ }\bibfield  {title} {\enquote {\bibinfo {title} {{Soliton
  Operators for the Quantized Sine-Gordon Equation}},}\ }\href {\doibase
  10.1103/PhysRevD.11.3026} {\bibfield  {journal} {\bibinfo  {journal} {Phys.
  Rev. D}\ }\textbf {\bibinfo {volume} {11}},\ \bibinfo {pages} {3026}
  (\bibinfo {year} {1975})}\BibitemShut {NoStop}%
\bibitem [{\citenamefont {Floreanini}\ and\ \citenamefont
  {Jackiw}(1987)}]{FloreaniniJackiw1987PRL}%
  \BibitemOpen
  \bibfield  {author} {\bibinfo {author} {\bibfnamefont {R.}~\bibnamefont
  {Floreanini}}\ and\ \bibinfo {author} {\bibfnamefont {R.}~\bibnamefont
  {Jackiw}},\ }\bibfield  {title} {\enquote {\bibinfo {title} {{Selfdual Fields
  as Charge Density Solitons}},}\ }\href {\doibase 10.1103/PhysRevLett.59.1873}
  {\bibfield  {journal} {\bibinfo  {journal} {Phys. Rev. Lett.}\ }\textbf
  {\bibinfo {volume} {59}},\ \bibinfo {pages} {1873} (\bibinfo {year}
  {1987})}\BibitemShut {NoStop}%
\bibitem [{\citenamefont {Putrov}\ \emph {et~al.}(2017)\citenamefont {Putrov},
  \citenamefont {Wang},\ and\ \citenamefont
  {Yau}}]{Putrov2016qdo1612.09298PWY}%
  \BibitemOpen
  \bibfield  {author} {\bibinfo {author} {\bibfnamefont {Pavel}\ \bibnamefont
  {Putrov}}, \bibinfo {author} {\bibfnamefont {Juven}\ \bibnamefont {Wang}}, \
  and\ \bibinfo {author} {\bibfnamefont {Shing-Tung}\ \bibnamefont {Yau}},\
  }\bibfield  {title} {\enquote {\bibinfo {title} {{Braiding Statistics and
  Link Invariants of Bosonic/Fermionic Topological Quantum Matter in 2+1 and
  3+1 dimensions}},}\ }\href {\doibase 10.1016/j.aop.2017.06.019} {\bibfield
  {journal} {\bibinfo  {journal} {Annals Phys.}\ }\textbf {\bibinfo {volume}
  {384}},\ \bibinfo {pages} {254--287} (\bibinfo {year} {2017})},\ \Eprint
  {http://arxiv.org/abs/1612.09298} {arXiv:1612.09298 [cond-mat.str-el]}
  \BibitemShut {NoStop}%
\bibitem [{\citenamefont {Narain}\ \emph {et~al.}(1987)\citenamefont {Narain},
  \citenamefont {Sarmadi},\ and\ \citenamefont {Witten}}]{Narain:1986am}%
  \BibitemOpen
  \bibfield  {author} {\bibinfo {author} {\bibfnamefont {K.~S.}\ \bibnamefont
  {Narain}}, \bibinfo {author} {\bibfnamefont {M.~H.}\ \bibnamefont {Sarmadi}},
  \ and\ \bibinfo {author} {\bibfnamefont {Edward}\ \bibnamefont {Witten}},\
  }\bibfield  {title} {\enquote {\bibinfo {title} {{A Note on Toroidal
  Compactification of Heterotic String Theory}},}\ }\href {\doibase
  10.1016/0550-3213(87)90001-0} {\bibfield  {journal} {\bibinfo  {journal}
  {Nucl. Phys.}\ }\textbf {\bibinfo {volume} {B279}},\ \bibinfo {pages}
  {369--379} (\bibinfo {year} {1987})}\BibitemShut {NoStop}%
\end{thebibliography}%

\end{document}